\documentclass[a4paper, 11pt, oneside]{article}
\usepackage{cmap}			 

\usepackage{amsmath,amsfonts,amssymb,amsthm,mathtools,braket}
\usepackage{icomma}
\usepackage[shortcuts]{extdash}
\usepackage{physics}
\usepackage[cm]{fullpage}

\usepackage{euscript}	 
\usepackage{mathrsfs}    

\usepackage{enumitem}

\usepackage{caption}
\usepackage{graphicx}                  
\graphicspath{{images/}{images2/}}     
\usepackage{wrapfig}                   

\usepackage{array,tabularx,tabulary,booktabs} 
\usepackage{longtable}                        
\usepackage{multirow}                         

\usepackage{tikz}
\usetikzlibrary{graphs,graphs.standard,decorations.markings}
\tikzset{
    set arrow inside/.code={\pgfqkeys{/tikz/arrow inside}{#1}},
    set arrow inside={end/.initial=>, opt/.initial=},
    /pgf/decoration/Mark/.style={
        mark/.expanded=at position #1 with
        {
            \noexpand\arrow[\pgfkeysvalueof{/tikz/arrow inside/opt}]{\pgfkeysvalueof{/tikz/arrow inside/end}}
        }
    },
    arrow inside/.style 2 args={
        set arrow inside={#1},
        postaction={
            decorate,decoration={
                markings,Mark/.list={#2}
            }
        }
    },
}

\usepackage{float}   
\usepackage{caption} 
\usepackage{indentfirst} 

\usepackage[unicode, pdftex]{hyperref}

\usepackage{amsmath,amssymb}
\usepackage[nosort,noadjust]{cite}

\makeatletter
\global\@topnum\z@
\@afterindentfalse
\makeatother

\newcommand{\tnu}{{\tilde\nu}}
\newcommand{\tk}{{\tilde k}}
\newcommand{\tl}{{\tilde l}}

\def\eq#1$$#2$${\begin{equation#1}#2\end{equation#1}}
\long\def\subeq#1{\begin{subequations}#1\end{subequations}}
\def\Split$$#1$${\begin{split}#1\end{split}}
\def\Align#1$$#2$${\begin{align#1}#2\end{align#1}}
\def\AlignAt#1$$#2$${\begin{alignat}{#1}#2\end{alignat}}
\def\Aligned#1{\begin{aligned}{}#1\end{aligned}}

\def\Gather#1$$#2$${\begin{gather#1}#2\end{gather#1}}
\def\Gathered#1{\begin{gathered}{}#1\end{gathered}}

\def\Multline#1$$#2$${\begin{multline#1}#2\end{multline#1}}

\def\Cases#1{\begin{cases}#1\end{cases}}
\def\d{\partial}
\def\bd{\bar\partial}

\let\mathcal=\mathscr

\def\Tr{\mathop{\rm Tr}\nolimits}
\def\Re{\mathop{\rm Re}\nolimits}
\def\Im{\mathop{\rm Im}\nolimits}

\def\Res{\mathop{\rm Res}\limits}

\def\cC{{\mathcal C}}
\def\cF{{\mathcal F}}

\def\cC{{\mathcal C}}
\def\cG{{\mathcal G}}
\def\cH{{\mathcal H}}

\def\cJ{{\mathcal J}}

\def\cM{{\mathcal M}}

\def\cO{{\mathcal O}}
\def\cR{{\mathcal R}}
\def\cS{{\mathcal S}}
\def\cT{{\mathcal T}}

\def\cV{{\mathcal V}}

\def\cZ{{\mathcal Z}}
\def\hH{{\hat H}}
\def\hG{{\hat G}}
\def\hK{{\hat K}}
\def\hM{{\hat M}}

\def\tPhi{{\tilde\Phi}}
\def\ve{\varepsilon}

\def\Arg{\mathop{\rm Arg}\nolimits}
\def\tr{\mathop{\rm tr}\nolimits}
\def\lcolon{\mathopen{\,:}}
\def\rcolon{\mathclose{:\,}}
\def\C{{\mathbb C}}
\def\R{{\mathbb R}}

\def\Z{{\mathbb Z}}
\def\cD{{\mathcal D}}
\def\bfC{{\mathbf C}}
\def\rmS{{\mathrm S}}

\def\P{{\mathbf P}}
\def\Q{{\mathbf Q}}
\def\a{{\mathbf a}}
\def\ba{\bar{\mathbf a}}

\def\T{\mathop{\rm T}\nolimits}

\def\e{{\rm e}}
\def\i{{\rm i}}

\def\bz{{\bar z}}
\def\bw{{\bar w}}

\def\bzeta{{\bar\zeta}}

\def\lrd{\mathop{\mathord{\mathop\d\limits^{\lower.5ex\hbox{$\scriptstyle\leftrightarrow$}}}}\nolimits}

\def\upstrut{\vrule height 1.05\ht\strutbox depth 0pt width 0pt\relax}
\def\average#1{\overline{\upstrut\vphantom{{#1}^\cdot}#1}}
\def\ren{{\text{ren}}}

\makeatletter

\def\section{\@startsection{section}{1}{\z@}%
                                   {-3.5ex \@plus -1ex \@minus -.2ex}%
                                   {2.3ex \@plus.2ex}%
                                   {\normalfont\normalsize\bfseries}}
\def\subsection{\@startsection{subsection}{2}{\z@}%
                                     {-3.25ex\@plus -1ex \@minus -.2ex}%
                                     {1.5ex \@plus .2ex}%
                                     {\normalfont\normalsize\bfseries\itshape}}
\def\@seccntformat#1{\csname the#1\endcsname.~~}
\long\def\@makecaption#1#2{%
  \vskip\abovecaptionskip
  \sbox\@tempboxa{\small#1. #2}%
  \ifdim \wd\@tempboxa >0.9\hsize
  {\leftskip=0.05\hsize\rightskip=0.05\hsize\relax\small
    #1. #2\par}
  \else
    \global \@minipagefalse
    \hb@xt@\hsize{\hfil\box\@tempboxa\hfil}%
  \fi
  \vskip\belowcaptionskip}
\def\Appendix{\appendix
  \def\@seccntformat##1{\expandafter\ifx\csname c@##1\endcsname\c@section Appendix~\fi\csname the##1\endcsname.~~}}

\let\over\@@over
\let\atop\@@atop
\let\above\@@above
\let\overwithdelims\@@overwithdelims
\let\atopwithdelims\@@atopwithdelims
\let\abovewithdelims\@@abovewithdelims

\makeatother

\numberwithin{equation}{section} 

\begin{document}

\title{Form factors of composite branch-point twist operators in the sinh-Gordon model on a multi-sheeted Riemann surface:\\
semiclassical limit}

\author{Michael Lashkevich$^{a,b}$ and Amir Nesturov$^c$\\
	\parbox[t]{.8\textwidth}{\normalsize\it\raggedright
	\begin{itemize}\itemsep=\smallskipamount
		\item[$^a$]Landau Institute for Theoretical Physics, 142432 Chernogolovka, Russia
		\item[$^b$]Kharkevich Institute for Information Transmission Problems, 19 Bolshoy Karetny per., 127994 Moscow, Russia
		\item[$^c$]Department of Physics, Faculty of Mathematics and Physics, University of Ljubljana,
Jadranska 19, Ljubljana SI-1000, Slovenia
	\end{itemize}
		ML:~lashkevi@landau.ac.ru, AN:~amir.nesturov@fmf.uni-lj.si}
}
\date{}
\maketitle

\begin{abstract}
Quantum sinh-Gordon model in $1+1$ dimensions is one of the simplest and best-studied massive integrable relativistic quantum field theories. We consider this theory on a multi-sheeted Riemann surfaces with a flat metric, which can be seen as a pile of planes connected to each other along cut lines. The cut lines end at branch points, which are represented by a twist operator $\cT_n$. Operators of such kind are interesting in the framework of the problem of computing von Neumann and Renyi entanglement entropies in the original model on the plane. The composite branch\-/point twist operators (CTO) are a natural generalization of the twist operators, obtained by placing a local operator to a branch point by means of a certain limiting procedure.

Correlation function in quantum field theory can be, in principle, found by means of the spectral decomposition. It allows one to express them in terms of form factors of local operators, i.e.\ their matrix elements in the basis of stationary states. In integrable models complete sets of exact form factors of all operators can be found exactly as solutions of a system of bootstrap equations. Nevertheless, identification of these solution to the operators in terms of the basic fields remains problematic. In this work, we develop a technique of computing form factors of a class of CTO determined in terms of the basic field in the semiclassical approximation.
\end{abstract}

\tableofcontents{}

\section{Introduction}
\label{sec:Introduction}

Any operator is uniquely defined by a complete set of its matrix elements with respect to the eigenstates of the Hamiltonian. In the case of a relativistic quantum field theory such eigenstates are characterized by sets of types, internal states and momenta of stable particles in the corresponding in\-/state. The matrix elements of local operators can be analytically continued to functions of complex momenta called form factors. Here we consider a model in the $(1+1)$\-/dimensional Minkowski space with one neutral particle of mass $m$ without internal states. In this case the form factor of a local operator $\cO(x)$ can be defined as
\begin{equation}
F_\cO(\theta_1,\ldots,\theta_N)=\langle{\rm vac}|\cO(0)|\theta_1,\ldots,\theta_N\rangle,
\label{ff-def}
\end{equation}
where $\theta_i$ ($\theta_1>\theta_2>\cdots>\theta_N$) is the rapidity defined in terms of the space\-/time momenta $p_i$ according to
\eq$$
p_i^0=m\cosh\theta_i,\qquad p_i^1=m\sinh\theta_i.
\label{theta-i-def}
$$
Due to the crossing symmetry and translational invariance a general matrix element is given by
\begin{equation}
\langle\theta'_1,\ldots,\theta'_M|\cO(x)|\theta_1,\ldots,\theta_N\rangle
=\e^{-\i x\left(\sum^N_{i=1}p_i-\sum^M_{i=1}p'_i\right)}
F_\cO(\theta_1,\ldots,\theta_N,\theta'_M-\i\pi,\ldots,\theta'_1-\i\pi).
\label{matel-ff}
\end{equation}
Denote for briefness
\begin{equation}
\ket{\Theta}=\ket{\theta_1,\ldots,\theta_n}.
\label{s-state}
\end{equation}
Correlation functions of local operators can be expressed in terms of form factors by means of  the spectral decomposition. For example, the vacuum two\-/point correlation function of two operators $\cO_1$, $\cO_2$ can be written as
\begin{equation}\label{expansion}
\langle\cO_1(x_1)\cO_2(x_2)\rangle=\int d\Gamma_\Theta\,\langle{\rm vac}|\cO_1(0)|\Theta\rangle\langle\Theta|\cO_2(0)|{\rm vac}\rangle\,
\e^{-\i P_\Theta(x_1-x_2)},
\end{equation}
where $P_\Theta$ is the total momentum of the state $\Theta$ and $\Gamma_\Theta$ is the measure in the space of states:
\begin{equation}
\int d\Gamma_\Theta=\sum^\infty_{N=0}{1\over N!}\int_{\R^N}{d^N\theta\over(2\pi)^N}.
\label{dGammas-def}
\end{equation}
In fact, the series converges fast at large distances, that is as the interval between the operators is much greater than the inverse mass of the particle, where the exponent in the integrand is fast oscillating (or decreasing in the Euclidean space). The smaller the distances are considered the more terms in the spectral decomposition should be taken into account, which demands larger computational resources for numerical calculations.

The form factor formalism is applicable to the problem of numerical calculation of the entanglement entropy. Imagine splitting the space axis into two subsets: the segment $A=[a,\,b]$ and its complement $B$. The space of states $\cH$ of the system can be decomposed into the tensor product $\cH_A\otimes\cH_B$ of spaces of states in the subsets $A$ and $B$. Then for the vacuum state $|0\rangle$ of the system the density matrix in the region $A$ can be defined:
\begin{equation}
\hat\rho_A=\Tr_B|0\rangle\langle0|.
\label{rhoA-def}
\end{equation}
Then the Renyi entropy of entanglement between the systems $A$ and $B$ is equal to
\begin{equation}
S_n={\log\Tr\hat\rho_A^n\over1-n}={\log\qty[\cZ_n(A)/\cZ_1^n]\over1-n}\qquad \qty(n\in\mathbb N).
\label{Sn-entropy}
\end{equation}
Here $\cZ_1$ is the partition function of the theory and $\cZ_n(A)$ is the so called bipartite partition function. In \cite{Calabrese:2004eu}, this function was defined as a partition function of the field theory on an Euclidean $n$-sheeted Riemann surface with two branch points of $n$th order at the ends of the segment $A$. Later, in \cite{Cardy:2007mb}, it was shown that it coincides with the partition function of $n$ copies (replicas) of the field theory connected by special operators positioned at the ends of the segment $A$, the branch point twist operators (BPTO) $\cT_n$ and $\cT_n^+$.\footnote{We will refer to them as to ``operators'' rather than ``fields'', leaving the term ``fields'' to the functions that enter the action and to the corresponding ``elementary'' operators.} Let $\cO^{(s)}(x)$ be any local operator on the $s$th replica (modulo $n$) and consider the product $\cO^{(s)}(x)\cT_n(x_0)$. If $x$ goes round the point $x_0$ counterclockwise and returns to the initial point this product turns into $\cO^{(s+1)}(x)\cT_n(x_0)$. The operator $\cT^+_n$ acts in an opposite direction across the stack of replicas. The bipartite partition functions is proportional to the pair correlation function:
\begin{equation}
\cZ_n(A)\propto\langle\cT^+_n(a)\cT_n(b)\rangle.
\label{cZn-corrfun}
\end{equation}

A natural generalization of the BPTO are the so called composite twist operators (CTO), which are obtained by positioning usual local operators at the branch points. As well as usual local operators, CTOs can be defined by a complete system of its form factors. Calculation of the correlation functions of composite twist operators is an interesting and important task.

The two-dimensional quantum  sinh-Gordon model is a well\-/studied example of an integrable model with the only particle and a simple $S$ matrix. Integrability itself implies that the form factors satisfy a well\-/known set of functional equations~\cite{Karowski:1978vz,Smirnov:1984sx,Smirnov:1992vz} called form factor axioms. Every solution to these equations is a complete set of form factors that defines a unique local operator in the theory. For the sinh\-/Gordon model on the plane such solutions can be found in~\cite{Fring:1992pt,Koubek:1993ke,Lukyanov:1997bp,Babujian:2002fi,Feigin:2008hs}. In the multi\-/sheeted theory the set of form factors of any operator $\cO^{(s)}(x)$ off branch points coincides with the set of form factors of the operator $\cO(x)$ for particles on the sheet $s$ and of the unit operators for particles on all other sheets. At branch points the system of form factors is different and is not reduced to the form factors on the plane. Correspondingly, the system of form factor axioms is modified. First it was found in another context in \cite{Niedermaier:1998rs}. Later it was rediscovered and adjusted to the BPTOs and CTOs in\cite{Cardy:2007mb,Castro-Alvaredo:2008usl}. In \cite{Castro-Alvaredo:2008usl} the form factors of the exponential CTOs $\cT_n\e^{\alpha\varphi}$ were found exactly.

It is important to compare the form factors obtained in the framework of the bootstrap program with those obtained by more straightforward methods directly from the Lagrangian. In the sinh\-/Gordon model it is possible to obtain the form factors of the power operators $\varphi^n$ and light exponential operators $\e^{\alpha\varphi}$ with $\alpha\lesssim1$ in the framework of the standard perturbation theory for small values of the coupling constant $b\sim\hbar^{1/2}$. They were shown~\cite{DeLuca:2016etx,Konik:2020gdi} to fit perfectly the bootstrap results. Nevertheless, the validity of the standard perturbation theory breaks for the heavy exponential operators, i.e.\ the operators with $\alpha\sim b^{-1}$, and their Fock descendants. The descendant operators are operators that contain a prefactor, which is a polynomial of the space\-/time derivatives of the field $\varphi$. In \cite{Lashkevich:2023hzk} a semiclassical approach to their study was developed. It was shown that the heavy operators correspond to nontrivial backgrounds in the classical sinh\-/Gordon model, and the form factors were studied within the perturbation theory against these backgrounds. It turned out that though the form factors of the heavy exponential operators have finite classical limit related to the asymptotics of the backgrounds, the form factors of their Fock descendants are essentially quantum quantities even in the leading order of the perturbation theory. Moreover, the non\-/chiral descendant operators demand a renormalization procedure, which is essentially that introduced by Al.~Zamolodchikov\cite{Zamolodchikov:1989zs} in the framework of the conformal perturbation theory.

This work generalizes the results of \cite{Lashkevich:2023hzk} to the sinh\-/Gordon theory on multi\-/sheeted Riemann surfaces. We obtain the form factors of CTOs in the semiclassical approximation. We study the heavy exponential operators, i.e.\ the operators $\cT_n\e^{\alpha\varphi}$ with $\alpha=\nu n/b$, while $\nu$ remains finite, and their descendants. In this approximation the form factors of exponential operators are momenta\-/independent and, in this sense, trivial. The nontrivial result of our paper is the form factors for several sets of descendant operators. The more general setting of the problem demands a deeper study of the functions related to the background solutions, which, in turn, allows us to drastically simplify the calculations. The description of the renormalization procedure becomes more explicit and general, so that its self\-/consistency and conformance with the conformal perturbation theory become more convincing.

The work is organized as follows. In section \ref{sec:model} we first recall the construction of the branch\-/point twist operators and composite twist operators. Then we describe the sinh-Gordon model and its radial classical solution and explain the quantization in the background of the radial classical solution at a branch point. The properties of the sinh\-/Bessel functions, which play the key role in the quantization procedure, are collected in section~\ref{sec:sinh-Bessel}. In section~\ref{sec:formfactors} we calculate form factors of a set of operators.

\textbf{Supplemental Material.} We attach a file named ``\texttt{shBessel-small-t.nb}'' to this paper as a supplement. It is a Wolfram Mathematica$^\circledR$ notebook that contains the calculation of the coefficients $c^{(s,s')}_{\nu,k}$, $c^{(s,s')\pm}_{\nu,kl}$, $J^{\pm\pm(s,s')}_{\nu,kl}$ in the expansions of the functions used in this paper for the values of the integers superscripts $s,s'\ge0$, $s+s'\le4$. The file is commented to facilitate reproducing, extending and using the results.

\section{Description of the model}
\label{sec:model}

\subsection{Twist operators}
\begin{figure}[t]
\begin{center}
\begin{tikzpicture}[scale=.75]
\newcommand{\Replica}[2][]{
\begin{scope}[#1]
  \draw (0,0) -- ++(12,0) -- ++(4,2) -- ++(-12,0) -- cycle;
  \draw (4.9,0.95) -- ++(6,0) -- ++(0.2,0.1) -- ++ (-6,0) -- cycle;
  \path (1.7,0.3) node {#2};
  \fill[blue] (5,1) circle (0.08) +(0,-0.4) node[blue] {\small $\cT^+_n(a)$}
  ++ (6,0) circle (0.08) +(0,-0.4) node[blue] {\small $\cT_n(b)$};
\end{scope}
}

\newcommand{\Redline}[1][]{
\begin{scope}[#1]
\draw[red,thin,dashed] (5.1,1.05) .. controls (3.3,-0.4) and (7,5.4) .. (5.1,3.93);
\end{scope}
}

\newcommand{\Bigredline}[1][]{
\begin{scope}[#1]
\draw[red!50!white,thin,dashed]
  (4.9,0.94) .. controls (6.8,2.4) and (7.4,4.5) .. (7.2,9);
\end{scope}
}

\Replica[shift={(0,0)}]{\scriptsize{replica $0$}}
\Replica[shift={(0,3)}]{\scriptsize{replica $1$}}
\Replica[shift={(0,6)}]{\scriptsize{replica $2$}}

\Redline[shift={(0.2,0)}]
\Redline[shift={(0.2,3)}]
\Redline[shift={(0.2,6)}]
\Redline[shift={(5.6,0)}]
\Redline[shift={(5.6,3)}]
\Redline[shift={(5.6,6)}]
\Bigredline[shift={(0.4,0)}]
\Bigredline[shift={(5.8,0)}]
\end{tikzpicture}
\end{center}
\caption{Schematic representation of a multi\-/sheeted surface with two branch points $a$ and $b$.}\label{fig1}
\end{figure}
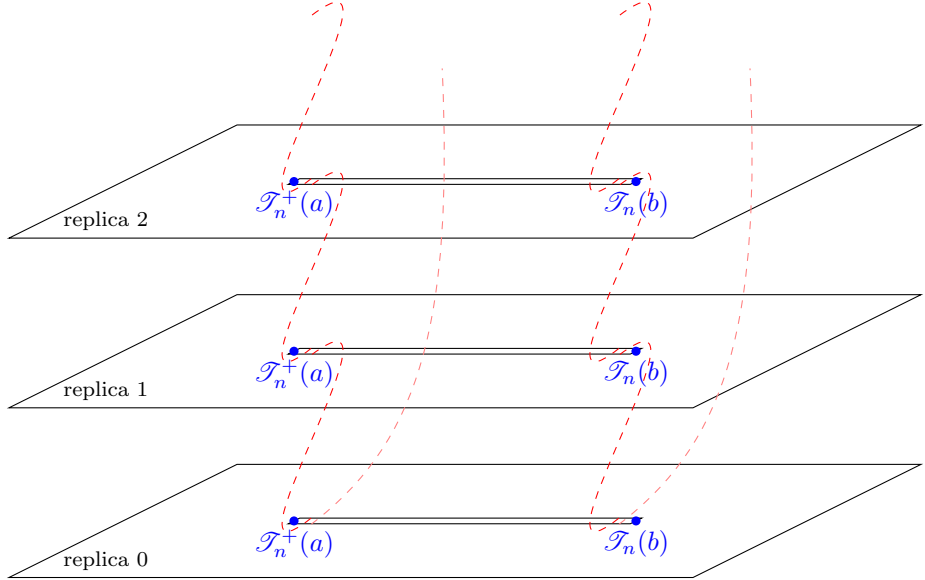

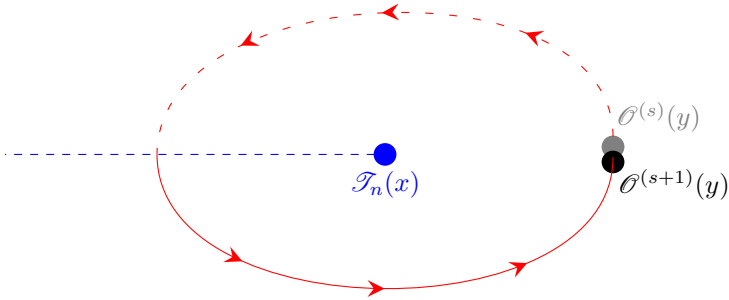
\begin{figure}[t]
\begin{center}
\begin{tikzpicture}
\begin{scope}[scale=1]
 \node (A) at (8,5.1) {};
 \node (B) at (8,4.9) {};
 \node (C) at (2,5) {};
 \node (T) at (5,5) {};
 \fill[blue] (T) circle(0.15) +(0,-0.4) node[blue] {\small $\cT_n(x)$};
 \fill[gray] (A) circle(0.15) +(0.6,0.4) node[gray] {\small $\cO^{(s)}(y)$};
 \fill[black] (B) circle(0.15) +(0.8,-0.3) node[black] {\small $\cO^{(s+1)}(y)$};
 \draw[blue,dashed] (T) -- +(-5,0);
 \draw[red, loosely dashed] (A) to[out=90,in=90] (C.center)
 [arrow inside={end=stealth,opt={red,scale=2}}{0.25,0.5,0.75}];
 \draw[red] (C.center) to[out=-90,in=-90] (A)
 [arrow inside={end=stealth,opt={red,scale=2}}{0.25,0.5,0.75}];
\end{scope}
\end{tikzpicture}
\end{center}
\caption{Dragging an operator $\cO$ around BPTO shifts it up one replica.}
\label{fig2}
\end{figure}

Consider a model on a Riemann surface with flat Euclidean metrics that consists of $n$ plane sheets connected along cut lines. In the simplest case of two branch points of the same multiplicity $n$ it is depicted in Figure~\ref{fig1}, but Riemann surfaces with multiple branch points of different multiplicities can be considered as well. Every cut has two ends with branch points at them. We can represent a field theory on such a surface as $n$ copies (replicas) of the underlying field theory on the plane with special operators, so called branch\-/point twist operators (BPTO) $\cT_n(x)$, $\cT^+_n(x)$ at the positions of the cuts' ends. Let $\cO(x)$ be any local operator in the theory on the plane. Denote as $\cO^{(s)}(x)$ its realization in the $s$th (modulo $n$) replica, assuming by default $s=0,\ldots,\,n-1$. We will often use the complex (or light\-/cone in the Minkowski space) coordinates:
$$
z=x^1+\i x^2=x^1-x^0,
\qquad
\bz=x^1-\i x^2=x^1+x^0,
$$
so that we will write $\cO(z,\bz)\equiv\cO(x)$. The operator $\cT_n(x)$ is defined as follows (see Figure~\ref{fig1}):
\eq$$
\Aligned{
\cO^{(s+1)}(z+w,\bz+\bw)\cT_n(z,\bz)
&=\cO^{(s)}(z+\e^{2\pi\i}w,\bz+\e^{-2\pi\i}\bw)\cT_n(z,\bz),
\\
\cO^{(s+1)}(z+w,\bz+\bw)\cT^+_n(z,\bz)
&=\cO^{(s)}(z+\e^{2\pi\i}z,\bz+\e^{-2\pi\i}\bw)\cT^+_n(z,\bz),
}\label{cOcT-circum}
$$
where the factors $\e^{\pm2\pi\i}$ mean that the operator is moved along a loop in the Euclidean plane around the point $x$ (see Figure~\ref{fig2}). In terms of simultaneous commutation relations this can be written as follows:
\begin{equation}
\cO^{(s)}(y)\cT_n(x)=
\begin{cases}
\cT_n(x)\cO^{(s+1)}(y),&\text{if $y^2=x^2$, $y^1<x^1$;} \\
\cT_n(x)\cO^{(s)}(y),&\text{if $y^2=x^2$, $y^1>x^1$,}
\end{cases}
\end{equation}
where the cut is assumed to be along the line $y^2=x^2$, $y^1<x^1$.

Now consider operators in the vicinity of the branch point $x$. Introduce the coordinates
\begin{equation}
\zeta=(z'-z)^{1/n},
\qquad
\bzeta=(\bz'-\bz)^{1/n}.
\label{zeta-coord-def}
\end{equation}
In a small vicinity of the point $x$ we may consider the theory as a conformal field theory. Thus one may define the operators on the $(\zeta,\bzeta)$ plane and take the limit $\zeta\to0$. Let $\Phi(\zeta,\bzeta)$ be a primary local operator of conformal dimensions $(\Delta,\bar\Delta)$. The transformation to the variables $(z',\bz')$ reads
\eq$$
\Phi(\zeta,\bzeta)=n^{\Delta+\bar\Delta}(z'-z)^{(1-1/n)\Delta}(\bz'-\bz)^{(1-1/n)\bar\Delta}\Phi^{(s)}(z',\bz'),
\label{Phi-transform}
$$
where the superscript $s$ depends on the argument of $\zeta$: $\pi{2s-1\over n}\le\Arg\zeta<\pi{2s+1\over n}$. Thus the primary composite twist operator (PCTO) can be defined as
\begin{equation}\label{CTF-general}
\cT_n\Phi(x)=n^{\Delta+\bar\Delta}\lim_{x'\to x}(z'-z)^{\qty(1-{1\over n})\Delta} \, (\bz'-\bz)^{\qty(1-{1\over n})\bar\Delta}
\e^{2\pi\i\qty(1-{1\over n})(\Delta-\bar\Delta)s}\cT_n(x)\Phi^{(s)}(x'),
\end{equation}
where the prefactor is assumed to be a shortest analytic continuation from the positive real semiaxis. With this assumption the definition is $s$\-/independent.

Below we will also use more general composite twist operators (CTOs), which will be defined as Fock descendants of the PCTOs. Let $\cO$ be any local operator and $\cT_n\cO$ be the corresponding CTO. Let $\varphi(x)$ be a neutral scalar field. In what follows it will be the basic field of the sinh\-/Gordon model. Define
\begin{equation}
\d^k\varphi(x')=\left(\d\over\d\zeta\right)^k\varphi(x'),
\qquad
\bd^k\varphi(x')=\left(\d\over\d\bzeta\right)^k\varphi(x').
\label{dkvarphi-def}
\end{equation}
Then the operators $\cT_n(\overline{\upstrut\d^k\varphi}_{r_0}\cO)(x)$, $\cT_n(\overline{\upstrut\bd^k\varphi}_{r_0}\cO)(x)$ are defined as
\eq$$
\Aligned{
\cT_n(\overline{\upstrut\d^k\varphi}_{r_0}\cO)(x)
&=\oint_{|\zeta|^n=r_0}{d\zeta\over2\pi\i\zeta}\,\d^k\varphi(\zeta,\bzeta)\,\cT_n\cO(x),
\\
\cT_n(\overline{\upstrut\bd^k\varphi}_{r_0}\cO)(x)
&=\oint_{|\zeta|^n=r_0}{d\zeta\over2\pi\i\zeta}\,\bd^k\varphi(\zeta,\bzeta)\,\cT_n\cO(x).
}\label{cT-dkvarphi-r0-def}
$$
The contours are assumed to be directed counterclockwise. Define the operators
\begin{equation}
\Aligned{
\cT_n(\d^k\varphi\,\cO)(x)
&=\left[\cT_n({\overline{\upstrut\d^k\varphi}_{r_0}}\cO)(x)+(\text{c.t.})\right]_{r_0\to0},
\\
\cT_n(\bd^k\varphi\,\cO)(x)
&=\left[\cT_n({\overline{\upstrut\bd^k\varphi}_{r_0}}\cO)(x)+(\text{c.t.})\right]_{r_0\to0}.
}\label{cT-dkvarphi-def}
\end{equation}
The renormalization counterterms $(\text{c.t.})$ are equal to zero in the conformal limit, but in the full theory consist of operators of lower scale dimensions that cancel contributions divergent in the limit $r_0\to0$. The renormalization construction is based on the conformal perturbation theory \cite{Zamolodchikov:1987ti} and will be described later for particular examples of operators.

The operators $\cT^+_n\cO(x)$ are defined in the same way with the only difference that in this case the shift to the upper replica corresponds to a clockwise round. Thus, the variables $\zeta$, $\bzeta$ should be defined as $\zeta=(\bz'-\bz)^{1/n}$, $\bzeta=(z'-z)^{1/n}$.

\subsection{Sinh\-/Gordon model, its classical limit and saddle\-/point approximation}
The sinh\-/Gordon model describes a real scalar field $\varphi(x)$ with the Euclidean action
\begin{equation}\label{action}
S[\varphi]={1\over8\pi}\int d^2 x\,\qty({(\d_\mu\varphi)^2\over2}+{m_0^2\over b^2}\cosh b\varphi).
\end{equation}
Here $m_0$ is a parameter of dimension $\text{(mass)}^{1+b^2}$, and $b$ is a non\-/dimensional constant. The only particle is associated with the field $\varphi$ and has the mass
\begin{equation}\label{particlemass}
m={4\sqrt\pi\over\Gamma\left(1+{p\over2}\right)\Gamma\left({1-p\over2}\right)} \left({\Gamma(1+b^2)\over\Gamma(1-b^2)}{m_0^2\over16}\right)^{(1-p)/2},
\qquad
p={b^2\over1+b^2}.
\end{equation}

The particle\-/particle $S$ matrix is given by
\eq$$
\cS(\theta)=-{\sin\pi p+\i\sinh\theta\over\sin\pi p-\i\sinh\theta},
\label{S-matrix}
$$
where $\theta=\theta_1-\theta_2$ is the difference of the rapidities of the colliding particles.

In the case of an $n$\-/sheeted Riemann manifold the integration in (\ref{action}) is assumed to be taken over the whole manifold with the plane metric. The model contains $n$ particles of the same mass, each lives on a certain sheet. If we consider two particles on the replicas $s$ and $s'$, they do not interact unless defined on the same sheet. Thus their $S$ matrix reads
\eq$$
\cS_{ss'}(\theta)=\Cases{\cS(\theta),&\text{if $s'=s$;}\\1,&\text{if $s'\ne s$.}}
\label{Sss-def}
$$

From now on we will consider the sinh\-/Gordon model on the $n$\-/sheeted Riemann surface $\cM_n$ with the only branch point $x=0$. To avoid repeated usage of the replica index $s$ we will assume that any operator $\cO(x)$ is understood as a function of polar coordinates $(r,\xi)$ with $0\le r<\infty$ and $\xi$ defined modulo $2\pi n$, so that
\begin{equation}
z=r\e^{\i\xi},
\qquad
\bz=r\e^{-\i\xi},
\qquad
\zeta=z^{1/n}=r^{1/n}\e^{\i\xi/n},
\qquad
\bzeta=\bz^{1/n}=r^{1/n}\e^{-\i\xi/n}.
\label{rxi-def}
\end{equation}
Thus the replica number $s$ is uniquely defined by the condition $\pi(2s-1)\le\xi<\pi(2s+1)\pmod{2\pi n}$. This corresponds to the cut along the negative real semiaxis. In other words we will understand the point $x$ as a point on the manifold $\cM_n$, and the integral in (\ref{action}) as an integral over $\cM_n$.

We will be interested in the correlation functions of the form
\eq$$
G^{(n)}_\cO(\{x_i\}_N)=\langle\varphi(x_N)\cdots\varphi(x_1)\cT_n\cO(0)\rangle,
\label{GcO-def}
$$
where $\cO$ is an arbitrary local operator and $\{x_i\}_N$ denotes the ordered set of the variables $x_1,\ldots,x_N$.

The spectrum of local operators in the sinh\-/Gordon theory consists of the exponential operators $\e^{\alpha\varphi}$ and their Fock descendants. The exponential operators on the plane are defined modulo the reflection property~\cite{Zamolodchikov:1995aa,Fateev:1997nn}
\eq$$
\e^{\alpha\varphi}=R_\alpha\e^{(Q-\alpha)\varphi}=R_{-\alpha}\e^{-(Q+\alpha)\varphi},
\qquad
Q=b+b^{-1},
\label{exp-reflection-plane}
$$
with a known reflection function $R_\alpha$. At the branch point the reflection property holds in a different form~\cite{Horvath:2021rjd}
\eq$$
\cT_n\e^{\alpha\varphi}=R^{(n)}_\alpha\cT_n\e^{(nQ-\alpha)\varphi}=R^{(n)}_{-\alpha}\cT_n\e^{-(nQ+\alpha)\varphi}
\label{exp-reflection-branchpoint}
$$
with certain (yet unknown) reflection functions $R^{(n)}_\alpha$. A proof of the reflection property in terms of the exact form factors will be given in our upcoming paper.

The Fock descendant operators are the operators of the form
\begin{equation}
\prod^K_{i=1}\d^{k_i}\varphi\prod^L_{j=1}\bd^{l_j}\varphi\,\e^{\alpha\varphi}.
\label{descendants-def}
\end{equation}
Here $K,\,L\ge0$ and $k_i,\,l_j>0$ are integers. Recall that at the branch points we assume $\d=\d/\d\zeta$, $\bd=\d/\d\bzeta$ and define descendants as in (\ref{cT-dkvarphi-r0-def}), (\ref{cT-dkvarphi-def}).

Let us start from the exponential operators. Define
\begin{equation}
\cT_nV_\nu(0)=\cG_\nu^{-1}\cT_n(\e^{b^{-1}n\nu\varphi})(0),
\qquad
\cG_\nu=\langle\cT_n(\e^{b^{-1}n\nu\varphi})\rangle.
\label{TVnu-def}
\end{equation}
We divide the exponential operator by its vacuum expectation value, since the problem of calculation of~$\cG_\nu$ is not the goal of the present paper.

Consider the correlation functions
\begin{equation}
G^{{(n)}}_\nu(\{x_i\}_N)=\langle\varphi(x_N)\cdots\varphi(x_1)\cT_nV_\nu(0)\rangle
={1\over\cZ}\int\cD\varphi\,\varphi(x_N)\cdots\varphi(x_1)\e^{-S[\varphi]+b^{-1}n\nu\varphi(0)},
\label{Gnu-def}
\end{equation}
where
\begin{equation}
\cZ=\int\cD\varphi\,\e^{-S[\varphi]+b^{-1}n\nu\varphi(0)}
\label{cZ-def}
\end{equation}
In the functional integral we assume that the field $\varphi$ is defined on the Riemann surface $\cM_n$.

In the present paper we consider the model for $b\ll0$. We may rescale the field $\varphi=b^{-1}\phi$. Then
\begin{equation}\label{classical-action}
S[b^{-1}\phi]={1\over8\pi b^2}\int d^2 x\,\qty({(\d_\mu\phi)^2\over2}+m^2\cosh\phi)={S_\text{cl}[\phi]\over b^2}.
\end{equation}
This means that $\hbar\sim b^2$ so that the limit $b\to0$ is the classical limit. Since in this limit $m=m_0$ we omit the subscript $0$ from now on. We will calculate all quantities in the leading approximation in the powers of $b$. Nevertheless, we will see that the nature of the objects we consider is essentially quantum, so that the resulting quantities involve quantum contributions.

For small values of $b$ the functional integral
\begin{equation}
G^{(n)}_\nu(\{x_i\}_N)={1\over b^N\cZ}\int\cD\phi\,\phi(x_N)\cdots\phi(x_1)\e^{-b^{-2}(S_\text{cl}[\phi]-n\nu\phi(0))}
\label{Gnu-phi}
\end{equation}
is governed by the saddle point defined by the equation
\eq$$
{\delta(S_\text{cl}[\phi]-n\nu\phi(0))\over\delta\phi(x)}=0.
\label{S-saddle-point}
$$
Explicitly it reads
\eq$$
\nabla^2\phi-m^2\sinh\phi=-8\pi n\nu\delta(x).
\label{shG-eq}
$$
We are interested in the radial solutions. Let us fix the asymptotics such solutions at small values of the radial coordinate $r$. Transform eq.~(\ref{shG-eq}) to the variables $\zeta=\tilde x^1+\i\tilde x^2$, $\bzeta=\tilde x^1-\i\tilde x^2$:
\eq$$
\tilde\nabla^2\phi-m^2n^2|\tilde x|^{2n-2}\sinh\phi=-8\pi n\nu\delta(\tilde x).
\label{shG-eq-zeta}
$$
For the asymptotics we may neglect the potential term, and thus we have
\eq$$
\phi=-4n\nu\log|\tilde x|+O(1)=-4\nu\log r+O(1)
\quad\text{as $r\to0$.}
\label{phi-small-r}
$$
The radial solution $\phi(x)=\phi_\nu(mr)$ with this property reads\cite{McCoy:1976cd,Cecotti:1992qh}
\begin{equation}
\phi_\nu(t)
=\sum^\infty_{N\in2\Z_{\ge0}+1}\lambda^N\phi^{(N)}(t),
\qquad
\lambda={\sin\pi\nu\over\pi},
\label{phinu-lambda-def}
\end{equation}
where
\begin{equation}
\phi^{(N)}(t)={4\over N}\int d^N\theta\,\prod^N_{i=1}{\e^{-t\cosh\theta_i}\over2\cosh{\theta_i-\theta_{i+1}\over2}}
\label{phi-r-theta}
\end{equation}
with $\theta_{N+1}=\theta_1$. We see that the function satisfies the `reflection' identities
\eq$$
\phi_\nu(t)=-\phi_{-\nu}(t)=\phi_{1-\nu}(t).
\label{phi-nu-refl}
$$
In fact, the asymptotic property (\ref{phi-small-r}) is valid in the region $-{1\over2}<\nu<{1\over2}$ only. In the region $0\le\nu\le1$ the asymptotics is given by
\begin{equation}\label{phinu-small}
\e^{\mp{1\over2}\phi_{\pm\nu}(t)}
=\beta_\nu t^{2\nu}+\beta_{1-\nu}t^{2-2\nu}+ O(t^2)
\quad\text{as $t\to0$,}
\end{equation}
where
\eq$$
\beta_\nu=2^{-6\nu}{\Gamma({1\over2}-\nu)\over\Gamma({1\over2}+\nu)}.
\label{beta-nu-def}
$$
It contains two concurrent contributions, and in the region $|\nu|<{1\over2}$ the first term dominates, while in the region ${1\over2}<|\nu|<1$ the second one dominates. There is no solution that have the asymptotic (\ref{phi-small-r}) with real $\nu$ such that $|\nu|>{1\over2}$. This is the classical counterpart of the reflection property (\ref{exp-reflection-plane}), (\ref{exp-reflection-branchpoint}). Thus we will always assume
\begin{equation}
|\nu|<\tfrac12.
\label{nu<1/2}
\end{equation}

We will also need the large $t$ asymptotic of the solution $\phi_\nu(t)$. It is dominated by the first term in the series (\ref{phi-r-theta}) and reads
\begin{equation}
\phi_\nu(t)\approx\sqrt{8\pi}\lambda t^{-1/2}\e^{-t}
\quad\text{as $t\to\infty$.}
\label{phi-nu-large-t}
\end{equation}

Now decompose the field $\varphi(x)$ into the classical background and quantum field $\chi(x)$:
\begin{equation}\label{decomposition}
\varphi(x)=b^{-1}\phi_\nu(mr)+\chi(x).
\end{equation}
So the next step is to expand the action (\ref{action}) in powers of the field $\chi$:
\begin{equation}
S[b^{-1}\phi_\nu+\chi]-S[b^{-1}\phi_\nu]=S_0[\chi]+S_{\text{int}}[\chi].
\label{action-chi-decomp}
\end{equation}
The first term in the r.h.s.\ is the quadratic contribution
\begin{equation}\label{deltaaction}
S_0[\chi]={1\over16\pi}\int d^2x\,\left((\d_\mu\chi)^2+m^2\chi^2\cosh\phi_\nu(mr)\right).
\end{equation}
The second term contains higher powers in $\chi$ and will be considered as an interaction action:
\begin{equation}\label{Sint}
\Gathered{
S_{\text{int}}[\chi]=\sum_{k=3}^\infty S^{(k)}[\chi],
\qquad
S^{(k)}[\chi]=b^{k-2}\int d^2x\,\chi^k(x) H^{(k)}_\nu(x),
\\
H^{(k)}_\nu(x)={m^2\over16\pi\,k!}\qty(\e^{\phi_\nu(mr)}+(-1)^k \e^{-\phi_\nu(mr)}).
}
\end{equation}
Each term $S^{(k)}$ is of the order $b^{k-2}$ and contains $k$th power of $\chi$, so that it is represented by a $k$\-/leg vertex in the Feynman diagram technique.

\subsection{Radial quantization}
\label{subsec:radquant}

The perturbation expansion based on the decomposition (\ref{action-chi-decomp})--(\ref{Sint}) is performed against the rotationally symmetric background of the radial solution $\phi_\nu(mr)$. Let us start from the linear approximation, which corresponds to the action $S_0$. In this approximation the field $\chi$ satisfies the equation of motion
\begin{equation}
\left(\d_r^2+r^{-1}\d_r+r^{-2}\d_\xi^2-m^2\cosh\phi_\nu(mr)\right)\chi(x)=0.
\label{chi-eom}
\end{equation}
The quantum field $\chi(x)$ should be represented as a decomposition in solutions to this equation with operator coefficients. Due to the rotational symmetry it is natural to decompose $\chi(x)$ in the eigenfunctions of the angular momentum.

In the vicinity of the origin this decomposition is known from the conformal field theory on the $(\zeta,\bzeta)$ plane in the radial quantization picture:
\begin{align}
\chi(x)
&\simeq\Q-2\i\P\log{\zeta\bzeta\over R^{2/n}}+\sum_{k\ne0}\left({\a_k\over\i k}\zeta^{-k}+{\ba_k\over\i k}\bzeta^{-k}\right)
\notag
\\
&=\Q-{4\i\over n}\P\log{r\over R}+\sum_{k\ne0}\left({\a_k\over\i k}\e^{-\i\tk\xi}+{\ba_k\over\i k}\e^{\i\tk\xi}\right)r^{-\tk}
\quad\text{as $r\to0$;}
\qquad
\tilde k={k\over n}.
\label{chi-decomp-assymp}
\end{align}
The sum is taken over integer values of $k$. We will often use the `tilde' notation as a synonym to dividing by $n$. The scale $R$ is arbitrary and drops out of all conformal correlation functions. The operators $\P$, $\Q$, $\a_k$, $\ba_k$ form a Heisenberg algebra with the commutation relations
\begin{equation}
[\P,\Q]=-\i,
\qquad
[\a_k,\a_l]=[\ba_k,\ba_l]=2k\delta_{k+l}.
\label{PQaba-commut}
\end{equation}
All other commutators are equal to zero. We will also need the vacuum states $|0\rangle_\nu$, $\langle\infty|$ defined as follows:
\begin{equation}
\begin{aligned}
&\P|0\rangle_\nu=\a_k|0\rangle_\nu=\ba_k|0\rangle_\nu=0,
\\
&\langle\infty|\Q=\langle\infty|\a_{-k}=\langle\infty|\ba_{-k}=0,
\qquad
k>0.
\end{aligned}
\label{vac-def}
\end{equation}
The subscript $\nu$ at the vacuum is not necessary mathematically, but it will be convenient to use it to mark the classical background.

In the radial quantization picture the radial variable $|\zeta|$ plays the role of imaginary time, while the angular variable $\Arg\zeta$ plays the role of a space coordinate. The infinite past is supposed to be at the point $\zeta=0$, while the infinite future is supposed to be at the point $\zeta=\infty$.

We extend the decomposition (\ref{chi-decomp-assymp}) to all distances in the form
\begin{equation}
\chi(x)=\Q f_0(x)+{4\i\over n}\P f_*(x)+\sum_{k\neq 0}\left({\a_k\over\i k}f_k(x)+{\ba_k\over\i k}\bar f_k(x)\right).
\label{chi-decomp}
\end{equation}
According to (\ref{chi-eom}), the functions $f_k(x)$ are solutions to the equation
\begin{equation}
\left(\d_r^2+r^{-1}\d_r+r^{-2}\d_\xi^2-m^2\cosh\phi_\nu(mr)\right)f_k(x)=0,
\qquad k\in\Z\cup\{*\},
\label{fn-equation}
\end{equation}
with fixed angular dependence:
\begin{equation}
f_k(x)=\e^{-\i\tk\xi}m^\tk u_k(mr),
\qquad
f_*(x)=u_*(mr).
\label{fk-form}
\end{equation}
We assume the asymptotic conditions
\begin{equation}
u_k(t)=t^{-\tk}(1+o(t)),
\qquad
u_*(t)=-\log t+O(1)
\quad\text{as $t\to0$}
\label{uk-t-small}
\end{equation}
to fit the expansion (\ref{chi-decomp-assymp}), and
\begin{equation}
u_k(t)=O\bigl(t^{-1/2}\e^{-t}\bigr)
\quad\text{for $k>0$ or $k=*$ as $t\to\infty$}
\label{uk-t-large}
\end{equation}
to ensure that correlation functions decrease properly at infinity. The functions $u_k(t)$ satisfy the sinh\-/Bessel equation~\cite{Lashkevich:2023hzk}
\begin{equation}
u''(t)+t^{-1}u'(t)-\left(\cosh\phi_\nu(t)+\kappa^2t^{-2}\right)u(t)=0,
\qquad
t>0,
\label{shB-equation}
\end{equation}
for
\begin{equation}
\kappa=\Cases{\tk,&\text{if $k\in\Z$;}\\0,&\text{if $k=*$.}}
\label{kappa-tk}
\end{equation}
This equation generalizes the Bessel equation and reduces to it for $\nu=0$.

In what follows we will also use the sinh\-/Bessel functions $K_{\nu,\kappa}(t)$, $I_{\nu,\kappa}(t)$, which are solutions to the sinh\-/Bessel equation (\ref{shB-equation}) with the asymptotic properties
\begin{equation}
K_{\nu,\kappa}(t)=\Cases{2^{\kappa-1}\Gamma(\kappa)t^{-\kappa}(1+o(1)),&\text{if $\kappa>0$;}\\
-\log t+O(1),&\text{if $\kappa=0$,}}
\qquad
I_{\nu,\kappa}(t)={t^\kappa\over2^\kappa\Gamma(\kappa+1)}(1+o(t))
\label{KI-t-small}
\end{equation}
as $t\to0$ and
\begin{equation}
K_{\nu,\kappa}(t)=\sqrt{\pi\over2}{\e^{-t}\over t^{1/2}}(K^\infty_{\nu,\kappa}+O(t^{-1})),
\qquad
I_{\nu,\kappa}(t)={1\over\sqrt{2\pi}}{\e^t\over t^{1/2}}\left((K^\infty_{\nu,\kappa})^{-1}+O(t^{-1})\right)
\label{KI-t-large}
\end{equation}
as $t\to\infty$. The Wronskian of these solutions is
\begin{equation}
K_{\nu,\kappa}(t)I'_{\nu,\kappa}(t)-K'_{\nu,\kappa}(t)I_{\nu,\kappa}(t)=t^{-1}.
\label{KI-Wronskian}
\end{equation}
The connection coefficients read
\begin{equation}
K^\infty_{\nu,\kappa}={\Gamma^2\left(\kappa+1\over2\right)\over\Gamma\left({\kappa+1\over2}+\nu\right)\Gamma\left({\kappa+1\over2}-\nu\right)}.
\label{Kinfty-explicit}
\end{equation}

It is evident that the functions $u_k$ only differ from the sinh\-/Bessel functions by a normalization:
\begin{equation}
u_k(t)=
\begin{cases}
K_{\nu,0}(t),&\text{if $k=*$;}
\\
2^{1-\tk}\Gamma^{-1}(\tk)K_{\nu,\tk}(t),&\text{if $k>0$;}
\\
2^{-\tk}\Gamma(1-\tk)I_{\nu,-\tk}(t),&\text{if $k\le0$.}
\end{cases}
\label{uk-KI-rel}
\end{equation}

The main point is that the correlation functions of the field $\chi(x)$ in the unperturbed theory (with respect to the quadratic action $S_0$) is given by a matrix element in the radial quantization picture:
\begin{equation}
\langle\chi(x_1)\cdots\chi(x_N)\cT_nV_\nu(0)\rangle_0
=\langle\infty|\T_r[\chi(x_1)\cdots\chi(x_N)]|0\rangle_\nu,
\label{chi-corr0-rad}
\end{equation}
where $\T_r$ is the radial ordering symbol:
\begin{equation}
\T_r\qty[\cO_1(x_1)\cdots\cO_N(x_N)]=\cO_{\sigma_1}(x_{\sigma_1})\cdots\cO_{\sigma_N}(x_{\sigma_N}),
\qquad
r_{\sigma_1}>\cdots>r_{\sigma_N},
\label{Trad-def}
\end{equation}
for integer spin local operators $\cO_i$. The consistency check of the radial quantization scheme and the proof of identity (\ref{chi-corr0-rad}) are given in Appendix~\ref{app:Consistency}.

\subsection{Correlation functions and form factors}

In the radial quantization picture the correlation function $G^{(n)}_\cO$ defined in (\ref{GcO-def}) is given by
\begin{equation}
G^{(n)}_\cO\qty(\{x_i\}_N)=\bra{\infty}\T_r\left[ \varphi(x_n)\ldots\varphi(x_1)\e^{-S_\text{int}}\right]\ket{\cO},
\label{Ncorr-modif}
\end{equation}
where $\ket\cO$ is a state that can be constructed from to the operator $\cT_n\cO$ in a unique way. Basically, this construction can be described as follows. The state $|0\rangle_\nu=|V_\nu\rangle$ defined in (\ref{vac-def}) on the background of the classical solution $\phi_\nu(mr)$ corresponds to the operator $\cT_nV_\nu$. The descendant operators (\ref{descendants-def}) are obtained recursively from the exponential operators by the procedure (\ref{cT-dkvarphi-r0-def}). This procedure is transferred to the radial picture straightforwardly:
\eq$$
\Aligned{
G^{(n)}_{\d^k\varphi\,\cO}(\{x_i\}_N)
&=\left[\oint_{|\zeta|^n=r_0}{d\zeta\over2\pi\i\zeta}{\d^k\over\d\zeta^k}G^{(n)}_\cO(\{x_i\}_N,\,x)+(\text{c.t.})\right]_{r_0\to0},
\\
G^{(n)}_{\bd^k\varphi\,\cO}(\{x_i\}_N)
&=\left[\oint_{|\zeta|^n=r_0}{d\zeta\over2\pi\i\zeta}{\d^k\over\d\bzeta^k}G^{(n)}_\cO(\{x_i\}_N,\,x)+(\text{c.t.})\right]_{r_0\to0}.
}\label{GcO-dkvarphi}
$$
Thus the calculation of each correlation function $G^{(n)}_\cO(\{x_i\}_N)$ reduces to the calculation of the correlation functions $G^{(n)}_\nu(\{x_i\}_{N'})$ with $N'\ge N$.

The form factors are related to the correlation functions according to the following reduction formula~\cite{Lashkevich:2023hzk}:
\eq$$
F^{(n)}_\cO(\{\theta_i\}_N)
=\left(-\sqrt{4\pi Z_\varphi}\right)^{-N}
\lim_{r_i\to\infty}\left(\prod^N_{i=1}r_i\int_Cd\xi_i\,\e^{\i mr_i\sinh(\theta_i+\i\xi_i)}\lrd_{r_i}\right)
G^{(n)}_\cO(\{x_i\}_N).
\label{FcO-GcO-rel}
$$
Here $Z_\varphi$ is the wave function renormalization constant. Since $Z_\varphi=1+O(b^4)$ we will omit it in what follows. The left\-/right derivative $\lrd$ is defined as
$$
f(t)\lrd_tg(t)=\tfrac12(f(t)g'(t)-f'(t)g(t)).
$$

There is a subtlety regarding the definition of the contour $C$. In the case of a field theory on the plane and a local operator $\cO$ the correlation function is periodic in $\xi_i$ with the period $2\pi$, so that $C$ can be assumed to be the segment $[-\pi,\pi]$. But for a twist operator corresponding to a branch point of a multi\-/sheeted manifold this periodicity is broken to the $2\pi n$ periodicity. A consistent way to overcome this difficulty demands adding two integrals over the variable $r$ on the banks of the cut to infinity. Instead, we continue the contour $C$ in the complex plane to $\pm\i\infty$, but accept the rule described below.

The correlation functions $G^{(n)}_\cO$ have the form of Fourier series in the variables $\xi_i$, which come from the functions $\e^{\mp\i\tk\xi_i}u_k(mr_i)$ in the series (\ref{chi-decomp}). The upper sign corresponds to the coefficient at $\a_k$, while the lower sign to that at $\ba_k$. Correspondingly we have integrals of the form
\eq$$
I^\pm_\alpha(t,\theta)=\int_{C^\pm}{d\xi\over2\pi}\,\e^{\i t\sinh(\theta+\i\xi)+\i\alpha\xi},
\qquad
\alpha=\mp\tk.
\label{Iint-def}
$$
Here the contour $C^+$ is chosen so that $\zeta\to\infty$ at its ends and $C^-$ so that $\bzeta\to\infty$ at its ends, namely
\eq$$
C^\pm=(\overrightarrow{-\pi\mp\i\infty,-\pi})\cup[\overrightarrow{-\pi,\pi}]
\cup(\overrightarrow{\pi,\pi\mp\i\infty}).
\label{Ccontour-def}
$$
The arrows indicate the direction of the contour. We will consider these integrals in the sense of analytic continuation from imaginary values of $\theta$ in the vicinity of $-{\i\pi\over2}$. Make a change $\xi\to\xi+\i\theta-{\pi\over2}$. Then we have
$$
I^\pm_\alpha(t,\theta)=\int_{C^\pm}{d\xi\over2\pi}\,\e^{t\cos\xi+\i\alpha\left(\xi+\i\theta-{\pi\over2}\right)}
=\e^{-\alpha\left({\i\pi\over2}+\theta\right)}\int_{C^\pm}{d\xi\over2\pi}\,\e^{t\cos\xi+\i\alpha\xi}.
$$
Rigorously speaking the contour $C^\pm$ should be shifted by $-\i\theta+{\pi\over2}$, but we assume this shift to be small enough and ignore it. We immediately recognize the $I$ Bessel function in the r.h.s.:
\begin{equation}
I^\pm_\alpha(t,\theta)=\e^{-\alpha\left({\i\pi\over2}+\theta\right)}I_{\mp\alpha}(t)
=\e^{\pm\tk\left({\i\pi\over2}+\theta\right)}I_\tk(t).
\label{Iint-Bessel-rel}
\end{equation}
This choice of the contour leads to the fact that the form factors are expressed in terms of the expressions $tI_\tk(t)\lrd_t u_k(t)$ and $tI_0(t)\lrd\phi_\nu(t)$, which have finite limit as $t\to\infty$. The choice of the contour is not crucial in the leading order in $b$, where such expressions with $k>0$ or $k=*$ only enter the form factors, but we fix it for future studies of subleading contributions.

We considered the integration contours as lying on the $s=0$ sheet of the manifold $\cM_n$. Due to the $\Z_n$ symmetry of the manifold we may translate it to other sheets. Instead of translating them we will shift the variables $\theta_i$, $\theta'_i$ in (\ref{ff-def}), (\ref{matel-ff}) assuming them to live on a set of $n$ lines for the in\-/states:
\eq$$
\theta\in\bigcup^{n-1}_{s=0}(\R-2\pi\i s).
\label{theta-set}
$$
It means that $\Re\theta$ is the rapidity of the particle, while $s=-\Im\theta/2\pi$ is the sheet number.

\section{Properties of the sinh\-/Bessel functions}
\label{sec:sinh-Bessel}

In this section we study the properties of the sinh\-/Bessel functions $K_{\nu,k}(t)$, $I_{\nu,k}(t)$. In \cite{Lashkevich:2023hzk} they were introduced and studied for integer values of the parameter $k$. The main instrument was the approach based on the Fredholm determinant formalism\cite{Tracy:1996CMP179}. Here we use this formalism to study these functions for non\-/integer values of $k$. This formalism provides an explicit expression for the functions $K_{\nu,k}(t)$ in terms of a series of multiple integrals with powers of the parameter $\lambda$ as coefficients. The functions $I_{\nu,k}(t)$ are defined indirectly, but we provide a construction of the series in powers of $t$ for both types of functions.

In this section we consider the letters $k$, $l$ as arbitrary real (or even complex after the analytic continuation) numbers, with no relation to the indices in sections \ref{sec:model} and \ref{sec:formfactors}. Physically they correspond to $\tk$, $\tl$ or $\kappa$ there.

\subsection{Definitions in terms of the Fredholm determinants}
\label{subsec:Tracy-Widom}

Consider the space $\cF=L_2[0,\infty)$ and define the operator $\hat K$ on $\cF$ with the kernel
\begin{equation}
K(u,v)={E(u)E(v)\over u+v},
\qquad
E(u)=\exp\left(-{1\over2}\sum_{k\in\mathbb K}t_k u^k\right).
\label{K-kernel-def}
\end{equation}
Here $t_k\in\C$ are parameters and $\mathbb K\subset\R$ is an arbitrary finite or countable set. In application to the theories on the multi\-/sheeted surface with one branch point of the order $n$ it is natural to set $\mathbb K={1\over n}\Z$, but generally it is not necessary. We will only assume that $0,\pm1\in\mathbb K$. Let us also define the functions
\begin{equation}
\phi({\vec t})=2\log\det(1+\hK)-2\log\det(1-\hK),
\qquad
\vec t=\{t_k\}_{k\in\mathbb K}
\label{phi-tk-def}
\end{equation}
in terms of the Fredholm determinants $\det(1\pm\hK)$. More explicitly
\begin{equation}
\phi(\vec t)=2\tr\log{1+\hK\over1-\hK}=\sum_{r\in2\Z_{\ge0}+1}{4\over r}\tr\hK^r
=\sum_{r\in2\Z_{\ge0}+1}{4\over r}\int d^ru\prod^r_{i=1}{\e^{-\sum_{k\in\mathbb K}t_k u_i^k}\over u_i+u_{i+1}},
\quad u_{r+1}=u_1.
\label{phi-tk-explicit}
\end{equation}
For those values of the parameters $t_k$, for which the integrals converge, the function $\phi(\vec t)$ is proved\cite{Tracy:1996CMP179} to satisfy the classical sinh\-/Gordon equation:
\begin{equation}
\d_1\d_{-1}\phi=\sinh\phi,
\label{shG-phi}
\end{equation}
where $\d_k=\d/\d t_k$. For an arbitrary function $\Phi(\vec t)$ let us define the function $\Phi^\text{r}(z,\bz)=\Phi(\vec t)$ restricted to the values
\begin{equation}
t_1=mz/2,
\qquad
t_{-1}=m\bz/2,
\qquad
t_0=-\log\lambda,
\qquad
t_k=0\quad (k\ne0,\pm1).
\label{basic-reduction}
\end{equation}
We will refer the function $\Phi^\text{r}$ as to the basic reduction of the function $\Phi$. For the function $\phi^\text{r}$ we have
\begin{equation}
\phi^\text{r}(z,\bz)=\phi_\nu(m|z|).
\label{phi-r-fin}
\end{equation}

For $k_1,\ldots,k_s\in\mathbb K$ define the functions
\begin{equation}
\Phi_{k_1\ldots k_s}(\vec t)=(-1)^s \d_{k_1}\cdots\d_{k_s}\phi(\vec t).
\label{Phidef}
\end{equation}
In what follows we only discuss the cases $s=1,2$, since we restrict ourselves to the descendant operators with no more than two derivatives. The restricted functions $\Phi^\text{r}_{k_1\ldots k_s}$ factorize
\begin{equation}
\Phi^\text{r}_{k_1\ldots k_s}(z,\bz)=\tPhi_{k_1\ldots k_s}(m|z|)\e^{-\i\sum_ik_i\xi}.
\label{tPhi-def}
\end{equation}
Note that all these functions have the same large $t$ asymptotics
\begin{equation}
\tPhi_{k_1\ldots k_s}(t)=\sqrt{8\pi}\lambda t^{-1/2}\e^{-t}(1+O(t))
\quad
\text{as $t\to\infty$.}
\label{tPhi-asymp}
\end{equation}
Besides, all these functions are symmetric with respect to transposition of subscripts and simultaneous change of their signs:
\begin{equation}
\tPhi_{k_1\ldots k_s}(t)=\tPhi_{k_{\sigma_1}\ldots k_{\sigma_s}}(t)=\tPhi_{-k_1\ldots-k_s}(t)
\qquad(\forall\sigma\in S_s).
\label{tPhi-symm}
\end{equation}

The $t_k$\-/derivative of equation (\ref{shG-phi}) provides a consistent equation for $\Phi_k$:
\begin{equation}
(\d_1\d_{-1}-\cosh\phi)\Phi_k=0.
\label{Phik-eq}
\end{equation}
After the basic reduction the radial part $\tPhi_k$ satisfies the sinh\-/Bessel equation (\ref{shB-equation}) with $\kappa=k$. Since the function $\tPhi_k(t)$ decreases at infinity, it is proportional to the function $K_{\nu,k}(t)$. The proportionality factor is easily established by a comparison of the asymptotics (\ref{KI-t-large}):
\begin{equation}
\tPhi_k(t)=4\lambda{K_{\nu,k}(t)\over K_{\nu,k}^\infty}.
\label{tPhik-Kk-rel}
\end{equation}

For small values of $t_{\pm1}$ the functions $\Phi_k^\text{r}$ have the following asymptotics:
\begin{equation}
\tPhi_k(t)=\Cases{t^{-k}\left(A_k+O\left(t^{2-4|\nu|}\right)\right) + t^k\left(A_{-k}+O\left(t^{2-4|\nu|}\right)\right),&\text{if $k\ne0$;}\\
-2A'_0 \log t+O(1),&\text{if $k=0$,}}
\label{tPhik-small-t}
\end{equation}
where
\begin{equation}
A_k={2^{k+1}\lambda\Gamma(k)\over K^\infty_{\nu,k}},
\qquad
A'_0=\lim_{k\to0}k A_k={2\over\pi}\tan\pi\nu.
\label{Ak-explicit}
\end{equation}

Let us fix the second solution $\tPhi^\vee_k(t)$ of the sinh\-/Bessel equation (\ref{shB-equation}) with $\kappa=k$ by the conditions
\begin{equation}
\tPhi_k(t)\lrd_t\tPhi^\vee_k(t)=t^{-1};
\qquad
\tPhi^\vee_k(t)=O(1)\quad\text{as $t\to0$ for $k\ge0$.}
\label{tPhi-tPhi-vee-def}
\end{equation}
It is assumed to be an analytic function of $k$. Evidently, it has the asymptotic properties
\begin{align}
\tPhi^\vee_k(t)
&=t^k(k^{-1}A_k^{-1}+O(1)),
\qquad
t\to0,
\label{tPhi-vee-small-t}
\\
\tPhi^\vee_k(t)
&={1\over\sqrt{8\pi}\lambda}t^{-1/2}\e^t(1+O(t)),
\qquad
t\to\infty.
\label{tPhi-vee-large-t}
\end{align}
Comparing with the asymptotics (\ref{KI-t-small}), (\ref{KI-t-large}) we obtain
\begin{equation}
I_{\nu,k}(t)=2\lambda\left(K^\infty_{\nu,k}\right)^{-1}\tPhi^\vee_k(t).
\label{I-tPhi-vee-rel}
\end{equation}

Now consider the second derivatives of equation (\ref{shG-phi}):
\begin{equation}
(\d_1\d_{-1}-\cosh\phi)\Phi_{kl} = \Phi_k\Phi_l\sinh\phi.
\label{Phikl-eq}
\end{equation}
The corresponding radial function $\tPhi_{kl}(t)$ satisfies the equation
\begin{equation}
\left(\d_t^2+t^{-1}\d_t-\cosh\phi_\nu(t)-(k+l)^2t^{-2}\right)\tPhi_{kl}(t)=\tPhi_k(t)\tPhi_l(t)\sinh\phi_\nu(t).
\label{tPhikl-equation}
\end{equation}
The small $t$ asymptotics of the functions $\tPhi_{kl}$ reads
\begin{subequations}
\label{tPhikl-small-t}
\begin{align}
\tPhi_{kl}(t)
&=t^{-k-l}\left(A_{kl}+O\left(t^{2-4|\nu|}\right)\right)+t^{k+l}\left(A_{-k.-l}+O\left(t^{2-4|\nu|}\right)\right)
\notag
\\
&\quad
+t^{2-4|\nu|-k+l}\left(B_{k.-l}+O\left(t^{2-4|\nu|}\right)\right)
\notag
\\
&\quad
+t^{2-4|\nu|+k-l}\left(B_{-k.l}+O\left(t^{2-4|\nu|}\right)\right),
\qquad
k+l\ne0;
\label{tPhikl-small-t-gen}
\\
\tPhi_{k.-k}(t)
&=-2A'_{k.-k}\log t+O(1)
\notag
\\
&\quad
+t^{2-4|\nu|-2k}\left(B_{kk}+O\left(t^{2-4|\nu|}\right)\right)+t^{2-4|\nu|+2k}\left(B_{-k.-k}+O\left(t^{2-4|\nu|}\right)\right).
\label{tPhikmk-small-t}
\end{align}
\end{subequations}
Here $A_{kl}$, $B_{kl}$, $A'_{k.-k}$ are constants. The $A$\-/constants read
\begin{align}
A_{kl}
&=A_{k+l}
+{2^{3k+3l-2}\lambda^3\over\pi\Gamma(k)\Gamma(l)}
\int_{p_i\ge0}dp_1\,dp_2\,{p_1p_2\sinh2\pi p_1\sinh\pi p_2\over\prod^2_{i=1}(\cosh^2\pi p_i-\sin^2\pi\nu)}
\notag
\\
&\quad\times
\prod_{\ve_1,\ve_2=\pm1}\Gamma\left(k+\i\ve_1p_1+\i\ve_2p_2\over2\right)\Gamma\left(l+\i\ve_1p_1+\i\ve_2p_2\over2\right),
\label{Akl-intrep}
\\
A'_{k.-k}
&=\lim_{l\to-k}(k+l)A_{kl}
={2\over\pi}\tan\pi\nu\,\left(1+{\sin^2\pi\nu\over\cos\pi\left(\nu+{k\over2}\right)\cos\pi\left(\nu-{k\over2}\right)}\right).
\label{Akmk-fin}
\end{align}
The derivation of these constants is given in Appendix~\ref{app:Intrep}. An alternative series representation (see Appendix~\ref{app:Akl-series}), more suitable for numeric calculations, reads
\begin{equation}
{A_{kl}\over A_{k+l}}=1+\sum^\infty_{n=0}(C_{k+2n.l}+C_{l+2n.0}-C_{2n.l}),
\qquad
C_{kl}=8\nu kl\left({1\over k+1}+{1\over k+l+1}\right){A_k A_l\over A_{k+l+2}}.
\label{Akl-seriesrep}
\end{equation}
For the $B$\-/constants we have
\begin{equation}
B_{kl}={\beta^{-2}_{|\nu|}A_kA_l\over 8(1-2|\nu|-k)(1-2|\nu|-l)}.
\label{Bkl-fin}
\end{equation}
They are found by direct integration of equation (\ref{tPhikl-equation}) for small values of $t$ with the use of expansions for the sinh\-/Bessel functions from the next subsection.

Note that for $kl\ge0$ one of the $A$\-/terms in (\ref{tPhikl-small-t}) is the leading one, while for $kl<0$ it holds if $\min(|k|,|l|)<1-4\nu^2$. Otherwise one of the $B$\-/terms is dominant.

In what follows we will need the functions
\begin{equation}
K_{\nu,kl}(t)={K_{\nu,k}^\infty K_{\nu,l}^\infty\over16\lambda^2}\tPhi_{kl}(t)
\label{Knu-kl-def}
\end{equation}
and the constants
\begin{equation}
K^\infty_{\nu,kl}={2^{k+l+1}\lambda\Gamma(k+l)\over A_{kl}}.
\label{Knu-kl-infty-def}
\end{equation}
In terms of this function equation (\ref{tPhikl-equation}) reads
\begin{equation}
\left(\d_t^2+t^{-1}\d_t-\cosh\phi(t)-(k+l)^2t^{-2}\right)K_{\nu,kl}(t)=K_{\nu,k}(t) K_{\nu,l}(t)\sinh\phi(t).
\label{Knu-kl-equation}
\end{equation}
The normalization (\ref{Knu-kl-def}) of the functions $K_{\nu,kl}(t)$ was chosen to avoid extra coefficients in this equation.

\subsection{Small $t$ expansions}
\label{subsec:sB-expansions}

The functions $K_{\nu,k}(t)$ and $I_{\nu,k}(t)$ satisfy the sinh-Bessel equation (\ref{shB-equation}) with $\kappa=k$ and have simple asymptotic properties (\ref{KI-t-small}) for small $t$. We can take these asymptotics as a base and apply the sinh\-/Bessel equation to obtain power series iteratively. In what follows we will use the notation
\begin{equation}
\delta^{(s,s')}_{\nu,k}=k/2+(1-2\nu)s+(1+2\nu)s'.
\label{delta-ss-k-def}
\end{equation}

For the function $I_{\nu,k}(t)$ with $\Re k\ge0$ this procedure results in a unique solution of the form
\begin{equation}
2^k\Gamma(k+1)I_{\nu,k}(t)={kA_k\tPhi^\vee_k(t)}=\sum_{s,s'\ge0}c^{(s,s')}_{\nu,k}t^{2\delta^{(s,s')}_{\nu,k}},
\qquad
c^{(0,0)}_{\nu,k}=1,
\qquad
c^{(s,s')}_{\nu,k}=c^{(s',s)}_{-\nu,k}.
\label{Igen}
\end{equation}
Several coefficients $c^{(s,s')}_{\nu,k}$ can be found in Appendix~\ref{app:Expansion}, and their full list for $s+s'\le4$ is given in Supplemental Material. This series defines the function $\tPhi^\vee_k(t)$ for all complex values of $k$ except isolated poles of the coefficients $A_k^{-1}c^{(s,s')}_{\nu,k}$. Thus for generic values of $k$, $\Re k>0$ the two solutions $\tPhi^\vee_{\pm k}(t)\sim t^{\pm k}+o(t)$ of the sinh\-/Bessel equation form a basis. The solution $\tPhi_k(t)=t^{-k}(A_k+o(t))$ is a linear combination of these solutions. Comparing the leading small $t$ asymptotics we have
$$
\tPhi_k(t)=kA_kA_{-k}\tPhi^\vee_{-k}(t)+C_k\tPhi^\vee_k(t).
$$
Due to the symmetry $\tPhi_k(t)=\tPhi_{-k}(t)$ and analyticity of this function in $k$ we obtain
\begin{equation}
\tPhi_k(t)=kA_kA_{-k}\left(\tPhi^\vee_k(t)-\tPhi^\vee_{-k}(t)\right).
\label{tPhi-k-tPhi-vee-k-rel}
\end{equation}
In terms of the sinh\-/Bessel functions it reads
\begin{equation}
K_{\nu,k}(t)={\pi\over2\sin\pi k}\left(I_{\nu,-k}(t)-{K^\infty_{\nu,k}\over K^\infty_{\nu,-k}}I_{\nu,k}(t)\right).
\label{Knu-k-Inu-k-rel}
\end{equation}
This formula generalizes the well\-/known formula that defines the function $K_k(t)$ in terms of $I_k(t)$ in the theory of the usual Bessel functions. It is convenient to rewrite it as follows. Let
\begin{equation}
K^*_{\nu,k}(t)
={\pi\over2\sin\pi k}I_{\nu,-k}(t)
=2^{k-1}\Gamma(k)\sum_{s,s'\ge0}c^{(s,s')}_{\nu,-k}t^{2\delta^{(s,s')}_{\nu,-k}}.
\label{Kgen}
\end{equation}
Then
\begin{equation}
K_{\nu,k}(t)=K^*_{\nu,k}(t)+\alpha_{\nu,k}I_{\nu,k}(t),
\qquad
\alpha_{\nu,k}=-{\pi\over2\sin\pi k}{K^\infty_{\nu,k}\over K^\infty_{\nu,-k}},
\qquad
k\ge0.
\label{Kgen+Igen}
\end{equation}
The coefficient $\alpha_{\nu,k}$ has poles at non\-/negative even integer values of $k$. On the other hand, the function $K^*_{\nu,k}(t)$ may have poles at the same values of $k$ due to possible poles of the coefficients $c^{(s,s')}_{\nu,-k}$. However, the function $K_{\nu,k}(t)$ must be finite for all $k\ge0$ due to its proportionality (\ref{tPhik-Kk-rel}) to the function $\tPhi_k(t)$ defined in terms of the Fredholm determinants. This means that the poles of $\alpha_{\nu,k}$ must be canceled by the poles of $c^{(s,s')}_{\nu,-k}$. On the examples $k=0,\,2$ we see that this leads to logarithmic terms:
\begin{align}
K_{\nu,0}(t)
&=\sum_{s,s'\geq0}\qty((\delta_\nu-\log t)c^{(s,s')}_{\nu,0}-\left.\d_k c^{(s,s')}_{\nu,k}\right|_{k=0})t^{2\delta^{(s,s')}_{\nu,0}},
\label{Knu0}
\\
K_{\nu,2}(t)
&=2t^{-2}+2\sum_{s>0}\left(c^{(s,0)}_{\nu,-2}t^{2\delta_{\nu,-2}^{(s,0)}}+c^{(0,s)}_{\nu,-2}t^{2\delta_{\nu,-2}^{(0,s)}}\right)
\notag
\\
&\quad
+\sum_{s,s'\geq0}\biggl({c^{(s,s')}_{\nu,2}\over8(1-4\nu^2)^2}\left(\delta_\nu+{5\over4}+{8\nu^2\over1-4\nu^2}-\log t\right)
\notag
\\
&\quad+2\,\d_k(k-2)c_{\nu,-k}^{(s+1,s'+1)}
-{\d_kc^{(s,s')}_{\nu,k}\over16(1-4\nu^2)^2}\biggr)\biggr|_{k=2}t^{2\delta_{\nu,2}^{(s,s')}}.
\label{Knu2}
\end{align}
Here
\begin{equation}
\delta_\nu=3\log2+\tfrac12\psi\qty(\tfrac12-\nu)+\tfrac12\psi\qty(\tfrac12+\nu),\qquad\psi(x)=(\log\Gamma(x))'.
\end{equation}
We expect that the contributions of the form $t^{2\delta^{(s,s')}_{\nu,-k}}\log t$ appear for every even positive integer value of $k$ in the terms with $s,s'\ge k/2$.

\subsection{Vertex integrals}\label{ssec:Jkl}
Assuming $k,l>0$ introduce the following integrals:
\subeq{\label{Js-def}
\begin{align}
\cJ_{kl}^+(t)
&=\int_t^\infty d\tau\,\tau I_{\nu,k+l}(\tau)K_{\nu,k}(\tau)K_{\nu,l}(\tau)\sinh\phi_\nu(\tau),
\label{Jp-def}
\\
\cJ_{kl}^-(t)
&=\int_t^\infty d\tau\,\tau I_{\nu,k-l}(\tau)K_{\nu,k}(\tau)K_{\nu,l}(\tau)\sinh\phi_\nu(\tau),
\label{Jm-def}
\\
\cJ^0_{kl}(t)
&=\int_0^t d\tau\,\tau I_{\nu,k-l}(\tau)K_{\nu,k}(\tau)I_{\nu,l}(\tau)\sinh\phi_\nu(\tau).
\label{J0-def}
\end{align}}
These integrals correspond to the 3-leg interaction vertex coming from the term $S^{(3)}$ in the expansion (\ref{Sint}) and enter expressions for form factors of descendant operators of the form (\ref{descendants-def}) with $K+L=2$.

The first two integrals can be written in terms of the only function:
\begin{equation}
j_{kl}(t)={1\over2}\int^\infty_td\tau\,\tau\tPhi^\vee_{k+l}(\tau)\tPhi_k(\tau)\tPhi_l(\tau)\sinh\phi_\nu(\tau),
\label{jkl-def}
\end{equation}
which is analytic in $k$, $l$. Namely,
\begin{equation}
\cJ^+_{kl}(t)={K^\infty_{\nu,k}K^\infty_{\nu,l}\over4\lambda K^\infty_{\nu,k+l}}j_{kl}(t),
\qquad
\cJ^-_{kl}={K^\infty_{\nu,k}K^\infty_{\nu,l}\over4\lambda K^\infty_{\nu,k-l}}j_{k.-l}(t).
\label{cJ-jkl-rel}
\end{equation}

Multiplying the sinh\-/Bessel equation (\ref{shB-equation}) with $\kappa=k+l$ for $u(t)=\tPhi^\vee_{k+l}(t)$ by $\tPhi_{kl}(t)$, multiplying the equation (\ref{tPhikl-equation}) by $\tPhi^\vee_{k+l}(t)$ and taking their difference, we obtain
\begin{equation}
(t(\tPhi^\vee_{k+l}(t)\tPhi_{kl}'(t)-\tPhi^{\vee\prime}_{k+l}(t)\tPhi_{kl}(t)))'=t\tPhi^\vee_{k+l}(t)\tPhi_k(t)\tPhi_l(t)\sinh\phi_\nu(t).
\label{tPhi-vee-tPhi-tPhi-Wronsian}
\end{equation}
Hence,
\begin{equation}
j_{kl}(t)
=\tau\left(\tPhi^\vee_{k+l}(\tau)\lrd_\tau\tPhi_{kl}(\tau)\right)\Bigr|_{\tau=t}^\infty.
\label{jint-Wronskian}
\end{equation}
It means that the integrals $\cJ^\pm_{kl}$ are determined by the asymptotics of the functions $\tPhi^\vee_k(t)$, $\tPhi_{kl}(t)$. The large $t$ asymptotic follows immediately from (\ref{tPhi-asymp}), (\ref{tPhi-tPhi-vee-def}):
\begin{equation}
\tau\left(\tPhi^\vee_{k+l}(\tau)\lrd_\tau\tPhi_{kl}(\tau)\right)\Bigr|_{\tau\to\infty}=-1
\label{tPhi-vee-tPhikl-infty}
\end{equation}

To analyze the small $t$ asymptotics define the functions $\tPhi^{\vee\pm}_{kl}(t)$ as solutions to the equations
\begin{equation}
\left(\d_t^2+t^{-1}\d_t-\cosh\phi_\nu(t)-(k\pm l)^2t^{-2}\right)\tPhi^{\vee\pm}_{kl}(t)=\tPhi^\vee_k(t)\tPhi^\vee_l(t)\sinh\phi_\nu(t).
\label{tPhikl-veepm-equation}
\end{equation}
We fix the solutions by the following series
\begin{equation}
\tPhi^{\vee\pm}_{kl}(t)
={1\over klA_kA_l}\sum_{(s,s')\in\rmS^+}c^{\pm(s,s')}_{\nu,kl}t^{2\delta^{(s,s')}_{\nu,k+l}},
\label{tPhikl-veepm-expansion}
\end{equation}
where $\rmS^+$ is the set of pairs of non\-/negative integers except $(0,0)$:
\begin{equation}
\rmS^+=\{(s,s')\,|\,s,s'\in\Z_{\ge0},\ s+s'>0\}.
\label{Splus-def}
\end{equation}
The coefficients $c^{\pm(s,s')}_{\nu,kl}$ of the series can be defined uniquely from the equation (\ref{tPhikl-veepm-equation}) by a recursive procedure for generic positive values of $k$, $l$. They satisfy the relations
\begin{equation}
c^{\pm(s,s')}_{\nu,kl}=c^{\pm(s,s')}_{\nu,lk}=-c^{\pm(s',s)}_{-\nu,kl}
\label{ckl-sym}
\end{equation}
and
\begin{equation}
c^{+(s,s')}_{\nu,k0}=c^{-(s,s')}_{\nu,k0}=(s-s')c^{(s,s')}_{\nu,k}.
\label{ck0-ck-coefs-rel}
\end{equation}
The last identity will be proved at the end of the subsection.

For $\ve_{1,2}=\pm$ introduce the functions
\begin{equation}
j^{\vee\ve_1\ve_2}_{kl}(t)
=\left.\tau\left(\tPhi^\vee_{k+l}(\tau)\lrd_\tau\tPhi^{\vee\ve_1\cdot\ve_2}_{-\ve_1k.-\ve_2l}(\tau)\right)\right|^t_0
={1\over2}\int^t_0d\tau\,\tau\tPhi^\vee_{k+l}(\tau)\tPhi^\vee_{-\ve_1k}(\tau)\tPhi^\vee_{-\ve_2l}(\tau)\sinh\phi_\nu(\tau)
\label{jvee-def}
\end{equation}
with the assumption of analytical continuation from the region $\Re k,\Re l\ge0$. Evidently,
\begin{equation}
j^{\vee\ve_1\ve_2}_{kl}(t)=j^{\vee\ve_2\ve_1}_{lk}(t)=j^{\vee\ve'_1\ve_2}_{k'l'}(t),
\qquad
\ve'_1=\ve_1\ve_2,
\quad
k'=\ve_2k,
\quad
l'=-\ve_2(k+l).
\label{jvee-sym}
\end{equation}
These functions admit simple expansions
\begin{equation}
j^{\vee\ve_1\ve_2}_{kl}(t)=-{\ve_1\ve_2\over kl(k+l)A_{-\ve_1k}A_{-\ve_2l}A_{k+l}}
\sum_{(s,s')\in\rmS^+}J^{\ve_1\ve_2(s,s')}_{\nu,kl}t^{2\delta^{(s,s')}_{\nu,(1-\ve_1)k+(1-\ve_2)l}}.
\label{jvee-expansion}
\end{equation}
The coefficients $J^{\ve_1\ve_2(s,s')}_{\nu,kl}$ can be obtained iteratively by using the equation (\ref{tPhikl-veepm-equation}) (see Appendix~\ref{app:Expansion} and Supplemental Material).

Now express the integral $j_{kl}(t)$ in terms of the integrals $j^{\vee\ve_1\ve_2}_{kl}(t)$. From the asymptotics (\ref{tPhik-small-t}), (\ref{tPhikl-small-t}) and the relation (\ref{tPhi-k-tPhi-vee-k-rel}) we obtain
\begin{equation}
\tPhi_{kl}(t)=v_{kl}\tPhi_{k+l}(t)
+klA_kA_{-k}A_lA_{-l}(\tPhi^{\vee+}_{-k.-l}(t)-\tPhi^{\vee-}_{-k.l}(t)-\tPhi^{\vee-}_{k.-l}(t)+\tPhi^{\vee+}_{kl}(t)),
\label{tPhikl-tPhi-veepm-rel}
\end{equation}
where
\begin{equation}
v_{kl}={A_{kl}\over A_{k+l}}={K^\infty_{\nu,k+l}\over K^\infty_{\nu,kl}}.
\label{vkl-def}
\end{equation}
Substituting it into (\ref{jint-Wronskian}) we obtain
\begin{equation}
j_{kl}(t)
=v_{kl}-1+{klA_kA_{-k}A_lA_{-l}\sum_{\ve_1,\ve_2=\pm1}\ve_1\ve_2j^{\vee\ve_1\ve_2}_{\nu,kl}(t)}.
\label{jkl-jvee-rel}
\end{equation}

The integrals $\cJ^0_{kl}(t)$ are also expressed in terms of the functions $j^{\vee\ve_1\ve_2}_{kl}$. Taking into account that they tend to zero as $t\to0$ and using (\ref{tPhi-k-tPhi-vee-k-rel}), we can write
\begin{equation}
\cJ^0_{kl}(t)
={8k\lambda^3\over K^\infty_{\nu,k-l}K^\infty_{\nu,-k}K^\infty_{\nu,l}}\left(j^{\vee++}_{k.-l}(t)-j^{\vee-+}_{k.-l}(t)\right).
\label{cJ0-jvee-rel}
\end{equation}

Using the expansion (\ref{jvee-expansion}) we can rewrite the integrals (\ref{Js-def}) as follows:
\subeq{\label{cJ-expansions}
\Align$$
\cJ^+_{kl}(t)
&=\cV^+_{kl}+{\Gamma(k)\Gamma(l)\over2\Gamma(k+l+1)}\sum_{(s,s)\in\rmS^+}\biggl(
J^{++(s,s')}_{\nu,kl}t^{2\delta^{(s,s')}_{\nu,0}}+{A_{-l}\over A_l}J^{+-(s,s')}_{\nu,kl}t^{2\delta^{(s,s')}_{\nu,2l}}
\notag
\\
&\quad
+{A_{-k}\over A_k}J^{-+(s,s')}_{\nu,kl}t^{2\delta^{(s,s')}_{\nu,2k}}
+{A_{-k}A_{-l}\over A_kA_l}J^{--(s,s')}_{\nu,kl}t^{2\delta^{(s,s')}_{\nu,2k+2l}}
\biggr),
\label{cJp-expansion}
\\
\cJ^-_{kl}(t)
&=\cV^-_{kl}+{2^{2l-1}\Gamma(k)\Gamma(l)\over\Gamma(k-l+1)}\sum_{(s,s)\in\rmS^+}\biggl(
{A_{-l}\over A_l}J^{++(s,s')}_{\nu,k.-l}t^{2\delta^{(s,s')}_{\nu,0}}+J^{+-(s,s')}_{\nu,k.-l}t^{2\delta^{(s,s')}_{\nu,-2l}}
\notag
\\
&\quad
+{A_{-k}A_{-l}\over A_kA_l}J^{-+(s,s')}_{\nu,k.-l}t^{2\delta^{(s,s')}_{\nu,2k}}
+{A_{-k}\over A_k}J^{--(s,s')}_{\nu,k.-l}t^{2\delta^{(s,s')}_{\nu,2k-2l}}
\biggr),
\label{cJm-expansion}
\\
\cJ^0_{kl}(t)
&=-{\Gamma(k)\over\Gamma(l+1)\Gamma(k+l+1)}\sum_{(s,s)\in\rmS^+}\biggl(
J^{++(s,s')}_{\nu,k.-l}t^{2\delta^{(s,s')}_{\nu,0}}+{A_{-k}\over A_k}J^{-+(s,s')}_{\nu,k.-l}t^{2\delta^{(s,s')}_{\nu,2k}}
\biggr).
\label{cJ0-expansion}
$$}
Here
\eq$$
\cV^\pm_{kl}={K^\infty_{\nu,k}K^\infty_{\nu,l}\over4\lambda K_{\nu,k\pm l}}(v_{k.\pm l}-1)
={K^\infty_{\nu,k}K^\infty_{\nu,l}\over4\lambda}\left({1\over K^\infty_{\nu,k.\pm l}}-{1\over K^\infty_{\nu,k\pm l}}\right).
\label{cVpm-kl-def}
$$

We give these expansions for the sake of the mathematical completeness, but in the next section we will need simpler estimations. Consider the integrals $\cJ^+_{kl}$ for $k,\,l\ge0$. Since in this case all the exponents $2\delta^{(s,s')}_{\nu,\kappa}$ ($\kappa=0,\,k,\,l$) in the r.h.s.\ are positive, the sum over $(s,s')$ in (\ref{cJp-expansion}) vanishes as $t\to0$. Thus
\begin{equation}
\cJ^+_{kl}(t)=\cV^+_{kl}+{O(t^{2-4|\nu|})},
\qquad
t\to0.
\label{cJp-cVp}
\end{equation}

If $k\ge l>0$, the exponent $2\delta^{(s,s')}_{\nu,-l}$ can be negative for small enough $s,s'$. Thus in the expansion (\ref{cJm-expansion}) for $\cJ^-_{kl}(t)$ the $J^{+-}$ terms may be large at small $t$. We obtain
\begin{equation}
\cJ^-_{kl}(t)={2^{2l-1}\Gamma(k)\Gamma(l)\over\Gamma(k-l+1)}
\sum_{(s,s')\in\rmS^+} J^{+-(s,s')}_{\nu,k.-l}t^{2\delta^{(s,s')}_{\nu,-2l}}+\cV^-_{kl}+o(1),
\qquad
t\to0.
\label{Jkl-series}
\end{equation}

The quantity $\cV^-_{kl}$ is a finite part of the integral $\cJ^-_{kl}(t)$ for generic values of the parameters $\nu$, $k$, $l$. For special values of $l$ the finite part may differ. Let $l_0=\delta^{(s,s')}_{\nu,0}$ for a particular value of the pair $(s,s')\in\rmS^+$. Then the finite part reads%
\footnote{Just this quantity was denoted by $\cV^-_{kl}$ in \cite{Lashkevich:2023hzk}, which is essential in the case $k=l=2$ there.}
$$
[\cJ^-_{kl_0}]^\text{fin}
=\left.\cV^-_{kl}+{2^{2l-1}\Gamma(k)\Gamma(l)\over\Gamma(k-l+1)}J^{+-(s,s')}_{\nu,k.-l}\right|_{l\to l_0}.
$$
In fact, we will not need to calculate this sum explicitly, but we will need the residues of the poles of $\cV^-_{kl}$. Look at the formula (\ref{tPhikl-tPhi-veepm-rel}) for the function $\tPhi_{kl}(t)$. This function is finite for all values of $l$ by definition. If the coefficient $v_{kl}$ has a pole at $l=-l_0$ it must be canceled by the contribution proportional to $c^{-(s,s')}_{\nu,-k.l}t^{2\delta^{(s,s')}_{\nu,2l}}$ in the term with $\tPhi^{\vee-}_{-k.l}(t)$. Hence,
\eq$$
\Res_{l=-l_0}v_{kl}+{A_kA_{l_0}\over A_{k-l_0}}\Res_{l=-l_0}c^{-(s,s')}_{\nu,-k.l}=0,
\qquad
l_0=\delta^{(s,s')}_{\nu,0}\ge0.
\label{vkl-Res}
$$
Though we assumed real $k\ge l_0$ the analyticity in $k$ removes this limitation. Thus we obtain
\eq$$
\Res_{l=l_0}\cV^-_{kl}=-2^{2l_0-1}{\Gamma(k)\Gamma(l_0)\over\Gamma(k-l_0)}\Res_{l=l_0}c^{-(s,s')}_{\nu,-k.-l},
\qquad
l_0=\delta^{(s,s')}_{\nu,0}\ge0.
\label{cVm-Res}
$$
Besides, using the explicit formulas for the coefficients we checked for $s+s'\le4$ that
\eq$$
\Res_{l=l_0}c^{-(s,s')}_{\nu,-k.-l}=-{l_0\over k-l_0}\Res_{l=l_0}c^{+(s,s')}_{\nu,-k.k-l},
\qquad
l_0=\delta^{(s,s')}_{\nu,0}\ge0.
\label{c-Res-rel}
$$

The integrals $\cJ^0_{kl}(t)$ tend to zero as $t\to0$. The estimation
\begin{equation}
\cJ^0_{kl}(t)={-{\Gamma(k)\over\Gamma(l+1)\Gamma(k+l+1)}}
\sum_{(s,s')\in\rmS^+} J^{++(s,s')}_{\nu,k.-l}t^{2\delta^{(s,s')}_0}+{O(t^{2k+2-4|\nu|})}
\label{Jkl0series}
\end{equation}
would be sufficient for the calculations of the form factors. Nevertheless, there is a simplification that allows one to avoid the direct use of the expansions (\ref{cJ-expansions}).

Indeed, the integrals $\cJ^-_{kl}$ and $\cJ^0_{kl}$ will enter all the expressions in the next section in the combination
\eq$$
F_{\nu,kl}(t)=K_{\nu,l}(t)\cJ^0_{kl}(t)+I_{\nu,l}(t)\cJ^-_{kl}(t).
\label{Fnu-k-def}
$$
It admits essential simplification. Define the function $I_{\nu,kl}(t)$ as a solution to the equation
\begin{equation}
(\d_t^2+t^{-1}\d_t-\cosh\phi_\nu(t)-(k-l)^2t^{-2})I_{\nu,kl}(t)=K_{\nu,k}(t)I_{\nu,l}(t)\sinh\phi_\nu(t)
\label{Inu-kl-equation}
\end{equation}
Due to (\ref{tPhi-k-tPhi-vee-k-rel}) we may choose the solution in the form
\Align$$
I_{\nu,kl}(t)
&=2^{k-l-1}{\Gamma(k)\over\Gamma(l+1)}klA_{-k}A_l\left(\tPhi^{\vee-}_{kl}(t)-\tPhi^{\vee+}_{-k.l}(t)\right)
\notag
\\
&=2^{k-l-1}{\Gamma(k)\over\Gamma(l+1)}\sum_{(s,s')\in\rmS^+}\biggl(c^{+(s,s')}_{\nu,-k.l}t^{2\delta^{(s,s')}_{\nu,-k+l}}
+{A_{-k}\over A_k}c^{-(s,s')}_{kl}t^{2\delta^{(s,s')}_{\nu,k+l}}\biggr),
\label{Inu-kl-expansion}
$$
Note by the way that the function $K_{\nu,kl}(t)$ can be expressed in terms of this $I_{\nu,kl}(t)$ according to
\eq$$
K_{\nu,kl}(t)={K^\infty_{\nu,k}K^\infty_{\nu,l}\over4\lambda K^\infty_{\nu,kl}}K_{\nu,k+l}(t)
+{\pi\over2\sin\pi l}\left(I_{\nu,k.-l}(t)
-{K^\infty_{\nu,k}K^\infty_{\nu,l}\over K^\infty_{\nu,-k}K^\infty_{\nu,-l}}I_{\nu,-k.l}(t)\right).
\label{Knu-kl-Inu-kl-rel}
$$
Now notice that in the same way as (\ref{jint-Wronskian}) we can derive
\subeq{\label{cJ0m-Wr-alt}
\Align$$
\cJ^0_{kl}(t)
&=\left.2\tau\left(I_{\nu,l}(\tau)\lrd_\tau I_{\nu,k.k-l}(\tau)\right)\right|^t_0,
\label{cJ0-Wr-alt}
\\
\cJ^-_{kl}(t)
&=\left.2\tau\left(K_{\nu,l}(\tau)\lrd_\tau I_{\nu,k.k-l}(\tau)\right)\right|^\infty_t.
\label{cJm-Wr-alt}
$$}
In the first line the contribution from $\tau=0$ vanishes, while in the second line that from $\tau=\infty$ is finite. By computing the exponents in the expansions we conclude that this finite constant is $\cV^-_{kl}$. Thus we obtain
\Align*$$
F_{\nu,kl}(t)
&=tK_{\nu,l}(t)(I_{\nu,l}(t)I'_{\nu,k.k-l}(t)-I'_{\nu,l}(t)I_{\nu,k.k-l}(t))
\\
&\quad
+I_{\nu,l}(t)\cV^-_{kl}-tI_{\nu,l}(t)(K_{\nu,l}(t)I'_{\nu,k.k-l}(t)-K'_{\nu,l}(t)I_{\nu,k.k-l}(t))
\\
&=I_{\nu,l}(t)\cV^-_{kl}+tI_{\nu,k.k-l}(t)(K'_{\nu,l}(t)I_{\nu,l}(t)-K_{\nu,l}(t)I'_{\nu,l}(t))
$$
The Wronskian in the parentheses is equal to $-t^{-1}$ and we finally obtain
\begin{equation}
F_{\nu,kl}(t)=I_{\nu,l}(t)\cV^-_{kl}-I_{\nu,k.k-l}(t).
\label{Fnu-Inu-kl-rel}
\end{equation}
This makes it possible to express the function $F_{\nu,kl}(t)$ in terms of the coefficients $c^{(s,s')}_{\nu,k}$ and $c^{\pm(s,s')}_{\nu,\mp k.k-l}$, which can be calculated by computer algebra systems much faster that the coefficients $J^{\ve_1\ve_2(s,s')}_{\nu,kl}$.

At last, let us prove the identity (\ref{ck0-ck-coefs-rel}). Its first part is evident so that we prove the second part. Take the limit $l\to0$ in the identity (\ref{Knu-kl-Inu-kl-rel}). Substituting the expansion (\ref{Inu-kl-expansion}) and using the first part of (\ref{ck0-ck-coefs-rel}) we obtain
\eq$$
K_{\nu,k0}(t)=-I_{k0}(t)\log t+\sum_{s,s'\ge0}\#t^{2\delta^{(s,s')}_{\nu,-k}}+O(t^k).
\label{Knu-k0-expansion}
$$
We are not interested in the coefficients in the terms that do not contain the factor $\log t$. On the other hand, from (\ref{Phidef}) and (\ref{Knu-kl-def}) we get
\Align$$
K_{\nu,k0}(t)
&={K^\infty_{\nu,k}\over4\lambda}\,\d_\nu{\lambda K_{\nu,\kappa}(t)\over K^\infty_{\nu,\kappa}}
={1\over4}\d_\nu K_{\nu,\kappa}(t)+\sum_{s,s'\ge0}\#t^{-2\delta^{(s,s')}_{\nu,k}}+O(t^k)
\notag
\\
&={2^{k-1}}\Gamma(k)\log t\sum_{(s,s')\in\rmS^+}(s'-s)c^{(s,s')}_{\nu,-k}t^{2\delta^{(s,s')}_{\nu,-k}}
+\sum_{s,s'\ge0}\#t^{2\delta^{(s,s')}_{\nu,-k}}+O(t^k).
\label{Knu-k0-Knu-k-rel}
$$
Comparing this with (\ref{Knu-k0-expansion}) and (\ref{Inu-kl-expansion}) we obtain the identity (\ref{ck0-ck-coefs-rel}) subject to $\delta^{(s,s)}_{\nu,2k}<0$. The analyticity of the coefficients in $k$ proves this identity for all values of $k$.

\section{Semiclassical form factors of the exponential operators and their descendants}
\label{sec:formfactors}

In this section we apply the formulas (\ref{Ncorr-modif})--(\ref{FcO-GcO-rel}) to several classes of operators and obtain their form factors. We describe in detail the calculation technique in the radial quantization picture.

\subsection{Operators $\cT_nV_\nu$ and $\cT_n(\d^k\varphi\,V_\nu)$}
\label{subsec:exp-dkvarphi}

Consider the correlation functions of the operator $\cT_nV_\nu$. In the leading order in the parameter $b$ it is fully defined by the classical solution:
\begin{equation}
G^{(n)}_\nu(\{x_i\}_N)
=\langle\infty|\T_r \left[\varphi(x_n)\ldots\varphi(x_1)\right]|0\rangle_\nu=\prod^N_{j=1}b^{-1}\phi_\nu(mr_j)+O(b^{2-N}).
\label{Gnu-semicl}
\end{equation}
The form factors are determined by the large distance asymptotic $|x_i|\to\infty$ of the correlation functions. According to (\ref{phi-nu-large-t}) we have
\begin{equation}
G^{(n)}_\nu(\{x_i\}_N)\simeq(\sqrt{8\pi}\lambda)^N\prod^N_{i=1}t_i^{-1/2}\e^{-t_i},
\qquad
r_i\to\infty.
\label{G-nu-assymp}
\end{equation}
Here and below we will often use the dimensionless variables
$$
t=mr,
\qquad
t_i=mr_i.
$$
Substituting it to the reduction formula (\ref{FcO-GcO-rel}) we obtain:
$$
F^{(n)}_\nu(\{\theta_i\}_N)
\simeq\left({\sqrt2\lambda\over b}\lim_{t\to\infty}t\int_Cd\xi\,\e^{\i t\sinh(\theta+\i\xi)}\lrd_t t^{-1/2}\e^{-t}\right)^N
=\left({\sqrt2\pi\lambda\over b}\lim_{t\to\infty}(I_0(t)+I'_0(t))t^{1/2}\e^{-t}\right)^N.
$$
Finally, we have
\begin{equation}
F^{(n)}_\nu(\{\theta_i\}_N)=\left(\sqrt{4\pi}\sin\pi\nu\over\pi b\right)^N+O(b^{2-N}).
\label{F-nu-final}
\end{equation}
This answer conforms with the result of~\cite{Horvath:2021rjd}. In what follows we omit the terms $O(b^\#)$ in the formulas for form factors assuming that they are calculated in the leading order.

To write down the correlation functions of descendant operators it will be convenient to introduce the following functions. Let $\hat\a$ be any monomial of the operators $\a_{-k}$, $\ba_{-k}$ and $\Q$. Define
\begin{equation}
g_{\hat\a}(\{x_i\}_N)
=b^{-s}{\langle\infty|\varphi(x_1)\cdots\varphi(x_N)\hat\a|0\rangle_\nu\over\langle\infty|\varphi(x_1)\cdots\varphi(x_N)|0\rangle_\nu},
\label{g-hata-def}
\end{equation}
where $s$ is the order of the monomial. These ratios are easily calculated. Indeed, each operator $\a_{-k}$ must be paired with the operator $\a_k$ from one of the fields $\varphi(x_i)$, substituting the classical solution $\phi_\nu(t_i)$ by $2\i f_k(x_i)$. The cases $\ba_{-k}$ and $\Q$ are similar. The summation over pairings is the summation over the subscript $i$. In what follows we will need the following particular cases:
\subeq{\label{g-hata-partcases}
\Align$$
g_{\a_{-k}}(\{x_i\}_N)
&=2\i\sum^N_{i=1}{f_k(x_i)\over\phi_\nu(t_i)}
\to\i\left(m\over2\right)^\tk{\varkappa_{\nu,\tk}\over\Gamma(\tk)}\sum^N_{i=1}\e^{-\i\tk\xi_i},
\label{g-ak}
\\
g_\Q(\{x_i\}_N)
&={4\over n}\sum^N_{i=1}{f_*(x_i)\over\phi_\nu(t_i)}
\to{N\over n}\pi\cot\pi\nu,
\label{g-Q}
\\
g_{\a_{-k}\a_{-l}}(\{x_i\}_N)
&=-4\sum^N_{i\ne j}{f_k(x_i)f_l(x_j)\over\phi_\nu(t_i)\phi_\nu(t_j)}
\to-\left(m\over2\right)^{\tk+\tl}{\varkappa_{\nu,\tk}\over\Gamma(\tk)}{\varkappa_{\nu,\tl}\over\Gamma(\tl)}
\sum^N_{i\ne j}\e^{-\i\tk\xi_i-\i\tl\xi_j},
\label{g-akl}
\\
g_{\a_{-k}\ba_{-l}}(\{x_i\}_N)
&=-4\sum^N_{i\ne j}{f_k(x_i)\bar f_l(x_j)\over\phi_\nu(t_i)\phi_\nu(t_j)}
\to-\left(m\over2\right)^{\tk+\tl}{\varkappa_{\nu,\tk}\over\Gamma(\tk)}{\varkappa_{\nu,\tl}\over\Gamma(\tl)}
\sum^N_{i\ne j}\e^{-\i\tk\xi_i+\i\tl\xi_j},
\label{g-akbl}
$$}
where we also give the expressions in the limit $t_i\to\infty$, which is always finite, and use the notation
\eq$$
\varkappa_{\nu,k}=\lambda^{-1}K^\infty_{\nu,k}.
\label{varkappa-def}
$$
The functions $g_{\ba_{-k}}$ and $g_{\ba_{-k}\ba_{-l}}$ are obtained from (\ref{g-hata-partcases}a,c) by the substitution $f_k\to\bar f_k$, $\xi_i\to-\xi_i$.

Now consider the operator $\cT_n(\d^k\varphi\,V_\nu)$. To define it we need to make an average of $\d^k\varphi(x)$ over a small circle in the $(\zeta,\bzeta)$ plane according to the rule (\ref{cT-dkvarphi-r0-def}), (\ref{cT-dkvarphi-def}). Then we have
\begin{equation}
|\d^k\varphi\,V_\nu\rangle=\lim_{r_0\to0}\average{\d^k\varphi}_{r_0}|0\rangle_\nu=\i\Gamma(k)\a_{-k}|0\rangle_\nu.
\label{dkvarphiVnu-def}
\end{equation}
Indeed, after taking the integral the only terms in (\ref{chi-decomp}) that survive are those with $\a_{-k}$ and $\ba_k$, but $\ba_k$ annihilates the vacuum. The correlation function reads
$$
G^{(n)}_{\d^k\varphi\,V_\nu}(\{x_i\}_N)
=\i\Gamma(k)\langle\infty|\T_r\left[ \varphi(x_n)\ldots\varphi(x_1) \right]\a_{-k} |0\rangle_\nu
=\i b\Gamma(k) \,g_{\a_{-k}}\qty(\{x_i\}_N) \,G^{(n)}_\nu\qty(\{x_i\}_N).
$$
From (\ref{g-ak}) we obtain
\begin{equation}
G^{(n)}_{\d^k\varphi\,V_\nu}(\{x_i \}_N)
=b\left(m\over2\right)^\tk\gamma_n(k)\varkappa_{\nu,\tk}
\sum^N_{i=1}\e^{-\i\tk\xi_i}\cdot G^{(n)}_\nu(\{x_i\}_N)\quad\text{as $r_i\to\infty$,}
\label{G-dkvarphiVu-Gnu-rel}
\end{equation}
where
\eq$$
\gamma_n(k)={\Gamma(k)\over\Gamma(k/n)}.
\label{gamma-n-k-def}
$$
From (\ref{Iint-Bessel-rel}) we immediately obtain for the form factor
\begin{equation}
F^{(n)}_{\d^k\varphi\,V_\nu}(\{\theta_i\}_N)
=\e^{{\i\pi\over2}\tk}b\left(m\over2\right)^\tk\gamma_n(k)\varkappa_{\nu,\tk}
\sum^N_{i=1}\e^{\tk\theta_i}\cdot F^{(n)}_\nu(\{\theta_i\}_N).
\label{FF1desc}
\end{equation}

The form factors of the operators
$\cT_n(\bd^k\varphi\,V_\nu)$ are obtained from this formula according to the rule
\begin{equation}
\d\leftrightarrow\bd\quad\Leftrightarrow\quad\theta_i\to-\i\pi-\theta_i.
\label{dbd-rule}
\end{equation}
This rule is general and applicable to all operators considered below.

\subsection{Operators $\cT_n(\d^k\varphi\,\d^l\varphi\,V_\nu)$}
\label{subsec:dkvarphi-dlvarphi}

In what follows we consider the operators of the form ${\cT_n}(\d^k\varphi\,{\tilde\d}^l\varphi\,V_\nu)$, where $\tilde\d$ is either $\d$ or $\bd$. The leading contribution to the form factors of these operators is of the order $b^{2-N}$. As it was shown in \cite{Lashkevich:2023hzk} it is necessary to calculate these form factors up to the first order in the term $S^{(3)}$ in the action (\ref{Sint}). In the order $b^{2-N}$ we have
\eq$$
G^{(n)}_{\d^k\varphi\,\tilde\d^l\varphi\,V_\nu}(\{x_i\}_N)
=\langle\infty|\T_r[\varphi(x_1)\cdots\varphi(x_N)\,
\average{\tilde\d^l\varphi}_{r_0}\average{\d^k\varphi}_{r_{01}}(1-S^{(3)})]|0\rangle_\nu
+\text{(c.t.)}\Bigr|_{r_{01}<r_0\to0}.
\label{G-dk-tdl-genform}
$$
This function consists of two terms corresponding to two terms of the factor $1-S^{(3)}$. In the first term we easily take the limit $r_{01}\to0$:
\eq$$
\average{\tilde\d^l\varphi}_{r_0}\average{\d^k\varphi}_{r_{01}}|0\rangle_\nu\Bigr|_{r_{01}\to0}
=\i\Gamma(k)\,\average{\tilde\d^l\varphi}_{r_0}\,\a_{-k}|0\rangle_\nu.
\label{dk-tdl-r01-lim}
$$
In the second term the situation is more subtle. Since we are interested in the case $|x_i|\to\infty$, consider the vicinity of the origin. Due to the $r$ ordering the integral in $S^{(3)}$ splits into three parts:
\Align$$
\T_r[\average{\tilde\d^l\varphi}_{r_0}\average{\d^k\varphi}_{r_{01}}S^{(3)}]|0\rangle_\nu
&=\average{\tilde\d^l\varphi}_{r_0}\average{\d^k\varphi}_{r_{01}}\int_{|y|<r_{01}}d^2y\,\lcolon\chi^3(y)\rcolon
bH^{(3)}_\nu(y)
|0\rangle_\nu
\notag
\\
&\quad
+\average{\tilde\d^l\varphi}_{r_0}\int_{r_{01}<|y|<r_0}d^2y\,\lcolon\chi^3(y)\rcolon
bH^{(3)}_\nu(y)\,\average{\d^k\varphi}_{r_{01}}|0\rangle_\nu
\notag
\\
&\quad
+\int_{|y|>r_0}d^2y\,\lcolon\chi^3(y)\rcolon
bH^{(3)}_\nu(y)\,\average{\tilde\d^l\varphi}_{r_0}\average{\d^k\varphi}_{r_{01}}
|0\rangle_\nu.
\label{S3-dk-tdl-contribs}
$$
It can be checked that the contribution of the first term is of the order $O(r_{01}^{2-4|\nu|})$ and may be neglected. In other words, we may set $r_{01}=0$:
\Align$$
\T_r[\average{\tilde\d^l\varphi}_{r_0}\average{\d^k\varphi}_{r_{01}}S^{(3)}]|0\rangle_\nu\Bigr|_{r_{01}\to0}
&=\i\Gamma(k)\,\average{\tilde\d^l\varphi}_{r_0}\int_{|y|<r_0}d^2y\,\lcolon\chi^3(y)\rcolon
bH^{(3)}_\nu(y)\a_{-k}|0\rangle_\nu
\notag
\\
&\quad
+\i\Gamma(k)\int_{|y|>r_0}d^2y\,\lcolon\chi^3(y)\rcolon
bH^{(3)}_\nu(y)\,\average{\tilde\d^l\varphi}_{r_0}\a_{-k}|0\rangle_\nu.
\label{S3-dk-tdl-fin}
$$
In what follows we write $\cdots\average{\d^k\varphi}_0\cdots$ instead of $\cdots\average{\d^k\varphi}_{r_{01}}\cdots|_{r_{01}\to0}$ for shorthand.

Now turn to the case $\tilde\d=\d$, which is the main subject of this subsection. In this case we also can take the limit $r_0\to0$ in both (\ref{dk-tdl-r01-lim}) and (\ref{S3-dk-tdl-fin}). Namely,
\eq$$
\average{\d^l\varphi}_{r_0}\average{\d^k\varphi}_0|0\rangle_\nu\Bigr|_{r_0\to0}
=-\Gamma(k)\Gamma(l)\a_{-l}\a_{-k}|0\rangle_\nu.
\label{dk-dl-lim}
$$
The integral over the region $|y|<r_0$ in (\ref{S3-dk-tdl-fin}) turns out to be of the order $O(r_0^{2-4|\nu|})$ and vanishes in the limit $r_0\to0$. Thus we have
\begin{equation}
\T_r[\average{\d^l\varphi}_{r_0}\average{\d^k\varphi}_0S^{(3)}]|0\rangle_\nu\Bigr|_{r_0\to0}
=-\Gamma(k)\Gamma(l)S^{(3)}\a_{-l}\a_{-k}|0\rangle_\nu.
\label{S3-dk-dl-lim}
\end{equation}
Expanding the operator $\lcolon\chi^3(y)\rcolon$ in $S^{(3)}$ according to (\ref{chi-decomp}) we find that only one term of the expansion contributes the correlation function:
\Align$$
S^{(3)}\a_{-l}\a_{-k}|0\rangle_\nu
&={bm^2\over{2^{\delta_{kl}}}\cdot8\pi}\int d^2x\,{\a_{-k-l}f_{-k-l}(x)\over-\i(k+l)}{\a_kf_k(x)\over\i k}{\a_lf_l(x)\over\i l}
\sinh\phi_\nu(m|x|)\cdot\a_{-l}\a_{-k}|0\rangle_\nu
\notag
\\
&=-\i{bn\over k+l}\int^\infty_0 dt\,tu_{-k-l}(t)u_k(t)u_l(t)\sinh\phi_\nu(t)\cdot\a_{-k-l}|0\rangle_\nu
\notag
\\
&= -4\i b{\Gamma(\tk+\tl)\over\Gamma(\tk)\Gamma(\tl)}\cV^+_{\tk\tl}\a_{-k-l}|0\rangle_\nu,
\label{S3-akal-action}
$$
where we used (\ref{cJp-cVp}). Thus we obtain in the leading order
\begin{equation}
|\d^k\varphi\,\d^l\varphi\,V_\nu\rangle
=-\Gamma(k)\Gamma(l)\left[\a_{-k}\a_{-l}+{\Gamma(\tk+\tl)\over\Gamma(\tk)\Gamma(\tl)}4\i b\cV^+_{\tk\tl}\a_{-k-l}\right]|0\rangle_\nu.
\label{kl-noren}
\end{equation}
The expression for the operators $\cT_n(\bd^k\varphi\,\bd^l\varphi\,V_\nu)$ is the same with the replacement $\a_\#\to\ba_\#$. Note that the result is finite and does not demand counterterms. We will see later that this is a general property of all chiral descendant operators.

For the correlation functions we have
\eq$$
{G^{(n)}_{\d^k\varphi\,\d^l\varphi\,V_\nu}(\{x_i\}_N)\over G^{(n)}_\nu(\{x_i\}_N)}
=-b^2\Gamma(k)\Gamma(l)\left(g_{\a_{-k}\a_{-l}}(\{x_i\}_N)
+{4\i\Gamma(\tk+\tl)\over\Gamma(\tk)\Gamma(\tl)}\cV^+_{\tk\tl}g_{\a_{-k-l}}(\{x_i\}_N)\right).
\label{G-kl-fin}
$$
Substituting this to (\ref{FcO-GcO-rel}) and using (\ref{Iint-Bessel-rel}) we obtain
\Multline$$
F^{(n)}_{\d^k\varphi\,\d^l\varphi\,V_\nu}(\{\theta_i\}_N)
=\e^{{\i\pi\over2}(\tk+\tl)}b^2\left(m\over2\right)^{\tk+\tl}\gamma_n(k)\gamma_n(l)
\\
\times
\left(\varkappa_{\nu,\tk}\varkappa_{\nu,\tl}\sum^N_{i\ne j}\e^{\tk\theta_i+\tl\theta_j}
-4\varkappa_{\nu,\tk+\tl}\cV^+_{\tk\tl}\sum^N_{i=1}\e^{(\tk+\tl)\theta_i}\right)
F^{(n)}_\nu(\{\theta_i\}_N).
\label{F-kl-fin}
$$
The form factors of the operators $\cT_n(\bd^k\varphi\,\bd^l\varphi\,V_\nu)$ are obtained according to the general rule (\ref{dbd-rule}).

\subsection{Renormalization procedure and operator $\cT_n(\d\bd\varphi\,V_\nu)$}
\label{subsec:dbdvarphi}

From now on we discuss the operators that demand renormalization. The renormalization procedure is known from the conformal perturbation theory \cite{Zamolodchikov:1990bk}. Recall the aspects of this theory that are relevant to our consideration. In the conformal perturbation theory the action $S$ of a field theory is split as follows
\begin{equation}
S=S_\text{CFT}+\mu\int_{\R^2}d^2x\,\Phi_\text{p}(x),
\label{CPT-S}
\end{equation}
where $S_\text{CFT}$ is the action of the conformal field theory that corresponds to the ultraviolet fixed point of the field theory under consideration, and $\Phi_\text{p}(x)$ is a primary spinless local operator of conformal dimension $\Delta_\text{p}<1$. Let $\{\cO_i\}$ be a complete set of linearly independent quasiprimary local operators in the ultraviolet conformal theory with conformal dimensions $(\Delta_i,\bar\Delta_i)$. The second term in (\ref{CPT-S}) is considered as a perturbation action. The terms of the perturbation expansion may contain ultraviolet (UV) divergences in the vicinity of the operators $\cO_i(x)$. With the spacial UV cutoff $r_0$ the term of the order $k$ looks like
\begin{equation}
(-\mu)^k\int_{|y_i-x|>r_0} d^{2k}y\,\Phi_\text{p}(y_1)\cdots\Phi_\text{p}(y_k)\cO_i(x) = \sum_{j:\,{\delta^{(k)}{}^j_i\le0}}U^{(k)}{}_i^j f^{(k)}{}^j_i(r_0)\cO_j(x)
+\text{(UV-finite terms)},
\label{CPT-k-order}
\end{equation}
where $U^{(k)}{}^j_i$ are $c$\-/number coefficients and
\begin{equation}
f^{(k)}{}^j_i(r)=\Cases{{r^{{2\delta^{(k)}{}^j_i}}\over2\delta^{(k)}{}^j_i},&{\delta^{(k)}{}^j_i<0};\\
\log{\rho\over r},&\delta^{(k)}{}^j_i=0,}
\qquad
\delta^{(k)}{}^j_i=\Delta_j-\Delta_i+k(1-\Delta_\text{p}).
\label{delta-exp-def}
\end{equation}
The scale $\rho$ is arbitrary. Correlation functions can be made UV finite by adding counterterms to each operator. Namely, the renormalized operator has the form
\begin{equation}
\cO^\ren_i(x)
=\cO_i(x)+\sum_{k>0}\>\sum_{i:\,{\delta^{(k)}{}^j_i\le0}}V^{(k)}{}^j_if^{(k)}{}^j_i(r_0)\cO_j(x){}\biggr|_{r_0\to0}
\label{CPT-cO-ren}
\end{equation}
with finite coefficients $V^{(k)}{}^j_i$, which can be expressed in terms of the coefficients $U^{(k)}{}^j_i$. Note that the operators $\cO_j$ that enter the r.h.s.\ of (\ref{CPT-k-order}), (\ref{CPT-cO-ren}) with nonzero coefficient are not arbitrary. They must exist in the operator product expansion of $k$ instances of $\Phi_\text{p}$ and $\cO_i$. In particular, they have the same spin: $\Delta_j-\bar\Delta_j=\Delta_i-\bar\Delta_i$.

The case $\delta^{(k)}{}^j_i=0$ is special. The appearance of the arbitrary dimensional constant $\rho$ leads to an ambiguity in the renormalization procedure. If either the theory or the operator depends on a continuous parameter $p$, and the quantity $\delta^{(k)}{}^j_i$ vanishes at a certain value $p_0$ of this parameter, this leads to a pole of correlation functions of the operator $\cO_i$ at $p=p_0$. This is why this case is called the resonant one.

In the case of the sinh\-/Gordon model the UV conformal field theory is the free massless field theory, while the perturbation is the hyperbolic cosine term:
\eq$$
\Gathered{
S_\text{CFT}=\int_{\cM_n}d^2x\,{(\d_\mu\varphi)^2\over16\pi}=\sum^n_{s=1}\int_{\R^2}d^2x\,{(\d_\mu\varphi^{(s)})^2\over16\pi},
\\
\Phi_\text{p}=\sum^n_{s=1}\cosh b\varphi^{(s)},
\qquad
\mu={m_0^2\over8\pi b^2},
\qquad
\Delta_\text{p}=-b^2.
}\label{CPT-shG}
$$
The conformal perturbation theory describes correlation functions at the distances $r\ll m^{-1}$, while the semiclassical approximation is valid for $r\gg\e^{-b^{-2}}m^{-1}$. The validity regions of these two expansions overlap, but the semiclassical theory describes larger distances. It means that though the structure of the renormalization terms in the semiclassical theory should be the same, the functions $f^{(k)}{}^j_i(r)$ may differ from those of the conformal perturbation theory by terms of the order $o(r^{2\delta^{(k)}{}^j_i})$. We will discuss it later on particular examples.

It can be checked that the exponential operators $\cT_nV_\nu$ do not need renormalization for $|\nu|<{1\over2}$. The chiral descendants, which are descendant operators of the maximal and the minimal spin on each level, do not need renormalization as well, since there are no operators of the same spin and lower conformal dimension to be admixed. In the rest of the section we consider non\-/chiral descendant operators and their renormalization.

Our first example is the operator $\cT_n(\d\bd\varphi\,V_\nu)$. This operator is interesting from the point of view that it is completely absent in the conformal limit $m\to0$, but cannot be discarded immediately in the massive case. Let us regularize it analogously to (\ref{cT-dkvarphi-r0-def}). The corresponding state in the radial quantization reads
\begin{equation}
|\average{\d\bd\varphi}_{r_0}\,V_\nu\rangle=\average{\d\bd\varphi}_{r_0}\,|0\rangle_\nu.
\end{equation}
From the equations of motion (\ref{shG-eq-zeta}) and (\ref{chi-eom}) we obtain
\begin{equation}
\d\bd\left(b^{-1}\phi_\nu(m|z|)+\chi(z,\bz)\right)
={n^2m^2\over4}|z|^{2-2/n}\left(b^{-1}\sinh\phi_\nu(m|z|)+\cosh\phi_\nu(m|z|)\chi(z,\bz)\right).
\end{equation}
After averaging over the circle $|z|=r_0$ and applying to the vacuum, only the term with $\Q$ survives in the decomposition of the quantum field $\chi$, and we get
\begin{align}
\average{\d\bd\varphi}_{r_0}|0\rangle_\nu
&={m^2n^2r_0^{2-2/n}\over4}\left(b^{-1}\sinh\phi_\nu(mr_0)+\Q\cosh\phi_\nu(mr_0)\right)|0\rangle_\nu
\notag
\\
&={m^2n^2r_0^{2-2/n}\over8b}\left(\e^{\phi_\nu(mr_0)}(1+b\Q)-\e^{-\phi_\nu(mr_0)}(1-b\Q)\right) |0\rangle_\nu
\notag
\\
&\approx{m^2n^2r_0^{2-2/n}\over8b}\left(\e^{\phi_\nu(mr_0)}|0\rangle_{\nu+b^2/n}+\e^{-\phi_\nu(mr_0)}|0\rangle_{\nu-b^2/n}\right).
\label{parbarpar-renorm}
\end{align}
The last line is valid for small values of $b^2$. Now let us write it in the leading order in $r_0$:
$$
\average{\d\bd\varphi(z,\bz)}_{r_0}|0\rangle_\nu
\approx{m^2n^2\over8b}\left( r_0^{2-2/n-4\nu}|0\rangle_{\nu+b^2/n}-r_0^{2-2/n+4\nu}|0\rangle_{\nu-b^2/n}\right).
$$
It means that
$$
\cT_n(\average{\d\bd\varphi}_{r_0}V_\nu(0))
\approx{m^2n^2\over8b}\left(r_0^{{2\delta^{(1)}{}^{\nu+b^2/n}_\nu}}\cT_nV_{\nu+b^2/2}(0)
+r_0^{{2\delta^{(1)}{}^{\nu-b^2/n}_\nu}}\cT_nV_{\nu-b^2/2}(0)\right),
$$
where $\delta^{(1)}{}^{\nu\pm b^2/n}_\nu={1-1/n\mp2\nu}$ are particular cases of the exponents $\delta^{(k)}{}^j_i$ defined in (\ref{delta-exp-def}). It is consistent with decomposition (\ref{CPT-k-order}) in the first order of the conformal perturbation theory. In the region $|\nu|<\tfrac12(1-1/n)$ the exponents are both positive and we can see that the operator $\cT_n(\d\bd\varphi\,V_\nu)=0$. Off this region at least one of the exponents is negative and the operator demands renormalization. Off the conformal perturbation theory applicability region the functions $f^{(k)}{}^j_i(r)$ deviate from the form given in (\ref{delta-exp-def}). Thus define the renormalized operator $\cT_n(\d\bd\varphi\,V_\nu)$ as follows
\eq$$
\left[\cT_n(\d\bd\varphi\,V_\nu)\right]^\ren
=\left.\cT_n(\average{\d\bd\varphi}_{r_0}V_\nu)
-{m^2r_0^{2-2/n}n^2\over8b}\left(\e^{\phi_\nu(mr_0)}\cT_nV_{\nu+b^2/n}+\e^{\phi_\nu(mr_0)}\cT_nV_{\nu-b^2/n}\right)\right|_{r_0\to0}
\label{dbdV-ren-def}
$$
independently of the values of $\nu$. We conclude that%
\footnote{The exception is the special case $n=1$, $\nu=0$, where there is no need for renormalization, and the quantum equation of motion $4\d\bd\varphi=b^{-1}m^2\sinh b\varphi$ holds\cite{Lashkevich:2023hzk}.}
\begin{equation}
[\cT_n(\d\bd\varphi\,V_\nu)]^\ren=0
\label{dbd-ren-fin}
\end{equation}
at least in the leading order in $b^2$. We conjecture that this identity is exact beyond the perturbation theory in $b$ and that all descendants containing $\d\bd\varphi$ drop out from the theory after the renormalization.

\subsection{Operators $\cT_n(\d^k\varphi\,\bd^l\varphi\,V_\nu)$ for $k>l$}
\label{subsec:dkvarphi-bdlvarphi}

These operators generally need renormalization, but in the semiclassical limit the renormalization for $k\ne l$ and $k=l$ is carried out differently. We start from the simpler case $k\ne l$ assuming $k>l$ for definiteness.

Again, let us start from the regularized operator $\cT_n\left(\average{\bd^l\varphi}_{r_0}\,\average{\d^k\varphi}_{r_{01}}\,V_\nu\right)$ with $r_0>r_{01}$. As we found above we may set $r_{01}=0$. After averaging $\bd^l\varphi$ over the circle only two terms in the expansion (\ref{chi-decomp}) survive:
\eq$$
\average{\bd^l\varphi}_{r_0}=\i l^{-1}\left(\average{(\bd^l\bar f_{-l})}_{r_0}\ba_{-l}-\average{(\bd^lf_l)}_{r_0}\a_l\right).
\label{av-bdl-r0}
$$
In the r.h.s.\ of (\ref{dk-tdl-r01-lim}) the term with $\a_l$ vanishes, while the derivative $\bd^l\bar f_{-l}(x)=l!+O(r)$ is finite as $r\to0$. Hence, the zeroth order contribution is finite:
\begin{equation}
\average{\bd^l\varphi}_{r_0}\,\average{\d^k\varphi}_0|0\rangle_\nu\Bigr|_{r_0\to0}
= -\Gamma(k)l^{-1}\average{(\bd^l\bar f_{-l})}_{r_0}\ba_{-l}\a_{-k}|0\rangle_\nu\Bigr|_{r_0\to0}
= -\Gamma(k)\Gamma(l)\ba_{-l}\a_{-k}|0\rangle_\nu.
\end{equation}
Now let us write down explicitly the r.h.s.\ of (\ref{S3-dk-tdl-fin}):
\begin{multline}
\T_r\bigg[\average{\bd^l\varphi}_{r_0}\average{\d^k\varphi}_0S^{(3)}\bigg]|0\rangle_\nu
\\
=-{\i b\Gamma(k)\Gamma(\tk-\tl)\over n\tl\Gamma(\tk)\Gamma(\tl)}\,
\bd^l\left(4\left(m\over2\right)^{2\tl}\bar f_{-l}(x)\cJ^-_{\tk\tl}(m|x|)
+2\tl\Gamma^2(\tl)f_l(x)\cJ^0_{\tk\tl}(m|x|)\right)\a_{l-k}|0\rangle_\nu\Bigr|_{|x|=r_0}
\\
=-{4\i b\Gamma(k)\Gamma(\tk-\tl)\over n\Gamma(\tk)}\left(m\over2\right)^\tl\,
\bd^l\left(\bzeta\over\zeta\right)^{l/2}F_{\nu,\tk\tl}\left(m(\zeta\bzeta)^{n/2}\right)\a_{l-k}|0\rangle_\nu\Bigr|_{|x|=r_0}.
\label{bdl-dk-S3}
\end{multline}
Substituting here (\ref{Fnu-Inu-kl-rel}) we obtain
\begin{multline}
\T_r\qty[\average{\bd^l\varphi}_{r_0}\average{\d^k\varphi}_0(1-S^{(3)})]|0\rangle_\nu
=\left(-\Gamma(k)\Gamma(l)+O\bigl(r_0^{2-4|\nu|}\bigr)\right)\a_{-k}\ba_{-l}|0\rangle_\nu
\\
+\i b\left(4\left(m\over2\right)^{2\tl}\gamma_n(k)\gamma_n(l)\Gamma(\tk-\tl)\cV^-_{\tk\tl}
+m^{2\tl}\sum_{(s,s')\in\rmS^+}{\cC^{(s,s')}_{\nu,kl}}(mr_0)^{{2\delta^{(s,s')}_{\nu,-2\tl}}}
+{o(r_0^0)}\right)\a_{l-k}|0\rangle_\nu,
\label{bdl-dk-1-S3-fin}
\end{multline}
where
\eq$$
\cC^{(s,s')}_{\nu,kl}
=-{2\Gamma(k)\over k-l}\left(n\delta^{(s,s')}_{\nu,0}\right)_lc^{+(s,s')}_{\nu,-\tk.\tk-\tl}
\label{cC-kl-def}
$$
and $(x)_k$ is the descending factorial:
\begin{equation}
(x)_k = x(x-1)(x-2)\ldots(x-k+1)={\Gamma(x+1)\over\Gamma(x-k+1)}.
\label{xk-fact-def}
\end{equation}
The divergent terms here are related to those in the conformal perturbation theory. Indeed, for $(s,s')\in\rmS^+$ the operator $\cT_n(\d^{k-l}\varphi\,V_{\nu+(s-s')b^2/n})$ should appear in the order $s+s'$ of the conformal perturbation theory. Let $\cO_i=\cT_n(\d^k\varphi\,\bd^l\varphi\,V_\nu)$, $\cO_j=\cT_n(\d^{k-l}\varphi\,V_{\nu+(s-s')b^2/n})$. Then the corresponding exponent defined in (\ref{delta-exp-def}) reads
\begin{equation}
\delta^{(s+s')}{}^j_i
=\delta^{(s,s')}_{\nu,{-2\tl}}+\eta^{(s,s')}b^2,
\qquad
\eta^{(s,s')}=s+s'-{(s-s')^2\over n}.
\label{delta-delta-compare}
\end{equation}
Notice that all terms in (\ref{bdl-dk-1-S3-fin}) provide contributions of the same order in $b^2$ in correlation functions. Thus we may assume $|0\rangle_{\nu+(s-s')b^2/n}\simeq|0\rangle_\nu$, and the renormalization subtracts the terms with $\delta^{(s,s')}_{\nu,{-2\tl}}<0$.

Off the special points, which will be discussed later, define the renormalized operator
\Align$$
\left[\cT_n(\d^k\varphi\,\bd^l\varphi\,V_\nu)\right]^\ren
&=\cT_n(\average{\bd^l\varphi}_{r_0}\d^k\varphi\,V_\nu)
\notag
\\
&\quad
-\i bm^{2\tl}\sum_{(s,s')\in\rmS^+}{\cC^{(s,s')}_{\nu,kl}}(mr_0)^{{2\delta^{(s,s')}_{\nu,-2\tl}}}
\cT_n(\d^{k-l}\varphi\,V_{\nu+(s-s')b^2/n})\Biggr|_{r_0\to0}.
\label{klbar-rendef}
$$
In the sum over $(s,s')$ we may not restrict to the terms with negative exponents, because the terms with positive exponents tend to zero in the limit $r_0\to0$. This results in
\begin{equation}
\bigl|[\d^k\varphi\,\bd^l\varphi\,V_\nu]^\ren\bigr\rangle
=-\Gamma(k)\Gamma(l) \left(\a_{-k}\ba_{-l}
+{4\i b\Gamma(\tk-\tl)\over\Gamma(\tk)\Gamma(\tl)}\left(m\over2\right)^{2\tl}\cV^-_{\tk\tl}\a_{l-k}\right)|0\rangle_\nu.
\label{klbar-afterren}
\end{equation}
After the renormalization we obtain the correlation functions
\eq$$
{G^{(n)}_{\d^k\varphi\,\bd^l\varphi\,V_\nu}(\{x_i\}_N)\over G^{(n)}_\nu(\{x_i \}_N)}
=-b^2\Gamma(k)\Gamma(l)\left(g_{\a_{-k}\ba_{-l}}(\{x_i\}_N)
+{4\i\Gamma(\tk+\tl)\over\Gamma(\tk)\Gamma(\tl)}\left(m\over2\right)^{2\tl}\cV^-_{\tk\tl}g_{\a_{-k+l}}(\{x_i \}_N)\right)
\label{G-kbl-fin}
$$
and the form factors
\Multline$$
F^{(n)}_{\d^k\varphi\,\bd^l\varphi\,V_\nu}(\{\theta_i\}_N)
=\e^{{\i\pi\over2}(\tk-\tl)}b^2\left(m\over2\right)^{\tk+\tl}\gamma_n(k)\gamma_n(l)
\\
\times
\left(\varkappa_{\nu,\tk}\varkappa_{\nu,\tl}\sum^N_{i\ne j}\e^{\tk\theta_i-\tl\theta_j}
-4\varkappa_{\nu,\tk-\tl}\cV^-_{\tk\tl}\sum^N_{i=1}\e^{(\tk-\tl)\theta_i}\right)
F^{(n)}_\nu(\{\theta_i\}_N).
\label{F-kbl-fin}
$$
Recall that this expression is valid subject to $k>l$.

Now consider the threshold values $\delta^{(\sigma,\sigma')}_{\nu,-2\tl}=0$ for some $(\sigma,\sigma')\in\rmS^+$. Though physically $l$ is an integer number, it is convenient to consider it as a continuous variable. Let
\eq$$
\tl_0=l_0/n=\delta^{(\sigma,\sigma')}_{\nu,0}
\label{l0-def}
$$
and consider the behavior of the functions $\cV^-_{kl}$ and $\cC^{(\sigma,\sigma')}_{\nu,\tk\tl}$ in the vicinity of $l=l_0$. At this point the operator $\cO_i$ is very close to the resonance with $\cO_j$ for small values of $b$. Notice that the operator $\cO_i$ here is generally not a quasiprimary one, but a combination of a quasiprimary operator of the level $(k,l)$ and spacial derivatives of quasiprimary operators of lower levels. Thus the threshold does not lead necessarily to a resonance pole.  Indeed, as we checked for $\sigma+\sigma'\le4$ there is a pole, if $|\sigma-\sigma'|\le1$, and there is no pole of the mentioned functions, if $|\sigma-\sigma'|\ge2$. This takes place in both cases $l\ne k$ and $l=k$. We expect it to be correct for larger values of $\sigma+\sigma'$. In Appendix~\ref{app:perturbations} we compare this with the results of the conformal perturbation theory in the case $l=k$ and $\sigma=1,2$, $\sigma'=0$. We show that the pole at $\delta^{(1)}{}^i_j=0$ always exists. There is no pole at $\delta^{(2)}{}^i_j=0$ for odd values of $k$, while for even values of $k$ this pole and one more neighboring pole are canceled by two neighboring zeros in the limit $b\to0$.

In the case $|\sigma-\sigma'|\ge2$ it is sufficient to redefine the finite part. Namely, in the final expressions (\ref{G-kbl-fin}), (\ref{F-kbl-fin}) for the correlation functions and form factors we have to substitute
\eq$$
\cV^-_{\tk\tl_0}\to\cV^-_{\tk\tl_0}
+4^{\tl_0-1}\left(\gamma_n(k)\gamma_n(l_0)\Gamma(\tk-\tl_0)\right)^{-1}\cC^{(\sigma,\sigma')}_{\nu,kl_0}.
\label{cVm-subst}
$$

The resonant poles exist if $|\sigma-\sigma'|\le1$, that is in the three cases:
\eq$$
\Aligned{
&\textrm{(a)}&\tl_0
&=2\sigma,
\qquad
&&\sigma=\sigma',
\\
&\textrm{(b)}&\tl_0
&=2\sigma'+1-2\nu,
\qquad
&&\sigma-\sigma'=1,
\\
&\textrm{(c)}&\tl_0
&=2\sigma+1+2\nu,
\qquad
&&\sigma-\sigma'=-1.
}\label{tk-resonance}
$$
Consider the vicinity of such a point. Since the renormalized operator should be finite, we expect that
\eq$$
\Res_{l=l_0}\left(\cV^-_{\tk\tl}+4^{\tl-1}\left(\gamma_n(k)\gamma_n(l)\Gamma(\tk-\tl)\right)^{-1}\cC^{(\sigma,\sigma')}_{\nu,kl}\right)=0.
\label{cV-cC-res-zero}
$$
Due to (\ref{cVm-Res}), (\ref{c-Res-rel}) this identity is valid at least for $\sigma+\sigma'\le4$. In this case the renormalized operator is not defined uniquely due to a logarithmic contribution to the divergence. It is defined up to adding the operator $\cT_n(\d^{k-l}\varphi\,V_\nu)$ with an arbitrary finite coefficient. As we will see below, in the case $l=k$ a more detailed description is available.

\subsection{Operators $\cT_n(\d^k\varphi\,\bd^k\varphi\,V_\nu)$}
\label{subsec:dkvarphi-bdkvarphi}

The case $l=k$ in the series $\cT_n(\d^k\varphi\,\bd^l\varphi\,V_\nu)$ is special due to two reasons. First, the limit $r_0\to0$ is divergent even in the zeroth order of the perturbation theory. Second, the first order contribution results in the operator $\Q$ instead of $\a_{l-k}$. The zero mode $\Q$ effectively acts as the generator of a shift in the variable $\nu$. This changes the structure of the divergent part.

Let us calculate the r.h.s.\ of (\ref{dk-tdl-r01-lim}). From (\ref{av-bdl-r0}) we obtain
\begin{multline}
\average{\bd^k\varphi}_{r_0}\average{\d^k\varphi}_0|0\rangle_\nu
=-k^{-1}\left(\average{(\bd^k\bar f_{-k})_{r_0}}\ba_{-k}-\average{(\bd^kf_k)}_{r_0}\a_k\right)\Gamma(k)\a_{-k}|0\rangle_\nu
\\
=\left(-\Gamma^2(k){}+O\bigl(r_0^{2-4|\nu|}\bigr)\right)\ba_{-k}\a_{-k}|0\rangle_\nu
+2\Gamma(k)\average{(\bd^kf_k)}_{r_0}|0\rangle_\nu.
\label{dk-bdk-r01-lim}
\end{multline}

Now let us calculate the r.h.s.\ of (\ref{S3-dk-tdl-fin}):
\begin{multline}
\T_r[\average{\bd^k\varphi}_{r_0}\average{\d^k\varphi}_0S^{(3)}]|0\rangle_\nu
={bm^2\over8\pi}\,\bd^k\Biggl(-\int_{|y|>|x|}d^2y\,\Q f_0(y){f_k(y) \a_k \over\i k}{{\bar f}_k(y)\ba_k\over\i k}
{{\bar f}_{-k}(z)\ba_{-k}\over k}\sinh\phi_\nu(m|y|)
\\
+\int_{|y|<|x|}d^2y\,{f_k(z)\a_k\over k}{f_k(y)\a_k\over\i k}\Q f_0(y)
{f_{-k}(y)\a_{-k}\over-\i k}\sinh\phi_\nu(m|y|)\Biggr)\Gamma(k)\a_{-k}|0\rangle_\nu\Biggr|_{|x|=r_0}
\\
=\Gamma(k)\,\bd^k\left({4(m/2)^{2\tk}\over\tk\Gamma^2(\tk)}{\bar f}_{-k}(x)\cJ^-_{\tk\tk}(m|x|)
+2f_k(x)\cJ^0_{\tk\tk}(m|x|)\right)b\Q|0\rangle_\nu\Biggr|_{|x|=r_0}.
\label{S3-dk-bdk-r01-lim}
\end{multline}
Combining (\ref{dk-bdk-r01-lim}) and (\ref{S3-dk-bdk-r01-lim}) we obtain
\begin{multline}
\T_r\left[{\average{\bd^k\varphi(z)}}_{r_0}\average{\d^k\varphi(z_1)}_0(1-S^{(3)})\right]|0\rangle_\nu
=\left(-\Gamma^2(k)+O\bigl(r_0^{2-4|\nu|}\bigr)\right)\a_{-k}\ba_{-k}|0\rangle_\nu
\\
+2\Gamma(k)\,\bd^k\left(f_k(x) \left(1-\cJ^0_{\tk\tk}(m|x|)b\Q\right)
-{2(m/2)^{2\tk}\over\tk\Gamma^2(\tk)}\bar f_{-k}(x)\cJ^-_{\tk\tk}(m|x|)b\Q\right)|0\rangle_\nu\Biggr|_{|x|=r_0}.
\label{kk-calc}
\end{multline}

Let us calculate the expression in the parentheses in the second term. It reads
$$
2\left(m\over2\right)^\tk\Gamma^{-1}(\tk)\e^{-\i\tk\xi}
\left(K_{\nu,\tk}(t)-F_{\nu,\tk0}(t)\,b\Q\right).
$$
Apply (\ref{Fnu-Inu-kl-rel}) to expression in the parentheses and expand it using (\ref{Igen}), (\ref{Knu-k-Inu-k-rel}) and (\ref{Inu-kl-expansion}):
\Multline$$
K_{\nu,\tk}(t)-F_{\nu,\tk0}(t)b\Q
=K_{\nu,\tk}(t)-I_{\nu,\tk}(t)\cV^-_{\tk\tk}\,b\Q+I_{\nu,\tk0}(t)b\Q
\\
={\alpha_{\nu,\tk}-\cV^-_{\tk\tk}\,b\Q\over2^\tk\Gamma(\tk+1)}t^\tk+2^{\tk-1}\Gamma(\tk)t^{-\tk}
+2^{\tk-1}\Gamma(\tk)\sum_{(s,s')\in\rmS^+}c^{(s,s')}_{\nu,-\tk}(1+(s-s')\,b\Q)t^{2\delta^{(s,s')}_{\nu,-k}}
+o(t^\tk).
\label{KmFQ}
$$
Here we used the identity (\ref{ck0-ck-coefs-rel}). Substituting this into (\ref{kk-calc}) we obtain
\begin{multline}
\T_r\left[{\average{\bd^k\varphi(z)}}_{r_0}\average{\d^k\varphi(z_1)}_0(1-S^{(3)})\right]|0\rangle_\nu
\\
=\left(-\Gamma^2(k)+O\bigl(r_0^{2-4|\nu|}\bigr)\right)\a_{-k}\ba_{-k}|0\rangle_\nu
+m^{2\tk}\bfC_k|0\rangle_\nu
+m^{2\tk}\sum_{(s,s')\in\rmS^+}(mr_0)^{{2\delta^{(s,s')}_{\nu,-2\tk}}}\bfC^{(s,s')}_k|0\rangle_\nu+{o(r_0^0)},
\label{kk-noren}
\end{multline}
where
\Align$$
\bfC_k
&=4^{1-\tk}n\,{\gamma_n^2(k)}\left(\alpha_{\nu,\tk}-\cV^-_{\tk\tk}\,b\Q\right)
=4^{1-\tk}n\,{\gamma_n^2(k)}\alpha_{\nu,\tk}\left(1+\tan\pi\nu\tan{\pi\tk\over2}\cdot b\Q\right),
\label{bfC}
\\
\bfC^{(s,s')}_k
&=\cC^{(s,s')}_{\nu,k}(1+(s-s')b\Q)
\label{bfC-ss-genform}
$$
with
\begin{equation}
\cC^{(s,s')}_{\nu,k}=2\Gamma(k)\left({n\delta^{(s,s')}_{\nu,0}}\right)_kc_{\nu,-\tk}^{(s,s')}.
\label{cC-ss-genform}
\end{equation}
We used an explicit formula
\begin{equation}
\cV^-_{kk}={\pi(K^\infty_{\nu,k})^2\tan\pi\nu \over4\cos\pi\!\left(\nu+{k\over2}\right)\cos\pi\!\left(\nu-{k\over2}\right)} = -\alpha_{\nu,k}\tan\pi\nu\tan{\pi k\over2},
\label{cVm-explicit}
\end{equation}
which follows from (\ref{Kinf-kmk-fin}) and (\ref{cVpm-kl-def}).

The terms containing $\bfC^{(s,s')}_k$ with ${\delta^{(s,s')}_{\nu,-2\tk}<0}$ correspond to divergent contributions to the operator $\cT_n(\d^k\varphi\,\bd^k\varphi\,V_\nu)$ proportional to $V_{\nu+(s-s')b^2/n}$. Thus for generic values of the parameters we may define the renormalized operator as follows:
\Multline$$
\left[\cT_n\left(\d^k\varphi\,\bd^k\varphi\,V_\nu\right)\right]^\ren
=\cT_n\left(\average{\bd^k\varphi}_{r_0}\average{\d^k\varphi}_0V_\nu\right)
-m^{2\tk}\sum_{(s,s')\in\rmS^+}\cC^{(s,s')}_{\nu,k}(mr_0)^{{2\delta^{(s,s')}_{\nu,-2\tk}}}\cT_nV_{\nu+(s-s')b^2/n}\Bigr|_{r_0\to0},
\\
\text{if}\quad\delta^{(s,s')}_{\nu,{-2\tk}}\ne0\quad\forall s,s'\in\rmS^+.
\label{dk-bdk-ren-def}
$$
In this case we obtain
\begin{equation}
|[\d^k\varphi\,\bd^k\varphi\,V_\nu]^\ren\rangle_\nu
=\left(-\Gamma^2(k)\a_{-k}\ba_{-k}+m^{2\tk}\bfC_k\right)|0\rangle_\nu.
\label{kk-afterren}
\end{equation}
The correlation functions of the renormalized operator are given by
\Multline$$
{G^{(n)}_{\d^k\varphi\,\bd^k\varphi\,V_\nu}(\{x_i\}_N)\over G^{(n)}_\nu(\{x_i \}_N)}
=-b^2\Gamma^2(k)
\\
\times\left(g_{\a_{-k}\ba_{-k}}(\{x_i\}_N)
+{4n\over\Gamma^2(\tk)}\left(m\over2\right)^{2\tk}\left(\cV^-_{\tk\tk}g_\Q(\{x_i\}_N)-b^{-2}\alpha_{\nu,\tk}-\gamma_{\nu,\tk}\right)\right).
\label{G-kbk-fin}
$$
The term $\gamma_{\nu,\tk}$ is added due to the fact that the term proportional to $|0\rangle_\nu$ may get corrections from the higher orders in the perturbation theory. Since it of the order $b^{-2}$ comparing the other terms, the higher orders in the perturbation theory should be taken into account. Up to now we cannot compute this term and leave it undetermined.

Thus the form factors read
\Multline$$
F^{(n)}_{\d^k\varphi\,\bd^k\varphi\,V_\nu}\qty(\{\theta_i\}_N)
=b^2\left(m\over2\right)^{2\tk}\gamma_n^2(k)
\\
\times\left(\varkappa_{\nu,\tk}^2
\sum_{i\neq j}\e^{\tk(\theta_i-\theta_j)}+ 4n(b^{-2}\alpha_{\nu,\tk}+\gamma_{\nu,\tk})-4\pi N\cV^-_{\tk\tk}\cot\pi\nu\right)
F^{(n)}_\nu\qty(\{\theta_i\}_N).
\label{dk-bdk-FF}
$$
We retain $\cV^-_{\tk\tk}$ here to facilitate comparison with expressions for the threshold cases below. We see that the correlation functions and form factors being of the order $b^{2-N}$ contain a contribution of the order $b^{-N}$, which is indistinguishable from that of the exponential operator $\cT_nV_\nu$.

Note that due to the second term in the parentheses in (\ref{dk-bdk-FF}) the operators under consideration acquire vacuum expectation values
\eq$$
\langle\cT_n(\d^k\varphi\,\bd^k\varphi\,V_\nu)\rangle=4n\left(m\over2\right)^{2\tk}\gamma_n^2(k)\alpha_{\nu,\tk}+O(b^2).
\label{dk-bdk-VEV-gen}
$$
They have poles at the poles of the coefficient $\alpha_{\nu,\tk}$. Below we argue that, in fact, the vacuum expectation values are finite at some of these points, but are of the order $b^{-2}$.

Now discuss the threshold points $k=l_0$ with $l_0$ defined in (\ref{l0-def}). In the r.h.s.\ of (\ref{kk-noren}) let us isolate the sum
\begin{equation}
|\sigma,\sigma';\tk\rangle_\nu=m^{2\tk}\left(\bfC_k
+(mr_0)^{{2\delta^{(\sigma,\sigma')}_{\nu,-2\tk}}}\bfC^{(\sigma,\sigma')}_k\right)|0\rangle_\nu.
\label{resonance-sum}
\end{equation}
We conjecture that
\begin{equation}
\Res_{k=l_0}\left(\bfC_k+\bfC^{(\sigma,\sigma')}_k\right)=0.
\label{Res-bfC-zero}
\end{equation}
This conjecture is consistent with (\ref{bfC})--(\ref{cC-ss-genform}). Indeed, for the known cases ($\sigma+\sigma'\le4$)
\eq$$
\cR_\nu^{(\sigma,\sigma')}
\equiv\Res_{\tk=\tl_0}\cC^{(\sigma,\sigma')}_{\nu,k}=-4^{1-\tl_0}n\gamma_n^2(l_0)\Res_{\tk=\tl_0}\alpha_{\nu,\tk}.
\label{Res-alpha-cC-rel}
$$
Now calculate the residue of $\bfC_k$. Since $\alpha_{\nu,\tk}$ has double zeros at odd positive values of $\tk$, the factor $\tan{\pi\tk\over2}$ does not lead to any extra poles. Notice that
$$
\Gathered{
\Res_{\tk=\tl_0}\alpha_{\nu,\tk}=0,\quad\text{if $|\sigma-\sigma'|>1$,}
\\
\tan\pi\nu\tan{\pi\tl_0\over2}=\sigma-\sigma',\quad\text{if $|\sigma-\sigma'|\le1$.}
}
$$
We conclude that
\eq$$
\Res_{\tk=\tl_0}\bfC_k=4^{1-\tl_0}n\gamma_n^2(l_0)(1+(\sigma-\sigma')b\Q)\Res_{\tk=\tl_0}\alpha_{\nu,\tk}
\label{Res-bfCk}
$$
Together with (\ref{Res-alpha-cC-rel}) it gives (\ref{Res-bfC-zero}).

Now let us calculate the vector $|\sigma,\sigma';\tk\rangle$ in the limit $\tk\to\tl_0$:
\begin{equation}
|\sigma,\sigma';\tk\rangle_\nu\Bigr|_{\tk\to\tl_0} = -2m^{2\tl_0}\cR^{(\sigma,\sigma')}_\nu\log mr_0\,|0\rangle_\nu+{}
4n\left(m\over2\right)^{2\tl_0}\gamma_n^2(l_0)\left(\alpha_\nu^{(\sigma,\sigma')}-\cV_\nu^{(\sigma,\sigma')}b\Q\right)|0\rangle_\nu,
\label{resonance-lim}
\end{equation}
where
\begin{align}
\alpha_\nu^{(\sigma,\sigma')}
&=\left.\alpha_{\nu,\tk}+{4^{\tk-1}\over n\gamma_n^2(k)}\cC^{(\sigma,\sigma')}_{\nu,k}\right|_{\tk\to\tl_0},
\label{alpha-brackets-def}
\\
\cV_\nu^{(\sigma,\sigma')}
&=\left.\cV^-_{\tk\tk}-(\sigma-\sigma'){4^{\tk-1}\over n\gamma_n^2(k)}\cC^{(\sigma,\sigma')}_{\nu,k}\right|_{\tk\to\tl_0}.
\label{Vtktk-brackets-def}
\end{align}
The quantities $\alpha_\nu^{(\sigma,\sigma')}$ and $\cV_\nu^{(\sigma,\sigma')}$ are finite due to the conjectured identity (\ref{Res-bfC-zero}).

We have to define the renormalized operator. There are two physically different situations in the case (a) $\sigma'=\sigma$ and in the cases (b,c) $|\sigma'-\sigma|=1$ in (\ref{tk-resonance}). In the case $\sigma'=\sigma$ the resonance is $\nu$\-/independent, and for integer values of $k$ it is approximate: $k-l_0=O(b^2)$. Hence the logarithm is nothing but an approximation to a difference of terms:
$$
\log{mr_0\over2}
={1\over2\eta^{(\sigma,\sigma)}b^2}\left(mr_0\over2\right)^{2\eta^{(\sigma,\sigma)}b^2}-{1\over2\eta^{(\sigma,\sigma)}b^2}.
$$
Thus it is physically reasonable to renormalize the operator by subtracting the first term only rather than the whole expression. This results in an additional term proportional to $b^{-2}V_\nu(x)$ in the renormalized operator.

In the cases $|\sigma'-\sigma|=1$ there is an exact resonance, but the resonance value of $\nu$ is shifted from that defined by the equation $k=l_0$ by a number of the order $b^2$. In these cases within the renormalization procedure the very logarithm should be subtracted.

To summarize we postulate
\Multline$$
[\cT_n(\d^k\varphi\,\bd^k\varphi\,V_\nu)]^\ren
=\cT_n\left(\average{\bd^k\varphi}_{r_0}\average{\d^k\varphi}_0V_\nu\right)
-m^{2\tk}\sum_{(s,s')\in\rmS^+,\ \delta^{(s,s')}_{\nu,-2\tk}<0}
(mr_0)^{{2\delta^{(s,s')}_{\nu,-2\tk}}}\cT_nV_{\nu+(s-s')b^2/n}
\\
+m^{2\tk}\cR_\nu^{(\sigma,\sigma')}\left(2\log{mr_0\over2}
+{\delta_{\sigma\sigma'}\over\eta^{(\sigma,\sigma)}b^2}+\delta_{\sigma\sigma'}\gamma^{(\sigma)}_\nu\right)
V_{\nu+(\sigma-\sigma')b^2/n}\Bigr|_{r_0\to0},
\label{dk-bdk-ren-resonance-def}
$$
where the constants $\gamma^{(\sigma)}_\nu=O(b^0)$ should be fixed by using the higher orders in the perturbation theory.

Hence, at the threshold point equation (\ref{kk-afterren}) is replaced by
\Multline$$
|[\d^k\varphi\,\bd^k\varphi\,V_\nu]^\ren\rangle
=-\Gamma^2(k)\left(\a_{-k}\ba_{-k}
-{4n\over\Gamma^2(\tk)}\qty({m\over2})^{2\tk}\left(\alpha_\nu^{(\sigma,\sigma')}-\cV_\nu^{(\sigma,\sigma')}b\Q\right)\right)|0\rangle_\nu
\\
+\delta_{\sigma\sigma'}m^{2\tk}\cR_\nu^{(\sigma,\sigma)}
\left({1\over2\sigma b^2}+\gamma^{(\sigma)}_\nu\right)|0\rangle_\nu.
\label{kk-afterren-resonance}
$$
Analogously to (\ref{dk-bdk-FF}) we obtain for the form factor
\Multline$$
F^{(n)}_{\d^k\varphi\,\bd^k\varphi\,V_\nu}\qty(\{\theta_i\}_N)
=b^2\left(m\over2\right)^{2\tk}\gamma_n^2(k)
\\
\times\left(\varkappa_{\tnu,\tk}^2
\sum_{i\neq j}\e^{\tk(\theta_i-\theta_j)}+ 4nb^{-2}\alpha_\nu^{(\sigma,\sigma')}-4\pi N\cV_\nu^{(\sigma,\sigma')}\cot\pi\nu\right)
F^{(n)}_\nu\qty(\{\theta_i\}_N)
\\
+\delta_{\sigma\sigma'}m^{2\tk}\cR_\nu^{(\sigma,\sigma)}
\left({1\over2\sigma b^2}+\gamma^{(\sigma)}_\nu\right)F^{(n)}_\nu(\{\theta_i\}_N).
\label{dk-bdk-FF-res}
$$
The last term, which admixes the operator $\cT_n(V_\nu)$, originates in the proximity to a resonance in the case (a) in (\ref{tk-resonance}). For $|\sigma-\sigma'|\ge1$ this formula only differs from (\ref{dk-bdk-FF}) by the substitution $\alpha_{\nu,\tk}\to\alpha^{(\sigma,\sigma')}_\nu$, $\cV^-_{\tk\tk}\to\cV^{(\sigma,\sigma')}_\nu$.

The zero\-/particle form factor is the vacuum expectation value of the operator. It reads
\eq$$
\langle\cT_n(\d^{2n\sigma}\varphi\,\bd^{2n\sigma}\varphi\,V_\nu)\rangle={m^{2\tk}\cR^{(\sigma,\sigma)}\over2\sigma b^2}+O(b^0).
\label{dk-bdk-VEV-k=2nsigma}
$$
We see that it is of the order $b^{-2}$. The calculation of contribution of the order $O(b^0)$ demands higher orders in the perturbation theory. In fact, it is ill\-/defined in our approximation, since a small (of the order $b^2$) variation of the parameter $\nu$ can affect it seriously.

Consider two particular cases, where the final answers can be written in a rather simple form.

\subsubsection{The case $\sigma=\sigma'=1$, $k=2n$}

In this case we have
\eq$$
\cR^{(1,1)}_\nu={(2n)!\,(2n-1)!\over16(1-4\nu^2)^2}
\label{cR11-fin}
$$
and
\begin{equation}
\alpha_\nu^{(1,1)}={1\over(1-4\nu^2)^2}\left(\delta_\nu+{1+4\nu^2\over1-4\nu^2}-{nH_{2n}\over2}\right),
\qquad
\cV_\nu^{(1,1)}=\cV^-_{22}={\pi\tan\pi\nu\over4(1-4\nu^2)^2},
\label{alpha-cV-11}
\end{equation}
where $H_m=\sum^m_{k=1}k^{-1}$ is the $m$th harmonic number.

The last term in (\ref{dk-bdk-FF-res}) leads to a nonzero vacuum expectation value of the operator:
\eq$$
\langle\cT_n(\d^{2n}\varphi\,\bd^{2n}\varphi\,V_\nu)\rangle={(2n)!\,(2n-1)!\over32(1-4\nu^2)^2}{m^4\over b^2}+O(b^0).
\label{d2n-bd2n-VEV}
$$

\subsubsection{The cases $\sigma=1$, $\sigma'=0$ and $\sigma=0$, $\sigma'=1$}

Since physically the variable $k$ is an integer, while the parameter $\nu$ is a continuous variable, the resonance point is reasonable to write in the form $\nu=\pm\nu_k$, where
\eq$$
\nu_k\equiv{1\over2}-{k\over2n},
\qquad
k=1,2,\ldots,2n-1.
\label{nu-k-def}
$$
We have
\begin{equation}
\cR^{(1,0)}_\nu=\cR^{(0,1)}_{-\nu}=-{n\over4}\Gamma^2(n(1-2\nu))\beta^{-2}_\nu
\label{cR10-cR01-fin}
\end{equation}
and
\eq$$
\Aligned{
\alpha^{(1,0)}_\nu=\alpha^{(0,1)}_{-\nu}
&={2^{2-4\nu}\pi\over\sin2\pi\nu}{\Gamma(1-\nu)\Gamma\left({1\over2}+\nu\right)\over\Gamma(\nu)\Gamma\left({1\over2}-\nu\right)}
\\
&\quad\times
\left(-2(n+1)\psi(1)-3\delta_\nu+\tfrac12(\psi(\nu)+\psi(1-\nu))+2n\psi(n(1-2\nu))\right),
\\
\cV^{(1,0)}_\nu=-\cV^{(0,1)}_{-\nu}
&=-\alpha^{(1,0)}_\nu-
{2^{3-4\nu}\pi\over\sin^22\pi\nu}{\Gamma(1-\nu)\Gamma\left({1\over2}+\nu\right)\over\Gamma(\nu)\Gamma\left({1\over2}-\nu\right)}.
}\label{alpha-cV-10-01-fin}
$$

\section{Conclusion}
\label{sec:conclusion}

In this work we extended the semiclassical approach to calculation of form factors \cite{Lashkevich:2023hzk} to the composite twist operators (CTOs) in the sinh\-/Gordon model on multi\-/sheeted Riemann surfaces.

Form factors of the exponential CTOs are already known exactly, and in the semiclassical limit $b\sim\hbar^{1/2}\to0$ they reduce to a product of simple constant one\-/particle contributions. However, the form factors of the descendant operators, obtained from the exponential ones by multiplication by products of space\-/time derivative of the basic field $\varphi$, are essentially quantum objects. Even in the leading order in $b$ they are determined by the quantum fluctuations about the classical background, and, generally, beyond the quadratic approximation. Furthermore, non\-/chiral descendant operators demand renormalization. We study the renormalization procedure in detail and demonstrate its consistency with that in the conformal perturbation theory introduced by Al.~Zamolodchikov \cite{Zamolodchikov:1989zs}.

We would like to give several comments that could suggest directions for the future studies:

\begin{itemize}

\item The form factors of local operators in an integrable model of quantum field theory can be obtained in two ways. First, they can be obtained exactly by solving a system of bootstrap equations. In this way the identification of a solution to a particular operator in terms of the basic field is a separate and complicated problem. Second, the form factors of a given operator defined in terms of the basic fields can be found approximately using a kind of perturbation theory. In our case, it is the perturbation theory on a classical background. Comparison of the results of the two methods can approach us to a solution of the identification problem. We are planning to perform such comparison in the upcoming paper.

\item Obtained form factors are expressed in terms of the asymptotic coefficients of the generalized sinh\-/Bessel functions. We present here a basic theory of these functions, but the problems formulated above demand a deeper theory and more explicit and general expressions for the coefficients. In particular, a more detailed study of the general functions $\Phi_{k_1\ldots k_s}$ is necessary to investigate higher descendant operators (with $K+L>2$ in (\ref{descendants-def})) and the subleading contributions in the parameter $b$.

\item It would be interesting to generalize the technique to the sine\-/Gordon model. Such generalization is not reduced to using solutions with imaginary values of the parameter $\nu$. To describe topological solitons (kinks and antikinks) rotationally non\-/invariant solutions should be taken into account.

\end{itemize}

It should be noted that the results of this paper simplify the derivation and extend the results of \cite{Lashkevich:2023hzk} in the case $n=1$. Another remark is that most of our results are not restricted to integer values of $n$. Though the non\-/integer values of $n$ have no clear interpretation in terms of the Minkowski space, in the case of the Euclidean space they correspond to spaces with a conic singularity. This may facilitate application of the results to computing the von Neumann entanglement entropy.

\section*{Acknowledgments}

The authors are grateful to  B.~Feigin and A.~Litvinov for discussions. The work was supported by the Russian Science Foundation under the grant 23\--12\--00333.

\Appendix

\section{Consistency checks of the radial quantization scheme}
\label{app:Consistency}

As stated before, in the radial quantization scheme the variables $t=mr$ and $\xi$ play the role of the (imaginary) time and space coordinates. So let us rewrite the quadratic in the field $\chi$ action (\ref{deltaaction}) in these coordinates:
\begin{equation}
S_0[\chi]=\int dt\,L,
\qquad
L={1\over16\pi}\int_0^{2\pi n} d\xi\,t \qty((\d_t\chi)^2+t^{-2}(\d_\xi\chi)^2+\chi^2\cosh\phi_\nu(t)).
\label{S0-t-xi-var}
\end{equation}
Thus,
$$
\rho={\i\,\delta L\over\delta(\d_t\chi)}=\i{t\,\d_t\chi\over8\pi}
$$
is the canonical momentum, and the Hamiltonian reads
\begin{equation}\label{chi-hamilton}
H_0[\chi,\rho]=\int^{2\pi n}_0d\xi\,(\i\rho\,\d_t\chi+L)
={1\over16\pi t}\int^{2\pi n}_0d\xi\,\qty((8\pi\rho)^2+(\d_\xi\chi)^2+t^2\chi^2\cosh\phi_\nu(t)).
\end{equation}

The following synchronous commutation relations should be satisfied:
\begin{equation}
[\chi(t,\xi),\chi(t,\eta)]=[\rho(t,\xi),\rho(t,\eta)]=0,
\qquad
[\rho(t,\xi),\chi(t,\eta)]=-\i\delta(\xi-\eta\bmod2\pi n).
\label{chi-rho-commut}
\end{equation}
Let us check them for the expansion (\ref{chi-decomp}). First, we have
\begin{align*}
[\chi(t,\xi),\chi(t,\eta)]
&=\sum_{k\in\Z}{{2\over k}}\qty(f_k(t,\xi)f_{-k}(t,\eta)+\bar f_k(t,\xi){\bar f}_{-k}(t,\eta))
\\
&=\sum_{k\in\Z}{4\over k}u_k(t)u_{-k}(t)\cos\tk(\xi-\eta)=0.
\end{align*}
The proof of the $[\rho,\rho]$ commutation relation is similar. For $\rho$ and $\chi$ we have
\begin{align*}
\i[\rho(t,\xi),\chi(t,\eta)]
&=-{t\over8\pi}[\d_t\chi(t,\xi),\chi(t,\eta)]
\\
&=-{t(u_*'(t)u(t)-u'(t)u_*(t))\over2\pi n}-\sum_{k\neq 0}{\e^{\i\tk(\xi-\eta)}\over4\pi k}t\qty(u_k'(t)u_{-k}(t)-u_k(t)u'_{-k}(t)).
\end{align*}
The Wronskians in the r.h.s.\ immediately follow from (\ref{KI-Wronskian}):
\begin{equation}
u_k'(t)u_{-k}(t)-u_k(t)u_{-k}'(t)=-{2k\over nt},
\qquad
u_*'(t)u_0(t)-u_*(t)u_0'(t)=-{1\over t}.
\end{equation}
Hence,
\begin{equation}
\i[\rho(t,\xi),\chi(t,\eta)]={1\over2\pi n}\sum_{k\in\Z}\e^{\i{k\over n}(\xi-\eta)}=\delta\left(\xi-\eta\bmod2\pi n\right),
\end{equation}
which proves the last relation in (\ref{chi-rho-commut}).

Let us prove the identity (\ref{chi-corr0-rad}): the correlation function for the quadratic action coincides with the radial ordered matrix element in the radial quantization scheme. Due to the Wick theorem it is sufficient to check this identity for the two\-/point correlation function
\begin{equation}
G(x,x')=\langle\chi(x)\chi(x')\rangle_0.
\label{G(x,x')-def}
\end{equation}
This function satisfies the equation
\begin{equation}
\left(-\nabla^2+m^2\cosh\phi_\nu(t)\right)G(x,x')=8\pi\delta_n(x-x').
\label{G(x,x')-diffeq}
\end{equation}
Here $\delta_n(x-x')$ is equal to $\delta(x-x')$ if $x$ and $x'$ lie on the same sheet and zero otherwise. In the coordinates $(t,\xi)$ it means that
\begin{equation}
\delta_n(x-x')=m^2t^{-1}\delta(t-t')\delta(\xi-\xi'\bmod2\pi n).
\label{deltan-rad}
\end{equation}
Let us search the solution in the form
\eq$$
G(x,x')=4\sum_{k\in\Z}G_{|k|}(t,t')\e^{\i\tk(\xi-\xi')}.
\label{G(x,x')-sumform}
$$
Substituting it to equation (\ref{G(x,x')-diffeq}) we obtain
\eq$$
\left(-\d_t^2-t^{-1}\d_t+\tk^2t^{-2}+\cosh\phi_\nu(t)\right)G_k(t,t')=t^{-1}\delta(t-t').
\label{Gk(t,t')-diffeq}
$$
It means that the function $G_k(t,t')$ as a function of the variable $t$ satisfies the sinh\-/Bessel equation (\ref{shB-equation}) with $\kappa=\tk$ for $t\ne t'$, and its derivative has a jump at $t=t'$:
\eq$$
-\left.\d_tG_k(t,t')\right|_{t=t'+0}+\left.\d_tG_k(t,t')\right|_{t=t'-0}=t^{\prime\,-1}.
\label{Gk(t,t')-jump}
$$
The function
\eq$$
G_k(t,t')=\Cases{K_{\nu,\tk}(t')I_{\nu,\tk}(t),&t<t';\\K_{\nu,\tk}(t)I_{\nu,\tk}(t'),&t>t',}
\label{Gk(t,t')-solution}
$$
satisfies both conditions. But the function $G(x,x')$ in the form (\ref{G(x,x')-sumform}) with this solution precisely coincides with the matrix element in the radial quantization picture:
\begin{equation}
G(x,x')=\langle\infty|\T_r[\chi(x)\chi(x')]|0\rangle_\nu.
\label{G(x,x')-final}
\end{equation}
This proves the statement (\ref{chi-corr0-rad}).

\section{Integral representation for the connection coefficients}
\label{app:Intrep}

\subsection{General formulas}
\label{app:Intrep-gen}

We are interested in the small $t$ asymptotics of the functions
\begin{equation}
\tPhi_{k_1\ldots k_s}(t)
=\sum^\infty_{n=1}\lambda^{2n+1}\tPhi^{(2n+1)}_{k_1\ldots k_s}(t),
\label{tPhi-ks-series}
\end{equation}
where
\begin{equation}
\tPhi^{(r)}_{k_1\ldots k_s}(t)
=4\int_{\{u_i\ge0\}} d^r u\,\prod^r_{i=1}{\e^{-{t\over2}(u_i+u_i^{-1})}\over{u_i+u_{i+1}}}
\cdot{u_1^{k_1}\prod^s_{a=2}\sum^r_{i=1}u_i^{k_a}}.
\label{tPhi-ks-def}
\end{equation}
We will assume $k_a$ to be complex numbers with positive real part. The values of the connection coefficients for general complex values of $k_a$ are obtained by analytic continuation. For small values of $t$ the main contribution to each integral comes from the region $u_i\gg1$ for all $i$. In this region we may rescale $tu_i/2\to u_i$ and omit $t^2u_i^{-1}/4$ in the exponents. We obtain
\begin{equation}
\tPhi^{(r)}_{k_1\ldots k_s}(t)=4\left(t\over2\right)^{-\sum k_a}\left(c^{(r)}_{k_1\ldots k_s}+o\left(t^0\right)\right),
\label{tPhi(r)-small-t}
\end{equation}
where
\begin{equation}
c^{(r)}_{k_1\ldots k_s}=\int_{\{u_i\ge0\}} d^r u\,\prod^r_{i=1}{\e^{-{1\over2}(u_i+u_{i+1})}\over u_i+u_{i+1}}
\cdot u_1^{k_1}\prod^s_{a=2}\sum^r_{i=1}u_i^{k_a}.
\label{c(r)nuks-def}
\end{equation}
The connection coefficients are given by
\begin{equation}
(K^\infty_{\nu,k_1\ldots k_s})^{-1}={A_{k_1\ldots k_s}\over2^{1+\sum k_a}\lambda\Gamma\left(\sum k_a\right)}
={2\over\Gamma\left(\sum k_a\right)}\sum^\infty_{n=1}\lambda^{2n}c^{(2n+1)}_{k_1\ldots ks}
\label{Kinf-c(r)}
\end{equation}
It will be convenient to express the answer in terms of the series in all integer powers of $\lambda$:
\begin{equation}
(K^\infty_{\nu,k_1\ldots k_s})^{-1}={1\over\lambda\Gamma\left(\sum k_a\right)}(N_{\nu,k_1\ldots k_s}-N_{-\nu,k_1\ldots k_s}),
\qquad
N_{\nu,k_1\ldots k_s}=\sum^\infty_{r=1}\lambda^rc^{(r)}_{k_1\ldots k_s}.
\label{Kinf-Nnu}
\end{equation}

Let us rewrite this in terms of operators on ${\cF=L_2}[0,\infty)$. Let $\hM$ be the multiplication by the variable operator and $\hK_0$ be the integral operator with the kernel
\begin{equation}
K_0(u,v)={\e^{-{1\over2}(u+v)}\over u+v}.
\label{K0-kernel-def}
\end{equation}
We will also use
\eq$$
\hG={1\over1-\lambda\hK_0},
\qquad
\hH={\lambda\hK_0\over1-\lambda\hK_0}=\hG-1.
\label{hGhH-def}
$$
Then the coefficient $c^{(r)}_{k_1\ldots k_s}$ is a sum of the traces
$$
\tr\left(\hM^{k_1}\hK_0^{l_1}\hM^{k_{\sigma_2}}\hK_0^{l_2}\cdots\hM^{k_{\sigma_s}}\hK_0^{l_s}\right),
$$
where $\sigma$ is an element of the set $P[2,\ldots,s]$ of permutations of numbers $2,\ldots,s$, and $\sum^s_{a=1}l_a=r$. The sum must include all noncoinciding traces. If we also sum up over all values of $r$, we obtain
\eq$$
N_{\nu,k_1\ldots k_s}=\sum_{\sigma\in P[2,\ldots,s]}\tr\left(\hM^{k_1}\hG\hM^{k_{\sigma_2}}\hG_{\sigma_2\sigma_3}
\cdots\hM^{k_{\sigma_{s-1}}}\hG_{\sigma_{s-1}\sigma_s}\hM^{k_{\sigma_s}}\hH\right)
=\sum_{\sigma\in P[2,\ldots,s]}
\tr\prod^{\substack{\curvearrowright\\[-2pt]s}}_{a=1}\hM^{k_{\sigma_a}}\hG_{\sigma_a\sigma_{a+1}},
\label{N-tr}
$$
where we assumed $\sigma_1=\sigma_{s+1}=1$ and
\eq$$
\hG_{ab}=\Cases{\hG,&\text{if $a<b$;}\\\hH,&\text{if $a\ge b$.}}
\label{Gab-def}
$$
In particular,
\eq$$
\Aligned{
N_{\nu,k}
&=\tr(\hM^k\hH),
\\
N_{\nu,kl}
&=\tr(\hM^k\hG\hM^l\hH)=N_{\nu,k+l}+\tr(\hM^k\hH\hM^l\hH).
}\label{N-examples12}
$$

To calculate these traces we will use the diagonalization of the operator $\hK_0$ given in Section 4E of~\cite{McCoy:1976cd}. It is diagonalized by the functions
\eq$$
f_p(u)=\left(p\sinh\pi p\over2\pi\right)^{1/2}\int^\infty_1d\xi\,\e^{-\xi u/2}P_{-{1\over2}+\i p}(\xi),
\qquad u,\ p\in[0,\,\infty),
\label{fp-def}
$$
where $P_\nu(x)$ is the Legendre function:
\eq$$
\hK_0f_p=\kappa_pf_p,
\qquad
\kappa_p={\pi\over\cosh\pi p}.
\label{hK-diag}
$$
The functions $f_p$ form an orthonormal basis:
\eq$$
\int^\infty_0 du\,f_p(u)f_{p'}(u)=\delta(p-p').
\label{fp-orthonorm}
$$
To calculate the traces we need the matrix elements of the operator $\hM^k$:
\eq$$
X^{(k)}_{pp'}=\langle f_{p'}|\hM^k|f_p\rangle=\int^\infty_0 du\,u^kf_p(u)f_{p'}(u).
\label{Xmatel-def}
$$
We need two formulas from \cite{Gradshteyn:2007}. Eq.\ 7.141(5) allows one to express the functions $f_p$ in terms of the Bessel functions:
\eq$$
f_p(x)=\left(2p\sinh\pi p\over\pi^2\right)^{1/2}x^{-1/2}K_{\i p}\left(x\over2\right).
\label{fp-Bessel}
$$
Then eq.\ 6.576(4) provides the following expression for the matrix element
\eq$$
X^{(k)}_{pp'}={2^{2k-2}\over\pi^2}{(pp'\sinh\pi p\sinh\pi p')^{1/2}\over\Gamma(k)}
\prod_{\ve_1,\ve_2=\pm1}\Gamma\left(k+\i\ve_1p+\i\ve_2p'\over2\right).
\label{Xmatel-final}
$$
From (\ref{fp-orthonorm}) we immediately obtain
\eq$$
\int^\infty_0dp''\,X^{(k)}_{pp''}X^{(l)}_{p''p'}=X^{(k+l)}_{pp'}.
\label{X-prod}
$$
Thus we have
\eq$$
N_{\nu,k_1\ldots k_s}
=\sum_{\sigma\in P[2,\ldots,s]}\int_{\{p_a\ge0\}}d^sp\,\prod^s_{a=1}
{(\lambda\kappa_{p_a})^{\Theta(\sigma_a-\sigma_{a+1})}\over1-\lambda\kappa_{p_a}}X^{(k_{\sigma_a})}_{p_{a-1}p_a},
\label{N-kappaX}
$$
where $\Theta(x)$ is again the Heaviside step function.

Let us write down the integral representations for $s=1,2$. Due to the evenness of the integrands it is convenient to continue the integrals to the whole real axis. For $s=1$ we have
\eq$$
N_{\nu,k}={\lambda\Gamma(k)\over2\Gamma^2\left(k+1\over2\right)}\int^\infty_{-\infty}dp\,{p\sinh\pi p\over\cosh\pi p-\sin\pi\nu}
\Gamma\left({k\over2}+\i p\right)\Gamma\left({k\over2}-\i p\right).
\label{N1-int}
$$
Therefore,
\eq$$
(K^\infty_{\nu,k})^{-1}={N_{\nu,k}-N_{-\nu,k}\over\lambda\Gamma(k)}
={1\over2\Gamma^2\left(k+1\over2\right)}\int^\infty_{-\infty}dp\,{p\sinh2\pi p\over\cosh^2\pi p-\sin^2\pi\nu}
\Gamma\left({k\over2}+\i p\right)\Gamma\left({k\over2}-\i p\right).
\label{Kinf1-int}
$$
The derivation of the explicit formula (\ref{Kinfty-explicit}) from the integral representation is given in the next subsection.

For $s=2$ we obtain
\Multline$$
N_{\nu,kl}=N_{\nu,k+l}+{2^{2k+2l-6}\lambda^2\over\pi^2\Gamma(k)\Gamma(l)}
\int_{\R^2}dp_1\,dp_2\,\prod^2_{i=1}{p_i\sinh\pi p_i\over\cosh\pi p_i-\sin\pi\nu}
\\
\times
\prod_{\ve_1,\ve_2=\pm1}\Gamma\left(k+\i\ve_1p_1+\i\ve_2p_2\over2\right)\Gamma\left(l+\i\ve_1p_1+\i\ve_2p_2\over2\right).
\label{N2-int}
$$
Thus we have
\Multline$$
(K^\infty_{\nu,kl})^{-1}
=(K^\infty_{\nu,k+l})^{-1}+{2^{2k+2l-5}\lambda^2\over\pi\Gamma(k)\Gamma(l)\Gamma(k+l)}
\int_{\R^2}dp_1\,dp_2\,{p_1p_2\sinh2\pi p_1\sinh\pi p_2\over\prod^2_{i=1}(\cosh^2\pi p_i-\sin^2\pi\nu)}
\\
\times
\prod_{\ve_1,\ve_2=\pm1}\Gamma\left(k+\i\ve_1p_1+\i\ve_2p_2\over2\right)\Gamma\left(l+\i\ve_1p_1+\i\ve_2p_2\over2\right).
\label{Kinf2-int}
$$
This proves (\ref{Akl-intrep}). An explicit formula in terms of the elementary functions in the case $l=-k$ is given in subsection~\ref{app:K-nuk-k}.

\subsection{Calculation of the integral (\ref{N1-int})}
\label{app:K-nuk}

Let us use the integral representation of the beta function. We have
\Align$$
{2\Gamma^2\left(k+1\over2\right)\over\lambda\Gamma(k)}N_{\nu,k}
=\cJ_{\nu,k}
&\equiv\int^\infty_{-\infty}dp\,{p\sinh\pi p\over\cosh\pi p-\sin\pi\nu}
\Gamma\left({k\over2}+\i p\right)\Gamma\left({k\over2}-\i p\right)
\notag
\\
&=\Gamma(k)\int^1_0dt\int^\infty_{-\infty}dp\,{p\sinh\pi p\over\cosh\pi p-\sin\pi\nu}
t^{{k\over2}+\i p-1}(1-t)^{{k\over2}-\i p-1}
\notag
\\
&=\Gamma(k)\int^1_0dt\,(t(1-t))^{k\over2}{d\over dt}F_\nu\left(\log\left(t\over1-t\right)\right),
\label{cJ1-def}
$$
where
\eq$$
F_\nu(x)=-\i\int^\infty_{-\infty}dp\,{\sinh\pi p\over\cosh\pi p-\sin\pi\nu}\e^{\i px}
={2\cosh\left(\nu+{1\over2}\right)x\over\sinh x}.
\label{Fnu-def}
$$
We consider these functions as analytic continuations from the region $|\Re\nu|<{1\over2}$. Hence, we obtain
$$
\cJ_{\nu,k}=2\Gamma(k)\int^1_0dt\,(t(1-t))^{k\over2}{d\over dt}(t(1-t))^{{1\over2}-\nu}{t^{2\nu+1}+(1-t)^{2\nu+1}\over2t-1}
$$
Take this integral by parts. For large enough $\Re k$ the boundary term vanishes and we have
\Align$$
\cJ_{\nu,k}
&=-2\Gamma(k)\int^1_0dt\,(t(1-t))^{{1\over2}-\nu}{t^{2\nu+1}+(1-t)^{2\nu+1}\over2t-1}{d\over dt}(t(1-t))^{k\over2}
\notag
\\
&=\Gamma(k+1)\int^1_0dt\,(t(1-t))^{{k+1\over2}-\nu-1}(t^{2\nu+1}+(1-t)^{2\nu+1})
={2\Gamma\left({k+3\over2}+\nu\right)\Gamma\left({k+1\over2}-\nu\right)\over k+1}.
\label{cJ1-fin}
$$
Thus
\eq$$
{N_{\nu,k}\over\lambda\Gamma(k)}
={\Gamma\left({k+3\over2}+\nu\right)\Gamma\left({k+1\over2}-\nu\right)
\over2\Gamma\left(k+3\over2\right)\Gamma\left(k+1\over2\right)}.
\label{N1-fin}
$$
Substituting it into (\ref{Kinf-Nnu}) we obtain (\ref{Kinfty-explicit}).

\subsection{Explicit expression for $K^\infty_{\nu,k.-k}$}
\label{app:K-nuk-k}

Let us calculate $K^\infty_{\nu;kl}$ in the limit $l\to-k$. According to (\ref{Kinf-Nnu})
\eq$$
(K^\infty_{\nu,k.-k})^{-1}=\lambda^{-1}\lim_{\delta\to0}\delta\left(N_{\nu,k.\delta-k}-N_{-\nu,k.\delta-k}\right).
\label{Kinf-kmk-Nnu}
$$
It means that we are interested in the residue of the pole of the function $N_{\nu;k,\delta-k}$ at $\delta=0$. The pole comes from two sources in (\ref{N2-int}). The first one is $N_{\nu,\delta}$ and its residue immediately follows from (\ref{N1-fin}). The second one is related with the deformation of the contour for negative values of $k$ or $l$. To describe this deformation let us change the variables
\eq$$
p_1=p+q,\qquad p_2=p-q.
\label{pi-pq-change}
$$
Denote the integral in (\ref{N2-int}) as $\cJ_{\nu,kl}$. Then $\cJ_{\nu,kl}=\int dp\,dq\,f_{\nu,kl}(p,q)$ with
\Align*$$
f_{\nu,kl}(p,q)
&={2(p^2-q^2)\sinh\pi(p+q)\sinh\pi(p-q)\over\cosh{\pi\over2}(p+q+\i\mu)\cosh{\pi\over2}(p+q-\i\mu)\cosh{\pi\over2}(p-q+\i\mu)\cosh{\pi\over2}(p-q-\i\mu)}
\\
&\quad\times
\Gamma\left(k+2\i p\over2\right)\Gamma\left(k-2\i p\over2\right)\Gamma\left(k+2\i q\over2\right)\Gamma\left(k-2\i q\over2\right)
\\
&\quad\times
\Gamma\left(l+2\i p\over2\right)\Gamma\left(l-2\i p\over2\right)\Gamma\left(l+2\i q\over2\right)\Gamma\left(l-2\i q\over2\right),
\qquad
\mu=\nu+\tfrac12.
$$
Here we used
$$
\cosh\pi p_i-\sin\pi\nu=\cosh\pi p_i+\cosh\i\pi\mu=2\cosh{\pi(p_i+\i\mu)\over2}\cosh{\pi(p_i-\i\mu)\over2}.
$$
Both contours for $p$ and $q$ should go above the poles at the points $-\i({k\over2}+n)$, $-\i({l\over2}+n)$ ($n\in\Z_{\ge0}$) and below the points $\i({k\over2}+n)$, $\i({l\over2}+n)$. In the limit $\delta=k+l\to0$ the contours are pinched between the poles at $-\i{k\over2}$ and $\i{l\over2}=-\i{k+\delta\over2}$ and between $\i{k\over2}$ and $-\i{l\over2}$. This pinching results in a pole of the integral. Due to the properties
$$
f_{\nu,kl}(p,q)=f_{\nu,kl}(p,-q)=f_{\nu,kl}(q,p),
$$
the contributions coming from pinching each contour between each pair of poles are the same, so that
\Align*$$
\cJ_{\nu,k.\delta-k}
&={32}\pi\i\int dp\,\Res_{q=\i(\delta-k)/2}f_{\nu,k.\delta-k}(p,q)+O(\delta^0)
\\
&={{32}\pi^3\over\delta}\Gamma(k)\Gamma(-k)
\int dp\prod_{\ve_1,\ve_2=\pm1}\cosh^{-1}{\pi\over2}\left(p+\ve_1{\i k\over2}+\ve_2\i\mu\right)+O(\delta^0).
$$
The last integral is taken straightforwardly by inserting the factor $\e^{\i\omega p}$ ($\omega>0$) into the integrand and summing up all the residues at the points $p=\i-\ve_1{\i k\over2}-\ve_2\i\mu+2\i n$ ($n\in\Z_{\ge0}$). In the limit $\omega\to0^+$ we obtain
$$
\cJ_{\nu,k.\delta-k}={{128}\pi^3\over\delta}{\Gamma(k)\Gamma(-k)\over\sin{\pi k\over2}\sin\pi\mu}
\left({\mu+{k\over2}\over\sin\pi(\mu+{k\over2})}-{\mu-{k\over2}\over\sin\pi(\mu-{k\over2})}\right)+O(\delta^0).
$$
Thus we have
\eq$$
N_{\nu,k.\delta-k}
={\lambda\over\delta}\left({\mu\over\sin\pi\mu}
-{\cos\pi\mu\over2\sin{\pi k\over2}\sin\pi\mu}
\left({\mu+{k\over2}\over\sin\pi(\mu+{k\over2})}-{\mu-{k\over2}\over\sin\pi(\mu-{k\over2})}\right)
\right)+O(\delta^0).
\label{Nnukmk-fin}
$$
From (\ref{Kinf-kmk-Nnu}) we finally obtain
\eq$$
(K^\infty_{\nu,k.-k})^{-1}={1\over\cos\pi\nu}
\left(1+{\sin^2\pi\nu\over\cos\pi\left(\nu+{k\over2}\right)\cos\pi\left(\nu-{k\over2}\right)}\right).
\label{Kinf-kmk-fin}
$$
This proves (\ref{Akmk-fin}).

\section{Series representation for the connection coefficient $K^\infty_{\nu,kl}$}
\label{app:Akl-series}

The functions $\Phi_{k_1\ldots k_s}$with different values of $k_i$ are related by a system of differential equations (see eqs.\ (3.22), (3.45) in~\cite{Lashkevich:2023hzk}). This results in algebraic relations between the coefficients at the leading small $t$ asymptotics. In particular, the coefficients $A_k$ satisfy the equation
\begin{equation}
A_{k+2}=4k(k+1)\qty(1-{4\nu^2\over(k+1)^2})A_k.
\label{Ak-rec}
\end{equation}
For integer values of $k$ this equation defines the values of $A_k$ with the initial conditions $A'_0=2\pi^{-1}\tan\pi\nu$, $A_1=4\nu$. This results in
\begin{equation}
A_k=2^{k+1}\lambda\Gamma(k){\Gamma\left({k+1\over2}+\nu\right)\Gamma\left({k+1\over2}-\nu\right)\over\Gamma^2\left({k+1\over2}\right)}
\label{Ak-fin}
\end{equation}
which is equivalent to (\ref{Kinfty-explicit}). The relation (\ref{Ak-rec}) does not prove (\ref{Ak-fin}) for generic values of $k$, but gives a good hint. Let us apply the same idea to obtain an alternative (to (\ref{Kinf2-int})) expression for $A_{kl}$.

The corresponding relation for $A_{kl}$ reads
\begin{equation}
A_{k+2.l}=4(k+l)(k+l+1)\left(1-{4\nu^2\over(k+l+1)^2}\right)A_{kl}
-8\nu kl\left({1\over k+1}+{1\over k+l+1}\right)A_kA_l.
\label{Akl-rec}
\end{equation}
We may use the relations
\begin{equation}
A_{0l}=\pi^{-1}\tan\pi\nu\,\d_\nu A_l,
\qquad
A_{1l}=2lA_l
\label{A0l-A1l}
\end{equation}
as an initial condition to obtain explicit expressions for integer values of $k$ and any complex values of $l$. Nevertheless, we did not succeed to obtain a solution to this equation in a closed form for all values of $k$. To solve this obstacle let us rewrite the relation (\ref{Akl-rec}) in the variables
\begin{equation}
D_{kl}={A_{kl}\over A_{k+l}},
\qquad
C_{kl}=8\nu kl\left({1\over k+1}+{1\over k+l+1}\right){A_k A_l\over A_{k+l+2}}.
\label{Dkl-Ckl-def}
\end{equation}
Then it reads
\begin{equation}
D_{k+2.l}=D_{kl}-C_{kl}.
\label{Dkl-rec}
\end{equation}
Assume $D_{kl}$ be finite as $k\to+\infty$: $D_{kl}\to d_{\nu,l}$. Then we obtain
\begin{equation}
D_{kl}=d_{\nu,l}+\sum_{n=0}^\infty C_{k+2n.l}.
\label{Dkl-solution}
\end{equation}
The series in the r.h.s.\ converges for $l>-1$, which justifies our conjecture. The series can be summed up for $l=0$ resulting in $d_{\nu,0}=1$. Next, use the symmetry $D_{kl}=D_{lk}$ for $k=0$ and get
\begin{equation}
d_{\nu,l}=1+\sum_{n=0}^\infty(C_{l+2n.0}-C_{2n.l}).
\label{f_nu_l}
\end{equation}
Finally, we obtain
\begin{equation}
{K_{\nu,k+l}^\infty\over K_{\nu,kl}^\infty}={A_{kl}\over A_{k+l}}=1+\sum_{n=0}^\infty(C_{k+2n.l}+C_{l+2n.0}-C_{2n.l})
\quad\text{for $l>-1$.}
\label{Knukl-series}
\end{equation}
The main advantage of this expression comparing to (\ref{Kinf2-int}) is that it is more suitable for precise numerical calculations.

\section{Expansions of the sinh\-/Bessel functions and related functions}
\label{app:Expansion}

Most of the functions we consider in the present paper possess a common feature: for small values of $t$ they admit expansions of the form
\begin{equation}
\sum_{k\in K}\sum_{s,s'\in\Z_{\ge0}}f_k^{(s,s')}t^{2\delta^{(s,s')}_{\nu,k}}
\end{equation}
with certain coefficients $f^{(s,s')}_k$ for appropriate finite set $K$. Below we give formulas for the coefficients in the functions we need in this paper mostly for $s'=0$ checked for $s\le4$. The full lists of all coefficients for $s+s'\le4$ are given in Supplemental Material.

Let us begin with the function $\phi_\nu(t)$ defined in (\ref{phinu-lambda-def}). We use the expansion \cite{Zamolodchikov:1994uw,Fateev:1998xb,Basor:1991ax,Gamayun:2013auu}
\begin{equation}
\e^{\phi_\nu(t)}=\e^{-\phi_{-\nu}(t)} = \beta^{-2}_\nu t^{-4\nu}\sum_{\mathclap{s,s'\geq0}} B_\nu^{(s,s')}t^{2\delta^{(s,s')}_{\nu,0}}.
\end{equation}
The first coefficients read
\begin{equation}
\begin{gathered}
B^{(0,0)}_\nu=1,
\qquad
B^{(1,0)}_\nu=-B^{(0,1)}_{-\nu}={\beta_\nu^{-2}\over8(1-2\nu)^2}
\\
B^{(2,0)}_\nu=3B^{(0,2)}_{-\nu}={3\beta_\nu^{-4}\over256(1-2\nu)^4},
\qquad
B^{(1,1)}_\nu=-{1-2\nu\over64(1-4\nu^2)^2},
\\
B^{(3,0)}_\nu={\beta_\nu^{-6}\over2^{10}(1-2\nu)^6},
\qquad
B^{(2,1)}_\nu=-{(1-\nu)^2\beta_\nu^{-2}\over64(3-2\nu)^2(1-2\nu)^3(1+2\nu)^2},
\\
B^{(1,2)}_\nu={(7-20\nu(1-\nu))\beta_\nu^2\over2^{10}(3+2\nu)^2(1-2\nu)^2(1+2\nu)^4},
\qquad
B^{(0,3)}_\nu=0.
\label{Bss'-fin}
\end{gathered}
\end{equation}
There are also general formulas in special cases\cite{Fateev:1998xb}:
\eq$$
B^{(s,0)}_\nu={(s+1)\beta_\nu^{-2s}\over2^{4s}(1-4\nu)^{2s}},
\qquad
B^{(0,3+s)}_\nu=0,
\qquad
s=0,\,1,\,2,\,\ldots
\label{Bss'-s0-0s}
$$

The sinh\-/Bessel functions $I_{\nu,k}(t)$ admit the expansion (\ref{Igen}). We have
\begin{equation}
c^{(s,0)}_{\nu,k}=c^{(0,s)}_{-\nu,k}={\beta_\nu^{-2s}\over2^{4s-1}(1-2\nu)^{2s-1}(1-2\nu+k)}.
\label{ck-coefs}
\end{equation}

The functions $\tPhi^{\vee\pm}_{kl}(t)$ admit the expansion (\ref{tPhikl-veepm-expansion}) with the coefficients $c^{\pm(s,s')}_{\nu,kl}$. The functions $\tPhi_{kl}(t)$ (and thus $K_{\nu,kl}(t)$) are obtained by (\ref{tPhikl-tPhi-veepm-rel}) and expressed in terms of the same coefficients. We have
\begin{align}
c^{+(s,0)}_{\nu,kl}
&=-c^{+(0,s)}_{-\nu,kl}={s(1-2\nu)(1-2\nu+k+l)+kl\over2^{4s-1}(1-2\nu)^{2s-1}(1-2\nu+k)(1-2\nu+l)(1-2\nu+k+l)}\beta_\nu^{-2s},
\label{cklp-s0-coeffs}
\\
c^{-(s,0)}_{\nu,kl}
&=-c^{-(0,s)}_{-\nu,kl}={s\beta^{-2s}_\nu\over2^{4s-1}(1-2\nu)^{2s-2}(1-2\nu+k)(1-2\nu+l)}.
\label{cklm-s0-coeffs}
\end{align}

Now turn to the coefficients in the expansions (\ref{cJ-expansions}). From the identity (\ref{jvee-sym}) we obtain
\begin{equation}
J^{++(s,s')}_{\nu,kl}=J^{++(s,s')}_{\nu,lk}=J^{++(s,s')}_{\nu,k.-k-l},
\qquad
J^{-+(s,s')}_{\nu,kl}=J^{+-(s,s')}_{\nu,lk},
\qquad
J^{--(s,s')}_{\nu,kl}=J^{+-(s,s')}_{\nu,-k.k+l}.
\label{J(s,s')-sym}
\end{equation}
Thus it is sufficient to write down the two sets of coefficients $J^{+\pm(s,s')}_{\nu,kl}$. From the evenness of the integrals $j^{\vee\ve_1\ve_2}_{kl}(t)$ in $\nu$ and oddness of $A_k$ we have
\begin{equation}
J^{\ve_1\ve_2(s,s')}_{\nu,kl}=-J^{\ve_1\ve_2(s',s)}_{-\nu,kl}.
\label{J(s,s')-odd}
\end{equation}
For $s'=0$ at least for $s\le4$ the coefficients have a rather compact form:
\subeq{\label{J-s0-coefs}
\Align$$
J^{++(s,0)}_{\nu,kl}
&=-{(s(1-2\nu)-k)(s(1-2\nu)-l)(s(1-2\nu)+k+l)-{2\over3}s(s^2-1)(1-2\nu)^3
\over2^{4s-2}(1-2\nu)^{2s-1}(1-2\nu+k+l)(1-2\nu-k)(1-2\nu-l)}\beta_\nu^{-2s},
\label{Jpp-s0-coefs}
\\
J^{+-(s,0)}_{\nu,kl}
&=-{s(s(1-2\nu)-k)(s(1-2\nu)+k+l)-{2\over3}s(s^2-1)(1-2\nu)^2
\over2^{4s-2}(1-2\nu)^{2s-2}(1-2\nu+k+l)(1-2\nu-k)(1-2\nu+l)}\beta_\nu^{-2s}.
\label{Jpm-s0-coefs}
$$}

\section{Comparison with the conformal perturbation theory}
\label{app:perturbations}

In subsections~\ref{subsec:dkvarphi-bdlvarphi}, \ref{subsec:dkvarphi-bdkvarphi} we claim that the threshold points $\delta^{(\sigma,\sigma')}_{\nu,-2\tl}=0$ with $|\sigma-\sigma'|\le1$ produce resonance poles, while those with $|\sigma-\sigma'|\ge2$ do not produce them in the semiclassical limit. Here we check this statement for $l=k$ and $\sigma=1,\,2$, $\sigma'=0$ in the first two orders of the conformal perturbation theory.

Assuming the notation of (\ref{CPT-shG}) consider a pair correlation function on $\cM_n$ of the form
\begin{multline}
\langle\cO'(\infty)\cO(0)\rangle
=\langle\cO'(\infty)\cO(0)\rangle^{(0)}+\langle\cO'(\infty)\cO(0)\rangle^{(1)}+\langle\cO'(\infty)\cO(0)\rangle^{(2)}+\cdots
\\
=\langle\cO'(\infty)\cO(0)\rangle^{(0)}-\mu\int_{\cM_n}d^2y\,\langle\cO'(\infty)\Phi_\text{p}(y)\cO(0)\rangle^{(0)}_\text{c}
\\
+{\mu^2\over2}\int_{\cM_n^2}d^2y_1\,d^2y_2\,
\langle\cO'(\infty)\Phi_\text{p}(y_1)\Phi_\text{p}(y_2)\cO(0)\rangle^{(0)}_\text{c}+\cdots.
\label{corrOO-expansion}
\end{multline}
The superscript $(s)$ means the contribution of the $s$th order in the parameter $\mu$ into the correlation function. The zeroth order is the correlation function in the free massless field theory. The subscript `c' means the connected part. We will omit subtracting the disconnected parts since they do not affect the divergences we are interested in. Let us assume
\begin{equation}
\cO'=\cT^+_n\!\left(\e^{-(\alpha+sb)\varphi}\right),
\qquad
\cO=\cT_n\!\left(\d^k\varphi\,\bd^k\varphi\,\e^{\alpha\varphi}\right).
\label{cOcO-def}
\end{equation}
Since within the conformal perturbation theory we must complete the plane by an infinite point, which turns it to a topological sphere, we have to put a CTO at infinity. The integer $s$ in the operators $\cO'$ will be chosen in such a way that it would allow us to extract the divergent contribution we want. Namely, we will assume $s$ to coincide with the perturbation order. The operator $\cO$ demands regularization of the form $\cT_n\!\left(\average{\bd^k\varphi}_{r_0}\average{\d^k\varphi}_{r_{01}}\e^{\alpha\varphi(0)}\right)$ with $r_0>r_{01}$.

Note that we omit here the infrared divergent contributions, which are known to be canceled in the structure functions of the operator product expansions~\cite{Zamolodchikov:1990bk}. In other words we assume all the integrals below to be analytically continued from the regions of the parameters, at which the integrals converge at large values of $|y_i|$.

The first order contribution to the correlation function reads
\begin{equation}
\langle\cO'(\infty)\cO(0)\rangle^{(1)}
=-\mu\int_{\cM_n}d^2y\,\left\langle\cT^+_n\!\left(\e^{-(\alpha+b)\varphi(\infty)}\right)\e^{b\varphi(y)}
\average{\bd^k\varphi}_{r_0}\average{\d^k\varphi}_{r_{01}}\cT_n\!\left(\e^{\alpha\varphi(0)}\right)\right\rangle^{(0)}.
\label{corrOO(1)-def}
\end{equation}
It should be stressed that the average in the r.h.s.\ is calculated on the $\cM_n$ surface, which is equivalent to integration over a plane with summation over the sheets.

Apply the transformation (\ref{zeta-coord-def})\--(\ref{CTF-general}). Let $z_i=x^1_i+\i x^2_i=\zeta_i^n$, $w_i=y^1_i+\i y^2_i=\eta_i^n$, $\mu'=\mu n^{2b^2}$. Then we have
\begin{align}
\langle\cO'(\infty)\cO(0)\rangle^{(1)}
&=-{\mu'\over2}\int d^2\eta\,|\eta|^{2(n-1)(1+b^2)}
\left\langle\cT^+_n\!\left(\e^{-(\alpha+b)\varphi(\infty)}\right)\e^{b\varphi(\eta,\bar\eta)}\average{\bd^k\varphi}_{r_0}
\average{\d^k\varphi}_{r_{01}}\cT_n\!\left(\e^{\alpha\varphi(0)}\right)\right\rangle^{(0)}_\text{plane}
\notag
\\
&=-{\mu'\over2}\oint_{|\zeta_1|^n=r_0}{d\zeta_1\over2\pi\i\zeta_1}\oint_{|\zeta_2|^n=r_{01}}{d\zeta_2\over2\pi\i\zeta_2}
\notag
\\
&\quad\times
\bd_1^k\d_2^k\int d^2\eta\,|\eta|^{2(n-1)(1+b^2)}
\left\langle\e^{-(\alpha+b)\varphi(\infty)}\e^{b\varphi(\eta,\bar\eta)}\varphi(x_1)\varphi(x_2)\e^{\alpha\varphi(0)}\right\rangle^{(0)}_{\text{plane}}.
\label{corrOO(1)-int}
\end{align}
The correlation function in the r.h.s.\ is taken for the operators on the $(\zeta,\bzeta)$ plane and reads
\begin{multline}
\left\langle\e^{-(\alpha+b)\varphi(\infty)}\e^{b\varphi(y)}\varphi(x_1)\varphi(x_2)\e^{\alpha\varphi(0)}\right\rangle^{{(0)}}_\text{plane}
\\
=\left(4(b\log|\eta-\zeta_1|^2+\alpha\log|\zeta_1|^2)
(b\log|\eta-\zeta_2|^2+\alpha\log|\zeta_2|^2)-2\log|\zeta_1-\zeta_2|^2\right)|\eta|^{-4\alpha b}.
\label{correxp-varphi-varphi(1)}
\end{multline}
Use the identity
\begin{multline*}
\oint_{|\zeta|^n=r_0}{d\zeta\over2\pi\i\zeta}\,\d_\zeta^k\log|\eta-\zeta|^2
=\oint_{|\zeta|^n=r_0}{d\zeta\over2\pi\i\zeta}\,\d_\zeta^{k-1}{1\over\eta-\zeta}
=(-1)^k\d_\eta^{k-1}\oint_{|\zeta|^n=r_0}{d\zeta\over2\pi\i\zeta}\,{1\over\eta-\zeta}
\\
=(-1)^{k-1}\d_\eta^{k-1}{\Theta(|\eta|^n-r_0)\over\eta},
\end{multline*}
where $\Theta(x)$ is the Heaviside step function. The same is valid for $\d_\bzeta^k$ with the substitution $\eta\to\bar\eta$ in the last expression. Applying this to the $\zeta_2$ we see that the terms containing $\log|\zeta_2|^2$ in (\ref{correxp-varphi-varphi(1)}) vanish and there is a regular limit $r_{01}\to0$. Then after setting $\zeta_2=0$ we see that the terms that contain $\log|\zeta_1|^2$ vanish too. We obtain
\begin{align}
\langle\cO'(\infty)\cO(0)\rangle^{{(1)}}
&=-2b^2\mu'\int d^2\eta\,|\eta|^{2(n-1)(1+b^2)-4\alpha b}
\,\d_{\bar\eta}^{k-1}{\Theta(|\eta|^n-r_0)\over\bar\eta}\cdot\d_\eta^{k-1}{1\over\eta}
\notag
\\
&=-2b^2\mu'r_0^{2\delta^{(1)}}I_k^{(1)}({(n-1)(1+b^2)-2n\nu}).
\label{corrOO(1)-I(1)-rel}
\end{align}
Here
\begin{equation}
\delta^{(1)}=-\tk+1-2\nu+b^2(1-n^{-1})=\delta^{(1,0)}_{\nu,-2\tk}+b^2(1-n^{-1})
\label{delta(1)-def}
\end{equation}
and
\begin{equation}
I_k^{(1)}(\rho)=\int d^2\eta\,|\eta|^{2\rho}\,\d_{\bar\eta}^{k-1}{\Theta(|\eta|-1)\over\bar\eta}\cdot\d_\eta^{k-1}{1\over\eta}.
\label{I(1)-def}
\end{equation}
The derivative $\d_\eta^{k-1}\eta^{-1}$ should be taken directly, while the derivative $\d_{\bar\eta}^{k-1}$ can be taken by parts:
\begin{equation}
I^{(1)}_k(\rho)=(k-1)!\int_{|\eta|\ge1}d^2\eta\,\eta^{\rho-k}\bar\eta^{-1}\,\d_{\bar\eta}^{k-1}\bar\eta^\rho
={\pi\,(k-1)!\,(\rho)_{k-1}\over k-1-\rho}.
\label{I(1)-fin}
\end{equation}
Finally, we have
\begin{equation}
\left\langle\cO'(\infty)\cO(0)\right\rangle^{{(1)}}
=2\pi\,(k-1)!\,(k-1+n\delta^{(1)})_{k-1}{b^2\mu'r_0^{2\delta^{(1)}}\over n\delta^{(1)}}.
\label{corrOO(1)-fin}
\end{equation}
In the leading order of $b^2$ we have $\delta^{(1)}=\delta^{(1,0)}_{\nu,-2\tk}$:
\begin{equation}
\left\langle\cO'(\infty)\cO(0)\right\rangle^{{(1)}}
=2\pi\,(k-1)!\,(n(1-2\nu)-1)_{k-1}{b^2\mu'r_0^{2\delta^{(1,0)}_{\nu,-2\tk}}\over n\delta^{(1,0)}_{\nu,-2\tk}}.
\label{corrOO(1)-fin-b0}
\end{equation}
The exponent and the position of the pole coincide with those of (\ref{kk-noren}) in the term with $s=1$, $s'=0$.

Now let us calculate the second order contribution:
\begin{align}
\langle\cO'(\infty)\cO(0)\rangle^{{(2)}}
&={\mu^2\over8}\int d^2 y_1\,d^2 y_2\,\left\langle\cT^+_n\!\left(\e^{-(\alpha+2b)\varphi(\infty)}\right)\e^{b\varphi(y_1)}\e^{b\varphi(y_2)} \average{\bd^k\varphi}_{r_0}\average{\d^k\varphi}_{r_{01}}\cT_n\!\left(\e^{\alpha\varphi(0)}\right)\right\rangle^{(0)}
\notag
\\
&={\mu^{\prime\,2}\over8}\oint_{|\zeta_1|^n=r_0}{d\zeta_1\over2\pi\i\zeta_1}\oint_{|\zeta_2|^n=r_{01}}{d\zeta_2\over2\pi\i\zeta_2}
\int d^2\eta_1\,d^2\eta_2\,|\eta_1|^{2(n-1)(1+b^2)}|\eta_2|^{2(n-1)(1+b^2)}
\notag
\\
&\quad\times
\left\langle\e^{-(\alpha+2b)\varphi(\infty)}\e^{b\varphi(\eta_1,\bar\eta_1)}\e^{b\varphi(\eta_2,\bar\eta_2)}
\,\bd^k\varphi(x_1)\,\d^k\varphi(x_2)\, \e^{\alpha\varphi(0)}\right\rangle^{(0)}_\text{plane}
\label{corrOO(2)-def}
\end{align}
In the correlation function on the plane in the r.h.s.\ only four terms given by
$$
4b^2|\eta_1|^{-4\alpha b}|\eta_2|^{-4\alpha b}|\eta_1-\eta_2|^{-4b^2}\left(\log|\eta_1-\zeta_1|^2+\log|\eta_2-\zeta_1|^2\right)
\left(\log|\eta_1-\zeta_2|^2+\log|\eta_2-\zeta_2|^2\right)
$$
contribute to the integral so that
\begin{multline*}
\langle\cO'(\infty)\cO(0)\rangle^{{(2)}}
={b^2\mu^{\prime\,2}\over2}\int d^2\eta_1\,d^2\eta_2\,|\eta_1\eta_2|^{2(n-1)(1+b^2)-4n\nu}\,|\eta_1-\eta_2|^{-4b^2}
\\
\times
\left(\d_{\bar\eta_1}^{k-1}{\Theta(|\eta_1|^n-r_0)\over\bar\eta_1}+\d_{\bar\eta_2}^{k-1}{\Theta(|\eta_2|^n-r_0)\over\bar\eta_2}\right)
\left(\d_{\eta_1}^{k-1}{\Theta(|\eta_1|^n-r_{01})\over\eta_1}+\d_{\eta_2}^{k-1}{\Theta(|\eta_2|^n-r_{01})\over\eta_2}\right).
\end{multline*}
Again, we may take the limit $r_{01}\to0$ and obtain
\begin{equation}
\langle\cO'(\infty)\cO(0)\rangle^{{(2)}}={b^2\mu^{\prime\,2}\over2}r_0^{2\delta^{(2)}}I_k^{(2)}((n-1)(1+b^2)-2n\nu,-2b^2),
\label{corrOO(2)-int2}
\end{equation}
where
\begin{equation}
\delta^{(2)}=-\tk+2(1-2\nu)+2b^2(1-2n^{-1})=\delta^{(2,0)}_{\nu,-2\tk}+2b^2(1-2n^{-1})
\label{delta(2)-def}
\end{equation}
and
\begin{equation}
I_k^{(2)}(\rho,\sigma)
=2\int d^2\eta_1\,d^2\eta_2\,|\eta_1|^{{2\rho}}|\eta_2|^{{2\rho}}|\eta_1-\eta_2|^{2\sigma}
\left(\d_{\eta_1}^{k-1}{1\over\eta_1}+\d_{\eta_2}^{k-1}{1\over\eta_2}\right)\d_{\bar\eta_1}^{k-1}{\Theta(|\eta_1|^2-1)\over\bar\eta_1}.
\label{Ik(2)-def}
\end{equation}
Taking the integral by parts for $\d_{\bar\eta_1}^{k-1}$ we obtain
\begin{align}
I_k^{(2)}(\rho,\sigma)
&=2{\,(k-1)!}\int_{|\eta_1|\ge1}d^2\eta_1\,d^2\eta_2\,(\eta_1^{-k}+\eta_2^{-k})\bar\eta_1^{-1}
\,\bd_1^{k-1}\left(|\eta_1\eta_2|^{2\rho}|\eta_1-\eta_2|^{2\sigma}\right)
\notag
\\
&=2{\,(k-1)!^2}\int_{|\eta_1|\ge1}d^2\eta_1\,d^2\eta_2\,|\eta_1\eta_2|^{2\rho}|\eta_1-\eta_2|^{2\sigma}(\eta_1^{-k}+\eta_2^{-k})
\notag
\\
&\quad\times
\sum^{k-1}_{l=0}{(\rho)_{k-l-1}(\sigma)_l\over(k-l-1)!\,l!}\bar\eta_1^{l-k}(\bar\eta_1-\bar\eta_2)^{-l}
\label{I(2)-expansion}
\end{align}
Let us change the variables $(\eta_1,\eta_2)\to(\eta_1,u=\eta_2/\eta_1)$. In these variables the integral over $\eta_1$ factors out and is easily taken. We obtain
\begin{equation}
I_k^{(2)}(\rho,\sigma)
={2\pi{\,(k-1)!^2}\over k-2-2\rho-\sigma}\sum_{l=0}^{k-1}{(\rho)_{k-1-l}(\sigma)_l\over(k-l-1)!\,l!}
\left(J_{0,l}(\rho,\sigma)+J_{k,l}(\rho,\sigma)\right),
\label{Ik(2)-J-rel}
\end{equation}
where
\begin{equation}
J_{k,l}(\rho,\sigma)=\int d^2u\,|u|^{2\rho}|1-u|^{2\sigma}u^{-k}(1-\bar u)^{-l}
\label{Jint-def}
\end{equation}
By using the technique of \cite{Dotsenko:1988lectures} we easily obtain
\begin{equation}
J_{k,l}(\rho,\sigma)
=2\pi{\Gamma(1+\rho)\Gamma(1+\sigma-l)\Gamma(k-1-\rho-\sigma)\over\Gamma(k-\rho)\Gamma(-\sigma)\Gamma(2-l+\rho+\sigma)}
={(1+\rho+\sigma)_k(1+\rho+\sigma)_l\over(\rho)_k(\sigma)_l}J_{0,0}(\rho,\sigma).
\label{Jint-fin}
\end{equation}
Substituting it into (\ref{Ik(2)-J-rel}) we obtain
$$
I_k^{(2)}(\rho,\sigma)
={2\pi\,(k-1)!\,J_{0,0}(\rho,\sigma)\over k-2-2\rho-\sigma}\left(1+{(1+\rho+\sigma)_k\over(\rho)_k}\right)
\sum_{l=0}^{k-1}{(\rho)_{k-1-l}(1+\rho+\sigma)_l\over(k-l-1)!\,l!}.
$$
The sum in the r.h.s.\ can be taken and we get
\begin{equation}\label{order2-v2}
I_k^{(2)}(\rho,\sigma)=2\pi\,(k-1)!\,{(1+2\rho+\sigma)_{k-1}\over k-2-2\rho-\sigma}
\left(1+{(1+\rho+\sigma)_k\over(\rho)_k}\right)J_{0,0}(\rho,\sigma).
\end{equation}

We have for the second correction to the correlation function
\begin{multline}
\langle\cO'(\infty)\cO(0)\rangle^{(2)}
=-{4\pi^2(k-1)!\,b^4\mu^{\prime\,2}r_0^{2\delta^{(2)}}\over n\delta^{(2)}}
(k-1+n\delta^{(2)})_{k-1}R_k
\\
\times
{\Gamma(1-2b^2)\Gamma\left({1\over2}(k+n\delta^{(2)})+b^2\right)\Gamma\left(b^2-{1\over2}(k+n\delta^{(2)})\right)
\over\Gamma(1+2b^2)\Gamma\left(1-{1\over2}(k+n\delta^{(2)})-b^2\right)\Gamma\left(1+{1\over2}(k+n\delta^{(2)})-b^2\right)},
\label{corrOO(2)-fin}
\end{multline}
where
\eq$$
R_k=1+(-1)^k{\left({1\over2}(k+n\delta^{(2)})-b^2\right)_k\over\left({1\over2}(k-n\delta^{(2)})-b^2\right)_k}.
\label{Rk-def}
$$
Consider the factor $R_k$. Notice that its denominator does not produce any poles in the correlation function, since they are canceled by the second gamma function in the denominator of (\ref{corrOO(2)-fin}). Furthermore, for odd values of $k$ it has a simple zero at $\delta^{(2)}=0$, so that the correlation function does not have a pole at $\delta^{(2)}=0$ in this case. For even values of $k$ the pole does not cancel exactly but for $|\delta|,\,b^2\ll1$ we obtain
$$
R_k\simeq -{n^2\delta^{(2)\,2}/k-2b^2\over n\delta^{(2)}/2+b^2}.
$$
The pole at $\delta^{(2)}=-2b^2/n$, as we noticed, is canceled in the correlation function, but the third gamma function in the numerator produces an extra pole at $\delta^{(2)}=2b^2/n$ in the vicinity. The two zeros at $\delta^{(2)}=\pm\sqrt{2k}b/n$ cancel the poles at $\delta^{(2)}=0,\,2b^2/n$ in the correlation function in the leading order in $b^2$. In both cases we get
$$
\left.R_k\right|_{b\to0}=-{2n\delta^{(2)}\over k-n\delta^{(2)}}={\delta^{(2,0)}_{\nu,-2\tk}\over\delta^{(1,0)}_{\nu,-2\tk}}.
$$

Finally, in the limit $b\to0$ we have
\begin{equation}
\langle\cO'(\infty)\cO(0)\rangle^{(2)}
={4\pi^2\,(k-1)!\,(2n(1-2\nu)-1)_{k-1}\over n^3(1-2\nu)^2}{b^4\mu^{\prime\,2}
r_0^{2\delta^{(2,0)}_{\nu,-2\tk}}\over\delta^{(1,0)}_{\nu,-2\tk}}.
\label{corrOO(2)-fin-b0}
\end{equation}
We see that in the leading order in $b^2$ the correlation function $\langle\cO'(\infty)\cO(0)\rangle$ with (\ref{cOcO-def}) has a pole at $\delta^{(1,0)}_{\nu,-2\tk}=0$ rather than at $\delta^{(2,0)}_{\nu,-2\tk}=0$ up to the second order of the conformal perturbation theory in consistency with (\ref{cC-ss-genform}) and (\ref{ck-coefs}).

Note that for a fixed $n$ the coefficients in (\ref{corrOO(1)-fin-b0}) and (\ref{corrOO(2)-fin-b0}) as functions of $\nu$, $k$ are proportional to $\beta_\nu^2\cC^{(1,0)}_{\nu,k}$ and $\beta_\nu^4\cC^{(2,0)}_{\nu,k}$ correspondingly. They do not have to coincide exactly, since we do not take into account the normalization of the operators accurately, but this proportionality deserves mention.


\providecommand{\href}[2]{#2}\begingroup\raggedright\endgroup

\end{document}